\documentclass[final,1p,sort&compress]{elsarticle}
\pdfoutput=1

\usepackage{natbib}
\usepackage{lineno}
\usepackage{amsmath,amsfonts,amssymb,amsthm}
\usepackage{hyperref} 
\usepackage{graphicx} 

\usepackage[english]{babel}

\usepackage[utf8]{inputenc}

\usepackage{bm}

\usepackage{relsize}

\usepackage{color}
\usepackage[usenames,dvipsnames,svgnames,table]{xcolor}

\usepackage[ruled]{algorithm}

\usepackage{bigstrut}

\usepackage{mathrsfs}

\usepackage{bibentry}
\newcommand{\ignore}[1]{}
\newcommand{\nobibentry}[1]{{\let\nocite\ignore\bibentry{#1}}}

\nobibliography*

\newcommand{\Par}[1]{\left(#1\right)}

\newcommand{\grad}{\nabla}

\newcommand{\vect}[1]{\mathbf{#1}}
\newcommand{\mat}[1]{\text{#1}}
\newcommand{\E}{\vect{E}} 

\newcommand{\dt}{\Delta t}
\newcommand{\dx}{\Delta x}
\newcommand{\dv}{\Delta v}
\newcommand{\eps}{\varepsilon}

\newcommand{\N}{\text{N}} 

\newcommand{\iu}{\text{j}}  

\makeatletter
\renewcommand*\env@matrix[1][\arraystretch]{%
  \edef\arraystretch{#1}%
  \hskip -\arraycolsep
  \let\@ifnextchar\new@ifnextchar
  \array{*\c@MaxMatrixCols c}}
\makeatother

\makeatletter
\providecommand{\doi}[1]{%
  \begingroup
    \let\bibinfo\@secondoftwo
      \urlstyle{rm}%
      \href{http://dx.doi.org/#1}{%
      doi:\discretionary{}{}{}%
      \nolinkurl{#1}%
}%
\endgroup
}
\makeatother

\newtheorem*{remark*}{Remark}

\journal{Journal of Computational Physics}

\begin{document}


\begin{frontmatter}

\title{Arbitrarily high order Convected Scheme solution \linebreak 
of the Vlasov-Poisson system}

\author[msu]{Yaman Güçlü \corref{cor}}
\ead{yguclu@math.msu.edu}

\author[msu,msu-e]{Andrew J. Christlieb}
\ead{christlieb@math.msu.edu}

\author[wisc]{William N.G. Hitchon}
\ead{hitchon@engr.wisc.edu}

\address[msu]{Department of Mathematics, 
Michigan State University,\\
619 Red Cedar Road, East Lansing, MI 48824, USA \\[0.3em]}

\address[msu-e]{Department of Electrical and Computer Engineering, 
Michigan State University,\\ 
428 South Shaw Lane, East Lansing, MI 48824, USA \\[0.3em]}

\address[wisc]{Department of Electrical and Computer Engineering, 
University of Wisconsin, \\
1415 Engineering Dr., Madison, WI 53706, USA \\[-3em]}

\begin{abstract}
The Convected Scheme (CS) is a `forward-trajectory' semi-Lagrangian meth\-od 
for solution of transport equations, which has been most often applied to the 
kinetic description of plasmas and rarefied neutral gases.
In its simplest form, the CS propagates the solution forward in time by 
advecting the so called `moving cells' along their characteristic 
trajectories, and by remapping them on the mesh at the end of the time step.
The CS is conservative, positivity preserving, simple to implement, and it is 
not subject to time step restriction to maintain stability.
Recently [\bibentry{Guclu2012}] a new methodology was introduced for reducing
numerical diffusion, based on a modified equation analysis: the remapping error
was compensated by applying small corrections to the final position of the 
moving cells prior to remapping.
While the spatial accuracy was increased from 2nd to 4th order, the new scheme 
retained the important properties of the original method, and was shown to be 
extremely simple and efficient for constant advection problems.

Here the CS is applied to the solution of the Vlasov-Poisson system, which 
describes the evolution of the velocity distribution function of a collection 
of charged particles subject to reciprocal Coulomb interactions.
The Vlasov equation is split into two constant advection equations, one in 
configuration space and one in velocity space, and high order time accuracy is 
achieved by proper composition of the operators.
The splitting procedure enables us to use the constant advection solver, 
which we extend to arbitrarily high order of accuracy in time and space: a new 
improved procedure is given, which makes the calculation of the corrections 
straightforward.
Focusing on periodic domains, we describe a spectrally accurate scheme based 
on the fast Fourier transform; the proposed implementation is strictly 
conservative and positivity preserving.
The ability to correctly reproduce the system dynamics, as well as resolving 
small-scale features in the solution, is shown in classical 1D-1V test cases, 
both in the linear and the non-linear regimes.

\end{abstract}


\cortext[cor]{Corresponding author.}

\end{frontmatter}


\section{Introduction}
	\label{sec:Introduction}

This paper addresses an approach to solving kinetic equations, which include
the Boltzmann equation, the Vlasov equation and the Fokker-Planck equation,
by means of a method which (in a terminology which has come into being since 
this method was developed) is a forward semi-Lagrangian scheme.

The kinetic equation and particularly its numerical solution have been 
discussed at length by us in~\cite{Hitchon1999}.
A detailed history of methods for solving the Vlasov equation is also given 
in~\cite{Rossmanith2011}, whose citation list covers algorithms from early 
particle methods~\cite{Hockney1981,Birdsall1988}, early semi-Lagrangian 
methods~\cite{Cheng1976}, and their evolution up to the year 2011.
Among the more recent developments, we point out some general trends that are 
common to both particle-in-cell (PIC) and mesh-based (or discrete-velocity) 
algorithms: energy conserving implicit 
formulations~\cite{Markidis2011,Chen2011,Chacon2013,Chen2014,Cheng2014}, 
and phase-space adaptivity~\cite{Wang2012,Innocenti2013,Hittinger2013}.

The kinetic equation employs seven independent variables (three space, three 
velocity and time), although it is certainly possible to employ a subset of 
these. 
Solution in this large domain can be computationally challenging, even in a 
reduced phase space.
The very wide difference in time scales involved adds to the burden.
Velocity (or some equivalent quantity) is always an independent variable in 
the problem.
The range of velocities which are required for meaningful plasma simulations 
causes problems for the choice of a mesh, if a mesh is used - good resolution 
is required over many orders of magnitude.
Further, the huge velocity range means some particles move a tiny distance in 
the physically relevant times, yet others move a distance comparable to or 
exceeding the system size.

These concerns can severely hamper mesh-based schemes, as can the fact that it 
can be difficult to achieve conservation of particles and energy, locally in 
phase space in these schemes - or even to maintain positivity of the solution.
On the other hand, the need for adequate statistics throughout all of velocity 
space and physical space is a challenge for fully Lagrangian (or `particle') 
approaches.
(For example, typically only a tiny fraction of the particles is to be found 
in the region of velocity space which accounts for ionization.)

The Convected Scheme (CS) is a method for solving conservation equations and 
characteristic equations, in the context of kinetic and fluid descriptions of 
time evolution of particle density.
The basic concept is to take an initial cell which `contains' (phase space or 
physical space) fluid, and allow the cell to convect with the flow experienced 
by the fluid.
After a time step, the location of such cells, relative to the fixed mesh into 
which the independent variables are divided, defines how the density has 
evolved during the time step.
If a moved cell overlaps a cell of the fixed mesh so that a fraction $F$ 
of the moved cell lies within the mesh cell,  then a fraction $F$ of the 
density associated with the moved cell will be put into that mesh cell.
The use of (positive) fractions makes the CS conservative and 
positivity preserving.
The CS is relatively simple to implement, and is not subject to any time step 
restrictions except those arising from the need to integrate the physical 
equations of motion and to resolve the evolution of the shape of the cell.
In particular the CS is not subject to the Courant criterion.

Recently the CS has been made high order accurate by means of small 
corrections to the displacement of the moved cell, or equivalently the flow 
velocity~\cite{Guclu2012}.
In this paper we show how the CS for a constant physical flow velocity can be 
made arbitrarily high order, easily and efficiently, while preserving the 
desirable features of the CS.
In an operator-splitting framework, kinetic equations such as the Vlasov or 
Boltzmann equations may be solved by combining `simple' advection solvers
at constant velocity; this is indeed what is done here.

As discussed in Section~\ref{sec:VlasovPoisson}, it is desirable that a 
solution of the Vlasov-Poisson system be efficient, high order, conservative 
and positivity preserving.
Although some authors have argued in the past that maintaining a positive 
solution may be less critical than providing a dissipation mechanism for fine 
`filamentation' structures~\cite{Arber2002}, we believe that positivity 
preservation is of paramount importance in conditions far from 
quasi-neutrality (e.g.\ plasma sheaths), as well as in the presence of 
collisions (Boltzmann-Poisson system).
Other properties that are popular among shock-capturing schemes, ranging from 
total variation boundedness to monotonicity preservation to min/max 
preservation, do not appear to be as valuable in this context.

Indeed, the high-order CS described in this paper defines a family of constant 
advection solvers that are strictly conservative and positivity-preserving.
For a given order of accuracy, our framework provides a closed form for the 
high order corrections in terms of higher spatial derivatives of the solution, 
and a unique positivity-preserving limiter.
The means of approximating such derivatives lead to different implementations 
of the high-order CS, with different dissipation and dispersion errors.
In order to address the issue of filamentation, we present numerical schemes 
having high-order diffusion as their leading truncation error.
As future work we plan to select `optimal' implementations of the high-order 
CS, and compare them to the state-of-the-art advection solvers for the 
Vlasov-Poisson system, building upon the work of Arber and 
Vann~\cite{Arber2002}, and of Filbet and Sonnendrücker~\cite{Filbet2003}.

As we investigate very high order of accuracy in phase-space, the effect of 
time accuracy on the final solution becomes more easily observable.
For non-collisional kinetic equations in periodic domains, high order 
symplectic time splitting can be easily implemented: following the example of 
Crouseilles, Faou and Mehrenberger~\cite{Crouseilles2011}, we explore the use 
of 4th and 6th-order optimized Runge-Kutta-Nyström time integrators by Blanes 
and Moan~\cite{BlanesMoan2002} in our simulations.
For moderate to high levels of accuracy, these schemes achieve much higher 
efficiency than Strang splitting; in combination with a spectrally accurate CS 
implementation, they allow us to pursue machine precision on some classical 
test-cases.

The paper is laid out as follows.
In Section~\ref{sec:VlasovPoisson} the Vlasov-Poisson system is presented, and 
its numerical challenges are discussed.
In Section~\ref{sec:OperatorSplitting}, the operator splitting procedure for 
the Vlasov-Poisson system is reviewed.
In Section~\ref{sec:HighOrderCS}, the high order Convected Scheme is derived.
In Section~\ref{sec:NumericalTests} the scheme is applied to 1D and 2D 
advection test-cases, to the solution of the 1D-1V linear Vlasov equation, 
and to the 1D-1V solution of the Vlasov-Poisson system.
Finally, in Section~\ref{sec:Conclusions} we present some conclusions and 
offer a path for future work.

\section{The Vlasov-Poisson system: background and numerical challenges}
	\label{sec:VlasovPoisson}
\newcommand{\gradx}{\grad_{\!\vect{x}}}
\newcommand{\gradv}{\grad_{\!\vect{v}}}

The Vlasov-Poisson system describes the time evolution of the velocity 
distribution functions $f_s(t,\vect{x},\vect{v})$ of S charged species subject 
to mutual electrostatic interactions, according to 
\begin{subequations}\label{eq:VlasovPoisson}
\begin{align}
\label{eq:VlasovPoisson.a}
&\Par{\partial_t + \vect{v} \cdot \gradx 
     -\frac{q_s}{m_s}\gradx\phi \cdot \gradv}
     f_s\!\Par{t,\vect{x},\vect{v}} \,=\, 0, \qquad (s = 1,\dots,\text{S}) \\
\label{eq:VlasovPoisson.b}
&\gradx^2\phi\Par{t,\vect{x}} \,=\, -\frac{1}{\eps_0} \sum_{s=1}^\text{S} q_s 
     \int_{\mathbb{R}^3} f_s\!\Par{t,\vect{x},\vect{v}} d\vect{v},
\end{align}
\end{subequations}
with given initial conditions and subject to appropriate boundary conditions.
In~\eqref{eq:VlasovPoisson}, the symbol $\partial_t$ represents a partial time 
derivative, while $\gradx$ and $\gradv$ are gradients taken with respect to 
the configuration variable $\vect{x}$ and the velocity variable $\vect{v}$, 
respectively.
Moreover, $\phi(t,\vect{x})$ is the electrostatic potential [V], $\eps_0$ is 
the vacuum permittivity [F/m], and $q_s$ and $m_s$ are the charge [C] and mass 
[kg] of a single particle of species $s$. 

It should be noted that, while the Vlasov equations~\eqref{eq:VlasovPoisson.a} 
constitute a set of hyperbolic conservation laws that we intend to step in 
time, the Poisson equation~\eqref{eq:VlasovPoisson.b} is purely elliptical and 
it depends on the instantaneous values of the first moment of the distribution 
functions.
This makes~\eqref{eq:VlasovPoisson} a system of non-linear integro-partial 
differential equations.

In a fully ionized plasma obtained from a single atomic species, often only 
two charged species are present, i.e.\ electrons and singly charged positive 
ions.
If one is interested in the electron dynamics only, which is much faster than 
the ion dynamics because of $m_e \ll m_i$, then the ions may be `frozen' and 
treated as a uniform neutralizing background with density $n_0$.
In such a situation the Vlasov-Poisson system~\eqref{eq:VlasovPoisson} is 
reduced to
\begin{subequations}\label{eq:VlasovPoisson-e}
\begin{align}
\label{eq:VlasovPoisson-e.a}
&\Par{\partial_t + \vect{v} \cdot \gradx 
     -\frac{q_e}{m_e}\gradx\phi \cdot \gradv}
     f_e\!\Par{t,\vect{x},\vect{v}} \,=\, 0, \\
\label{eq:VlasovPoisson-e.b}
&\gradx^2\phi\Par{t,\vect{x}} \,=\, \frac{q_e}{\eps_0}
     \left[
     n_0 - \int_{\mathbb{R}^3} f_e\!\Par{t,\vect{x},\vect{v}} d\vect{v} 
     \right].
\end{align}
\end{subequations}
Despite the fact that the large time-scale separation between electrons and 
ions has now been removed, obtaining a mesh-based numerical solution to 
\eqref{eq:VlasovPoisson-e} remains a challenging task:
\begin{itemize}
\item High dimensionality (up to six dimensions) may easily lead to an 
      unmanageable number of phase-space cells when the mesh is refined; for 
      smooth solutions, coarser meshes can be effective if \emph{higher order} 
      methods are employed;
\item High characteristic speeds for the far ends of the velocity mesh 
      restrict the maximum allowed time step of standard explicit schemes; 
      \emph{semi-Lagrangian} methods are preferable, as they are not subject 
      to the CFL limit;
\item The exact solution to the system~\eqref{eq:VlasovPoisson-e} has 
      infinitely-many invariants, and it is impossible to match them all in a 
      numerical solution; nevertheless, the numerical scheme should be 
      \emph{charge-conservative}, in order to avoid numerical instabilities 
      driven by the Poisson solve;
\item By definition, the velocity distribution function $f_e\!\Par{t,\vect{x},
      \vect{v}}$ is non-negative, and initial conditions necessarily satisfy this 
      property; to maintain $f_e \ge 0$ under all circumstances, the numerical 
      scheme should be \emph{positivity-preserving}.
\end{itemize}
These considerations suggest the need for an efficient semi-Lagrangian scheme 
that is high order, charge-conservative, and positivity-preserving.
Not surprisingly, constructing such a method in a fully six-dimensional 
phase-space is not a trivial task.
Noting that high order accuracy in time is not among the requirements above, 
one may think of simplifying the problem by some form of operator splitting 
between configuration space (the $\vect{x}$ coordinates) and velocity space 
(the $\vect{v}$ coordinates).
At the price of introducing a time-splitting error, this would allow one to 
construct much simpler schemes.

\section{Operator splitting for the Vlasov-Poisson system}
    \label{sec:OperatorSplitting}

Given the considerations presented in Section~\ref{sec:VlasovPoisson}, we 
intend to design a semi-Lagrangian scheme that is high order, 
charge-conservative, and positivity-preserving.
This is achieved by combining an operator splitting procedure, reviewed in 
this section, with an \emph{arbitrarily} high-order CS for constant advection, 
which we will derive in Section~\ref{sec:HighOrderCS}.
The resulting scheme is fundamentally different from earlier low-order 
versions of the CS, which were applied to the motion in the full phase 
space~\cite{Hitchon1989,Parker1993,Hitchon1994,Bretagne1994,
Matsunaga1999,Christlieb2000,Feng2000}.

In Section~\ref{sec:OperatorSplitting.Strang} we describe the classical 
solution procedure by Cheng and Knorr~\cite{Cheng1976}, based on 2nd-order 
Strang splitting.
In Section~\ref{sec:OperatorSplitting.ODEs-review} we give a brief overview of 
higher order split ODE integrators that are suited to the solution of 
Hamiltonian systems, with the primary goals of clarifying the terminology and 
pointing out the fundamental similarities between these methods.
In Section~\ref{sec:OperatorSplitting.VP-HighOrder} we review the few 
implementations of high-order splitting for the Vlasov-Poisson system that 
have been carried out to date, and we discuss the subtleties connected to the 
electric field update.

\subsection{Strang splitting for Vlasov-Poisson}
\label{sec:OperatorSplitting.Strang}
The operator splitting strategy was first applied to  
the Vlasov-Poisson system by Cheng and Knorr in 1976~\cite{Cheng1976}, who time 
advanced the solution by means of Strang splitting~\cite{Strang1968}:
\begin{enumerate}\itemsep3pt
\item $\Delta t/2$ step on $\Par{\partial_t + \vect{v}\cdot\gradx} f_e = 0$;
\item Compute $n_e = \int_{\mathbb{R}^3} f_e\, d\vect{v}$,
      solve $\gradx^2\phi = \frac{q_e}{\eps_0}\Par{n_e-n_0}$,
      and evaluate $\vect{E} = -\gradx{\phi}$;
\item $\Delta t$ step on
      $(\partial_t + \frac{q_e}{m_e}\vect{E}\cdot\gradv) f_e = 0$;
\item $\Delta t/2$ step on $\Par{\partial_t + \vect{v}\cdot\gradx} f_e = 0$.
\end{enumerate}
Clearly, steps 1 and 4 constitute a constant advection equation along 
$\vect{x}$, for each point in phase-space, with an advection velocity 
$\vect{v}$ that does not depend on the independent variables $t$ and 
$\vect{x}$.
Similarly, step~3 is a constant advection equation along $\vect{v}$, for each 
point in phase space, with an advection velocity 
$\frac{q_e}{m_e}\vect{E}(\vect{x})$ that does not depend on the independent 
variables $t$ and $\vect{v}$.

If the advection schemes are at least second order accurate in time, the 
overall procedure is second order accurate as well.
Further, Rossmanith and Seal~\cite{Rossmanith2011} pointed out that the 
intermediate electric field $\vect{E}$ is also second order accurate in time, 
despite being computed after advection in the $\vect{x}$ variables only.

In their pioneering work~\cite{Cheng1976}, Cheng and Knorr implemented the 
advection steps by tracing the characteristics backward in time, and 
interpolating the values from the grid using either cubic splines or 
trigonometric polynomials.
An exponential low-pass filter was employed to stabilize the solution.
The resulting scheme was conservative, but not positivity-preserving.

Over the past 15 years, much work on semi-Lagrangian solvers using the same 
Strang splitting procedure, and with a variety of spatial discretizations, has 
been carried out by Sonnendrücker and his collaborators (see for 
example~\cite{Sonnendrucker1999,Besse2003,Gutnic2004,Crouseilles2010}).
High order semi-Lagrangian methods with Strang splitting in time were 
proposed in several recent works: in 2010 Qiu and Christlieb~\cite{Qiu2010} 
used WENO (weighted essentially non-oscillatory) reconstruction to obtain a 
conservative shock-capturing scheme; in 2011 Rossmanith and 
Seal~\cite{Rossmanith2011} and Qiu and Shu~\cite{Qiu2011} used a discontinuous 
Galerkin (DG) representation of the solution in phase-space.
Further, in 2012 Seal~\cite{Seal2012_Thesis} proposed a `hybrid' scheme also 
based on Strang splitting, where an Eulerian DG method in configuration space 
(with multi-rate Runge-Kutta time stepping) was combined with semi-Lagrangian 
DG in velocity space.
(The rationale for this approach lies in the ability of Runge-Kutta DG to 
handle complex geometries.)
A similar hybrid scheme was presented in 2013 by Guo and Qiu~\cite{Guo2013}, 
who employed semi-Lagrangian WENO in velocity space instead of DG.

%
%

In the aforementioned works, Strang splitting has proven to be successful due 
to its implementation simplicity, and to its peculiar geometric properties.
In fact, in the limit of an exact solution of the advection stages (which 
essentially means negligible remapping error), Strang splitting preserves the 
symplectic nature of the Poisson structure of the Vlasov-Poisson system and 
hence it is a \emph{symplectic} time integrator~\cite{Hairer2006}.
Accordingly, the resulting numerical flow in phase-space is 
\emph{incompressible}.
The total energy of the system, which coincides with the Hamiltonian 
functional
\begin{equation}\label{eq:VP-Hamiltonian}
  H[f_e](t)\ =\ \int_{\vect{x}\in\mathbb{R}^3} \int_{\vect{v}\in\mathbb{R}^3} 
  \frac{m_e v^2}{2} f_e(t,\vect{x},\vect{v}) \,d\vect{x}\, d\vect{v}\ +\
  \int_{\vect{x}\in\mathbb{R}^3} 
  \frac{\eps_0}{2}\left| \E[f_e](t,\vect{x}) \right|^2 d\vect{x}\, ,
\end{equation}
is not strictly conserved, but its variations in time remain bounded and do not 
show exponential growth or decay typical of non-symplectic algorithms.
Therefore, we can say that Strang splitting is \emph{energy-stable}.
Other important properties of Strang splitting in this limiting case are 
\emph{time reversibility} and the exact conservation of 
\emph{linear and angular momentum}.

\subsection{A review of high order splitting methods for ordinary differential equations}
\label{sec:OperatorSplitting.ODEs-review}
In the same way that Strang splitting was applied to the Vlasov-Poisson 
system, higher order splitting algorithms can be used.
As will be shown in Section~\ref{sec:OperatorSplitting.VP-HighOrder}, such 
algorithms can be derived from standard splitting methods for ODEs, applied to 
the equation of motion of a characteristic trajectory in phase-space.
For this reason, in this section we review high-order splitting methods for 
ODEs, which are good candidates for the time integration of the Vlasov-Poisson 
system.

\subsubsection{Composition methods}
The natural way to increase the order of a splitting method is the use of a 
\emph{composition} technique, which consists of the repeated application of 
(usually) one base scheme within the time step, and its adjoint, using 
positive and negative substeps.
If we consider a generic ODE integrator $\Phi_h$, where the subscript 
$h$ indicates its parametric dependence on the time-step size 
$h \equiv \Delta t$, then an $s$-stage composition method is of the form 
\begin{equation}\label{eq:composition}
  \Psi_h \,=\, \Phi_{\gamma_s h} \circ \dots
               \Phi_{\gamma_2 h} \circ \Phi_{\gamma_1 h},
\end{equation}
where the real numbers $\gamma_i$, with $i=1,2,\dots,s$, define the size 
and sign of the fractional time-steps $\gamma_i h$.
(Consistency requires that $\sum_i \gamma_i = 1$).
A proper choice of the number of stages $s$ and of the coefficients $\gamma_i$ 
permits one to annihilate the leading error terms of the original scheme, and 
possibly minimize a combination of the higher order terms.

It should be noted that, given an ODE system $\dot{\vect{u}}=F(\vect{u})$, a 
splitting scheme may be interpreted as the composition of the exact evolution 
of a partition $F=F_1+F_2$ of the vector field $F$.
This close connection between composition and splitting methods was first 
formalized by McLachlan~\cite{McLachlan1995}.

Symmetric composition schemes, where $\gamma_{s-i}=\gamma_{i+1}$ $\forall i$, are particularly important because they maintain the structure of the original method: in particular, they preserve the time-reversibility and symplecticity of the base scheme when applied to Hamiltonian problems~\cite{Hairer2006}.
For example, we consider a 2nd-order self-adjoint ODE integrator $\Phi_h$; 
self-adjoint means that $(\Phi_h)^{-1} = \Phi_{-h}$, and implies that the 
integrator is symmetric and non-dissipative, with error terms of even order 
in $\Delta t$.
We can then apply a symmetric composition to the base scheme $\Phi_h$, to 
construct a self-adjoint method of even order 4, 6, and so on.
This approach was used to construct high order symplectic integrators based on compositions of the Störmer-Verlet algorithm.

In 1990 Yoshida~\cite{Yoshida1990} recursively applied the `triple-jump' composition to Störmer-Verlet to obtain schemes of arbitrary (even) order of accuracy.
Such a procedure minimizes the number of stages for a given order (3 for 4th-order, 7 for 6th-order, and so on), but it results in very large error constants.
Shortly afterwards, Suzuki~\cite{Suzuki1990} proposed a similar procedure, based on a 5-stage `fractal' composition scheme, that leads to integrators with smaller error constants, and ultimately much higher efficiency.
After a decade, McLachlan~\cite{McLachlan2002} optimized Suzuki's technique to minimize the principal error term for a given number of stages.
The resulting `optimal' integrators possess a large number of stages (19 for 4th-order), but higher efficiency.
By combining this approach with the processing technique by Blanes~\cite{Blanes1999}, which we describe in the next section, McLachlan could improve efficiency even further, and reduce the number of stages (13 for 6th order).

\subsubsection{Processing techniques}
In many circumstances, the efficiency of splitting and composition methods can be increased by using the `processing' technique, which was originally proposed in 1969 by Butcher for Runge–Kutta methods~\cite{Butcher1969}.
This consists of composing a given ODE integrator
$\psi_h$ (the `kernel') of order $\N$ with a `pre-processor' $\pi_h^{-1}$ and a `post-processor' $\pi_h$, so as to obtain a method $\hat{\psi}_h$ of order $\hat{\N} > \N$:
\begin{equation}\label{eq:processing}
  \hat{\psi}_h \,=\, \pi_h \circ \psi_h \circ \pi_h^{-1}.
\end{equation}
When such a composition exists, the method $\psi_h$ is said to be of `effective' order $\hat{\N}$.
The processing technique has been most successful in the context of geometric numerical integration~\cite{Hairer2006,Blanes2008}, where constant time-step sizes are widely employed.
In fact, if $\Delta t$ is not changed, $n$ consecutive steps of~\eqref{eq:processing} can be applied very efficiently as 
\begin{equation}\label{eq:processing.n-steps}
  \bigl(\hat{\psi}_h\bigr)^n \,=\, 
  \pi_h \circ \bigl(\psi_h\bigr)^n \circ \pi_h^{-1},
\end{equation}
where the pre-processor $\pi_h^{-1}$ is computed at the beginning of the integration, the kernel $\psi_h$ is applied once per time-step, and the post-processor $\pi_h$ is evaluated whenever output is required.
Thanks to~\eqref{eq:processing.n-steps}, very efficient processed methods $\hat{\psi}_h$ can be designed, where many of the order conditions are satisfied by the processor; as a result, the kernel $\psi_h$ will involve far fewer function evaluations than a conventional integrator~\cite{Blanes1999}.

Optimization of a processed method naturally leads to expensive processors: if intermediate results are frequently required, one can approximate the post-processor $\pi_h$ with a cheaper integrator $\tilde{\pi}_h$, which satisfies the order conditions but minimizes the workload rather than the truncation error~\cite{Blanes2004}.
Since the pre-processor $\pi_h^{-1}$ is not approximated, and the approximate post-processor $\tilde{\pi}_h$ is never used for advancing the solution, the local error introduced by this procedure is usually negligible.
Moreover, one can take the post-processing `offline' and store only low-order results: at a later time, either the full $\pi_h$ or the approximated $\tilde{\pi}_h$ can be used, depending on the need.

\subsubsection{Partitioned Runge-Kutta and Runge-Kutta-Nyström schemes}

The discussion so far has not taken into consideration the specific structure 
of the ODE equations to be integrated.
Not surprisingly, for some classes of ODE systems the number of order 
conditions is reduced, and/or the number of error terms to be minimized is 
smaller; in such cases, more efficient schemes can be designed.
Sometimes these `new' schemes coincide with other classes of ODE integrators.
We consider an autonomous Hamiltonian system
\begin{equation*}
  \left\{
  \begin{aligned}
    \frac{d\vect{q}}{dt} &= +\frac{\partial H}{\partial\vect{p}},\\
    \frac{d\vect{p}}{dt} &= -\frac{\partial H}{\partial\vect{q}},
  \end{aligned}
  \right.
\end{equation*}
where $(\vect{q},\vect{p})$ are the generalized coordinates (position, 
momentum), and $H(\vect{q},\vect{p})$ is the Hamiltonian of the system 
(i.e., in most cases, its total energy).
Because of its natural partitioning, we could integrate the system above by 
means of a partitioned Runge-Kutta (PRK) 
method~\cite{HairerNorsettWanner1993}, where two different Butcher tableaux 
are employed for $\vect{q}$ and $\vect{p}$.
Further, symplectic PRK schemes can be designed, provided that their 
coefficients satisfy certain constraints~\cite{AbiaSanzSerna1993}, but such 
methods are implicit in general.
Nevertheless, if the Hamiltonian is separable as 
\begin{equation*}
  H(\vect{q},\vect{p}) = V(\vect{q}) + T(\vect{p}),
\end{equation*}
where $V(\vect{q})$ and $T(\vect{p})$ are the potential and kinetic energy, 
respectively, then one can design PRK methods that are symplectic \emph{and} 
explicit.
It turns out that all such methods can be recast so that stage updates on 
$\vect{q}$ and $\vect{p}$ are interleaved, and each update only depends on the 
previous value.
If we also notice that each stage is solved exactly, because of the separable 
Hamiltonian, then we can conclude that \emph{all explicit symplectic PRK 
integrators are splitting methods}.

The importance of such a realization lays in the fact that one may use the 
theory of PRK methods, and in particular their order conditions and 
symplecticity constraints, to design symplectic time-splitting schemes of 
arbitrary order of accuracy.
In particular, if one admits a larger number of stages than what is strictly 
needed to satisfy the order conditions, then a thorough \emph{optimization} of 
such methods can be performed~\cite{BlanesMoan2002}.

If, in addition to being separable, the Hamiltonian has a quadratic kinetic 
energy $T(\vect{p}) = \frac{1}{2}\vect{p}^T \mat{M}\,\vect{p}$ with $\mat{M}$ 
a symmetric constant matrix, then 
the 1st-order ODE system can be recast as a 2nd-order one,
\begin{equation*}
  \frac{d^2\vect{q}}{dt^2} = 
  {-\mat{M}} \frac{\partial V}{\partial\vect{q}},
\end{equation*}
to which Runge-Kutta-Nyström (RKN) methods~\cite{HairerNorsettWanner1993} can 
be applied.
Similarly to the previous observation, one finds that \emph{explicit 
symplectic RKN schemes are time-splitting schemes}.
Since the number of order conditions for RKN schemes is lower than in the 
PRK case, for a given number of stages one can design a more accurate 
splitting method, or can achieve the same accuracy with a smaller number of 
stages~\cite{BlanesMoan2002}.

\subsection{High order splitting for Vlasov-Poisson}
\label{sec:OperatorSplitting.VP-HighOrder}

Despite their widespread use in the ODE community, the introduction of high 
order splitting methods to the simulation of the Vlasov-Poisson system is 
relatively recent, and it attracted interest only very slowly.
To the authors' knowledge, Watanabe and Sugama were the first researchers to 
evolve the Vlasov-Poisson system using high-order splitting methods: they 
presented preliminary results for 4th-order splitting in their 2001 pioneering 
work~\cite{Watanabe2001}, and they gave an extensive comparison between 
splitting of order 1, 2, 4 and 6 in their 2003 contribution to the 1st 
Vlasovia workshop (later reported in~\cite{Watanabe2005}).

Most importantly,~\cite{Watanabe2001} presented a simple theoretical connection 
between splitting methods for Hamiltonian ODEs (for which extensive theory 
already existed at the time) and time splitting for the Vlasov-Poisson system.
A similar discussion was independently presented in 2002 by Mangeney, Califano, 
Cavazzoni and Travnicek, who applied time splitting to the Vlasov-Maxwell 
system~\cite{Mangeney2002}.
The underlying idea is that applying a splitting scheme to the Vlasov equation 
is equivalent to applying the same splitting scheme to the integration of the 
Lagrangian trajectories.
In fact, according to the method of characteristics, the exact solution to the 
Vlasov equation is
\begin{equation*}
  f\Par{t,\vect{X}(t),\vect{V}(t)} \,=\, f\Par{0,\vect{x}_0,\vect{v}_0}, 
\end{equation*}
where $\Par{\vect{X}(t),\vect{V}(t)}$ is the phase-space trajectory originating 
at $\Par{\vect{x}_0,\vect{v}_0}$, which is determined by the Hamiltonian
\begin{equation}\label{eq:SingleParticleHamiltonian}
  h[f](\vect{q},\vect{p}) = \frac{\|\vect{p}\|^2}{2m} + q \phi[f](\vect{q}),
  \qquad
  \text{with $(\vect{q},\vect{p})=(\vect{x},m\vect{v})$,}
\end{equation}
and satisfies the equations of motion
\begin{equation*}
\begin{cases}
  \biggl.
  \dot{\vect{X}} = \vect{V} \\
  \dot{\vect{V}} = \dfrac{q}{m} \E[f]\Par{\vect{X}}
\end{cases}
\qquad \text{with} \quad \E(\vect{x}) = -\grad_\vect{x}\phi(\vect{x}).
\end{equation*}
Since the Hamiltonian~\eqref{eq:SingleParticleHamiltonian} is separable, a 
(symplectic) splitting method with $s$ stages for the equations of motion will 
be in the form
\begin{equation*}
\begin{split}
  \vect{X}_0 &= \vect{X}(t-\Delta t) \\
  \vect{V}_0 &= \vect{V}(t-\Delta t)
\end{split}
\qquad
\begin{split}
  \vect{X}_k &= \vect{X}_{k-1} + (a_k\Delta t) \vect{V}_{k-1} \\
  \vect{V}_k &= \vect{V}_{k-1} + (b_k\Delta t) \frac{q}{m} \E[f]\Par{\vect{X}_k}
\end{split}
\qquad
\begin{split}
  \vect{X}(t) &= \vect{X}_s \\
  \vect{V}(t) &= \vect{V}_s
\end{split}
\end{equation*}
where the non-dimensional coefficients $a_k$ and $b_k$ for each of the $s$ 
stages $(k=1,2,\dots,s)$ completely determine the numerical scheme.
For consistency, $\sum_k a_k \equiv 1$ and $\sum_k b_k \equiv 1$.

The substitution of the scheme above into the semi-Lagrangian relation 
\begin{equation*}
  f\Par{t,\vect{x},\vect{v}} = 
  f\Par{t-\Delta t,\vect{X}(t-\Delta t),\vect{V}(t-\Delta t)}, 
\end{equation*}
yields a splitting scheme for the Vlasov-Poisson equation
\begin{equation}\label{eq:splitting.VlasovPoisson}
\begin{cases}
  f_k^*(\vect{x},\vect{v}) \,=\, 
  f_{k-1}\Bigl(\vect{x} - (a_k\Delta t) \vect{v}, \vect{v} \Bigr) \\[1em]
  f_k(\vect{x},\vect{v}) \,=\, 
  f_k^*\Par{\vect{x}, \vect{v} - (b_k\Delta t) \dfrac{q}{m} \E[f_k^*](\vect{x})}
\end{cases}
\qquad (k=1,2,\dots,s)\, ,
\end{equation}
with $f_0(\vect{x},\vect{v}) = f(t-\Delta t,\vect{x},\vect{v})$ and 
$f(t,\vect{x},\vect{v}) = f_s(\vect{x},\vect{v})$.
We point out that the electrostatic field $\E$ must be recomputed after 
each $\vect{x}$-sweep.
A similar formulation was also implemented by Pohn, Shoucri and Kamelander in 
2005~\cite{Pohn2005}.

It should be noted that all the aforementioned works made use of the 
`triple-jump' 4th-order splitting 
method~\cite{CreutzGocksch1989,ForestRuth1990,Yoshida1990,Suzuki1990,CandyRozmus1991},
which is far less efficient than its modern variations and does not compete 
well with Strang splitting when a low accuracy level is required.
Given the modest results obtained, these pioneering experiments with 
high-order splitting did not attract much interest in the Vlasov-Poisson 
community and went almost forgotten until very recently.

In 2009 Schaeffer~\cite{Schaeffer2009} proposed an original 4th-order 
splitting for the linear Vlasov equation; this does not share the 
Hamiltonian structure of the Vlasov-Poisson system, as the electric field 
$\E(t,\vect{x})$ is not self-consistent and can have an arbitrary dependence 
on space and time.
Nevertheless, the extension of such an algorithm to the Vlasov-Poisson system 
may be possible.

In 2011 Rossmanith and Seal~\cite{Rossmanith2011} proposed a semi-Lagrangian 
solver based on the Discontinuous Galerkin formulation, which used the 
usual triple-jump 4th-order splitting but with a notable difference:
instead of re-evaluating the electric field after each advection step in 
configuration space (i.e.\ along $\vect{x}$), they extrapolated the electric 
field forward in time using knowledge of the moments of the distribution 
function.
Such a procedure, known as the `Cauchy-Kovalevsky' (CK) or `Lax-Wendroff' 
method~\cite{LaxWendroff1960}, was also investigated by Respaud and 
Sonnendrücker for application to \emph{unsplit} semi-Lagrangian 
methods~\cite{Respaud2011}, as it permits one to compute the characteristic 
trajectories very accurately by means of a truncated Taylor series, e.g.
\begin{equation*}
  \vect{X}(t+\Delta t) \,=\, \vect{X}(t) + \Delta t \vect{V}(t) + 
  \frac{\Delta t^2}{2}\frac{q}{m}\vect{E}(t,\vect{X}) + 
  \frac{\Delta t^3}{6}\frac{q}{m}\dot{\vect{E}}(t,\vect{X}) + 
  \dots
\end{equation*}
after which follows remapping (or conservative interpolation) in the full 
phase-space.

In~\cite{Rossmanith2011}, the exact time-average of a 4th-order time 
extrapolation of $\E$ was used for each of the advection steps in velocity 
space, and the resulting scheme was proven to be 4th-order accurate in time.
Such a modified scheme does not rely on splitting the Hamiltonian of the 
Vlasov-Poisson system, and because of the asymmetry introduced by the CK 
procedure, it is neither symplectic nor time reversible; nevertheless, this 
fact may go unnoticed in practice, if the error introduced by the remapping 
phase is predominant.
In general, the combination of a high-order splitting method with the CK 
procedure drastically reduces the required number of solutions of the Poisson 
equation, 
from one per stage to only one per time-step; nevertheless, it is unclear 
whether the computational cost of the algorithm is reduced, because of the 
need for calculating higher moments of the distribution function, which also 
adds considerable complication to the implementation.

The first detailed characterization of 4th-order splitting methods for the 
Vlasov-Poisson system was made in 2011 by Crouseilles, Faou and 
Mehrenberger~\cite{Crouseilles2011}.
After a discussion of the specific Poisson structure of the Vlasov-Poisson 
system, which is similar to that of Runge-Kutta-Nyström (RKN) systems, they 
obtained order conditions that are the same as RKN systems up to 4th order.
On this basis, they implemented the highly efficient RKN splitting schemes by 
Blanes and Moan~\cite{BlanesMoan2002}, and their numerical results showed 
dramatic improvement over the triple-jump 4th-order splitting.

Indeed, in the present work we follow the last example~\cite{Crouseilles2011}: 
we implement the splitting procedure~\eqref{eq:splitting.VlasovPoisson}, with 
coefficients $a_k$ and $b_k$ given by an optimized RKN method for 
ODEs~\cite{BlanesMoan2002}, and we recompute the electric field after each 
$\vect{x}$-sweep.

\section{Arbitrarily high order Convected Scheme for the constant advection equation}
	\label{sec:HighOrderCS}
In the previous section a review was given of splitting methods for advancing 
in time a numerical solution to the Vlasov-Poisson (VP) system.
Whether a low or high order splitting is employed, the fundamental stages of 
the solution procedure are `simple' constant advection steps, either in the 
configuration variables $\vect{x}$ or in the velocity variables $\vect{v}$.
If one opts for a Cauchy-Kowalewski procedure, a Taylor representation of the 
electric field is precomputed at the beginning of the time-step.
If one prefers a symplectic method based on Hamiltonian spitting, then the 
electric field simply needs to be recalculated after each $\vect{x}$-advection 
stage, by solving a Poisson's equation.

With the VP system in mind, in this section we design an arbitrarily high 
order Convected Scheme (CS) for the solution of the constant advection 
equation on a uniform mesh: the resulting algorithm has no time-step 
restriction (being semi-Lagrangian), it is mass conservative, and it is 
positivity preserving.
While this new scheme is not strictly monotonicity preserving, its inherent 
high-order numerical diffusion is able to effectively dissipate those 
filamentation features (typical of non-linear solutions to the VP system) 
which have decreased below the cell size, without introducing spurious 
oscillations.

We start by Taylor expanding the exact solution to the uniform advection 
equation, as a series of spatial derivatives, and we equate this series to 
an expanded form of the CS.
If the exact solution has at least $\N-1$ smooth derivatives, we obtain higher 
order corrections to the flow velocity by equating terms at each order.
Such corrections make the scheme capable of matching the exact solution, 
up to a local truncation error $O\Par{\Delta x^\N}$.
A recursion relation is given, as well as closed-form expressions, which 
are valid up to any order $\N$.
The resulting high order CS is conservative and it preserves the positivity of 
the solution.
For moderate order $\N$, we propose to approximate the required derivatives by 
means of central differences; for high $\N$ and periodic domains, we propose 
to use a filtered fast Fourier transform (FFT), which leads to a scheme with 
spectral convergence behavior.

\subsection{Semi-Lagrangian and finite-difference formulations: conservation and positivity}
\label{sec:HighOrderCS.formulations}

The Convected Scheme (CS) can be described as a Forward Semi-Lagrangian scheme,
although this terminology was not current when the scheme was first used.
The CS was built on the concept of a `moving cell' (MC), the fundamental 
vehicle of transport of mass during a time step $\Delta t$.
Basically, mass is `uploaded' from a mesh cell into a MC, which is then 
advected according to the flow field, and finally the MC `downloads' its mass
onto one or more contiguous mesh cells, according to a simple remapping rule 
that ensures mass conservation and positivity preservation.

Several variations on the CS exist, depending on the life span, trajectory 
integration, and shape evolution of a MC.
In all cases, the spatial density profile within a MC is just a constant 
function.
Extending our previous work~\cite{Guclu2012}, here we consider the simple 
`cell-centered semi-Lagrangian' version of the CS: the MC only exists within 
one time step, it does not change in shape, volume, or orientation, and only 
the trajectory of its center is tracked.

Such an algorithm is especially simple when applied to the 1D constant 
advection equation 
\begin{equation}\label{eq:ConstantAdvection}
    \Par{\frac{\partial}{\partial t} + u\, \frac{\partial}{\partial x} }
    n(t,x) = 0,
\end{equation}
where the density $n(t,x)$ is a function of time $t$ and of the spatial 
coordinate $x$, and $u$ is the (constant) advection velocity.

To simplify matters even further, we consider a spatial grid with uniform 
spacing $\Delta x$, and we let $x_i$ be the location of the center of 
cell~$\mathscr{C}_i$.
If we let $n_i^k \approx n(t_k,x_i)$ be the numerical solution at time instant 
$t_k$ and mesh point $x_i$, then the mass in cell $\mathscr{C}_i$ is 
simply~$n_i^k \Delta x$.
The CS update consists of the following simple semi-Lagrangian algorithm:
\begin{enumerate}
    \item Load mass $\{n_i^k\Delta x\}$ from the mesh cells 
          $\{\mathscr{C}_i\}$ into an array of moving cells (MCs);
    \item Transport each MC according to the trajectory of its center point, 
          to the location $X_i = x_i+u\Delta t$;
    \item Remap the mass $n_i^k\Delta x$ contained in each MC onto the fixed 
          mesh, according to the geometrical overlapping fractions.
\end{enumerate}

We now discuss the `area remapping rule' employed by the CS.
By normalizing the dimension $x$ with respect to the cell size~$\Delta x$,
we can represent the `shape' of both the MCs and the mesh cells as rectangular 
functions with unit height and width.
If we consider a MC at position $X_i$ originating from the mesh 
cell~$\mathscr{C}_i$, we can calculate its overlapping fraction over any 
target cell~$\mathscr{C}_j$ as 
\begin{equation}\label{eq:OverlappingFractions}
    F_{ij} =  \int_{-\infty}^{+\infty}
              \text{rect}\Par{\frac{x-X_i}{\Delta x}}\,
              \text{rect}\Par{\frac{x-x_j}{\Delta x}}\, \frac{dx}{\Delta x}
         \ =\ \mathcal{H}\Par{\frac{x_j-X_i}{\Delta x}},
\end{equation}
where $\text{rect}(z)$ is the rectangular function and $\mathcal{H}(z)$ is the 
hat function:
\begin{equation*}
    \text{rect}(z) :=
    \begin{cases}
        1           & \text{if $|z|<\frac{1}{2}$,}\\
        \frac{1}{2} & \text{if $|z|=\frac{1}{2}$,}\\
        0           & \text{otherwise;}
    \end{cases}
    \qquad \text{and} \qquad
    \mathcal{H}(z) :=
    \begin{cases}
        1-|z| & \text{if $|z|<1$,}\\
        0     & \text{otherwise.}
    \end{cases}
\end{equation*}

Notable properties of the hat function are positivity, i.e.\ 
$\mathcal{H}(z)\ge 0$ $\forall z$, and partition of unity, i.e.\ 
$\sum_i \mathcal{H}(i-z)=1$ for any~$z\in\mathbb{R}$.
Therefore, the overlapping fractions are positive and sum up to one:
\begin{equation}
    F_{ij} \ge 0 \quad \forall\ i,j
    \qquad \text{and} \qquad
    \sum_j F_{ij} = 1 \quad \forall\ i.
\end{equation}
Because of this, it follows immediately that the CS is always 
\emph{positivity-preserving}, i.e.\ a positive solution at time $t_k$ implies 
a positive solution at time~$t_{k+1}$, and that it is \emph{mass conservative} 
on an infinite (or periodic) domain:
\begin{align}
    \label{eq:Positivity-SL}
    n_j^{k+1} &= \sum_i n_i^k F_{ij} \ge 0 \qquad \forall j, \\
    \label{eq:Conservation-SL}
    \sum_j n_j^{k+1} &= \sum_j \sum_i n_i^k\, F_{ij}
                     = \sum_i n_i^k\, \sum_j F_{ij} = \sum_i n_i^k.
\end{align}

The benefits of this simple `area remapping rule', namely mass conservation, 
positivity preservation, and ease of implementation, are counter-balanced by 
the large amount of numerical diffusion, $O(\Delta x^2)$, that is introduced 
by the scheme whenever the final location $X_j$ does not coincide with the 
center of a mesh cell.
The focus of the following section is to derive a high-order version of the CS 
by computing small corrections $\delta X_i$ to the final locations~$X_i$: 
because the remapping rule is left untouched, it is clear from 
\eqref{eq:Positivity-SL} and~\eqref{eq:Conservation-SL} that this procedure 
can neither affect positivity nor conservation.

The high-order corrections will be derived from a Taylor expansion of $n(t,x)$ 
about the point $(t_k,x_i)$; if the solution lacks a sufficient degree of 
smoothness, or if it is under-sampled by the grid, the calculated corrections 
may be very far off from the `exact' answer.
We point out that `wrong' high-order corrections are especially troublesome 
if they are not `small' compared to~$\Delta x$.
We also note that there is no numerical diffusion if $X_j$ falls on the center 
of a mesh cell (in which case $\delta X_j=0$); therefore, our correction 
should not push the center of a MC beyond the closest cell centers.
On the basis of these considerations, we can devise a simple limiting strategy 
on $\delta X_j$ that amounts to requiring that
\begin{equation}\label{eq:PositivityConstraint.SL-form}
  \text{if   $X_j \in (x_i,x_{i+1})$ 
        then $X_j\!+\!\delta X_j \in [x_i,x_{i+1}]$.}
\end{equation}

The analysis that follows in section \ref{sec:HighOrderCS.theory} will benefit  
from an equivalent formulation of the CS, which is more typical of finite 
difference schemes: instead of focusing on MCs and the corresponding mass 
transport, we can write the numerical solution at time~$t_{k+1}$ as an 
explicit function of the solution at time~$t_k$.
For this purpose, it is useful to decompose the Courant number 
$C := u\Delta t/\Delta x$ into its integer and fractional parts, as
\begin{equation}\label{eq:CourantNumber}
    C := \frac{u\,\Delta t}{\Delta x} = S + \alpha,
    \quad \text{with $S\in\mathbb{Z}$ and $\alpha\in(-1,1)\subset\mathbb{R}$},
\end{equation}
because no remapping error is associated with shifting the solution by an 
integer number of cells~$S$ on a uniform mesh.
The CS update for the constant advection equation~\eqref{eq:ConstantAdvection} 
can then be written in `finite difference form' as
\begin{equation}\label{eq:CS.FD-form}
    n_{i+S}^{k+1}\ =\ 
    \begin{cases}    
    U_{i-1}^k n_{i-1}^k + \Par{1-U_i^k} n_i^k  &\quad \text{if $\alpha\ge 0$},
    \\[0.2em] 
    \Par{1+U_i^k} n_i^k - U_{i+1}^k n_{i+1}^k  &\quad \text{otherwise},
    \end{cases}
\end{equation}
where the non-dimensional quantity $U$ is the fractional normalized 
displacement, expressed in units of the cell width.
If no high-order corrections are applied to the scheme, this is simply the 
fractional Courant number, $U(t,x)\equiv\alpha$.
In the presence of high order corrections, we find the normalized displacement
\begin{equation}\label{eq:U(x,t)}
    U(t,x)\,=\, \bigl[u+\tilde{u}(t,x)\bigr]\frac{\Delta t}{\Delta x} - S
          \,=\, \alpha + \tilde{\alpha}(t,x),
\end{equation}
where~$\tilde{u}(t,x)$ is an `anti-diffusive correction' to the velocity field, 
and its normalized equivalent 
$\tilde{\alpha}(t,x) := \tilde{u}(t,x)\Delta t / \Delta x$
represents an `anti-diffusive correction' to the normalized displacement.
(Strictly, $\tilde{u}$ and~$\tilde{\alpha}$ include antidiffusive as well as 
higher order corrections.)
The derivation of the high-order normalized displacement~\eqref{eq:U(x,t)} 
is detailed in Section~\ref{sec:HighOrderCS.theory}, for any required order of accuracy.

Regardless of the correction~$\tilde{\alpha}$, mass conservation on an 
infinite (or periodic) domain follows immediately from~\eqref{eq:CS.FD-form}: 
in the case of~$\alpha \ge 0$ we have in fact
\begin{equation*}
\begin{split}
    \sum_i n_i^{k+1} = \sum_i n_{i+S}^{k+1}
    &= \sum_i \left[ U_{i-1}^k n_{i-1}^k + \Par{1-U_i^k} n_i^k \right] = \\
    &= \underbrace{\sum_i U_{i-1}^k n_{i-1}^k - \sum_i U_i^k n_i^k}_{=0}
       + \sum_i n_i^k = \sum_i n_i^k ,
\end{split}
\end{equation*}
and similarly for the case of~$\alpha < 0$.

Positivity preservation is a more delicate matter, as it can only be 
guaranteed if, when $\alpha \ge 0$, we have $U_i^k \ge 0$ and $(1-U_i^k)\ge 0$ 
$\forall i$.
And similarly if $\alpha < 0$, we must have $U_i^k \le 0$ and $(1+U_i^k)\ge 0$ 
$\forall i$.
This lack of `automatic' positivity preservation can be explained by noticing 
that the integer shift $S$ is computed a priori from~\eqref{eq:CourantNumber}, 
and it is not allowed to change when we include the correction 
$\tilde{\alpha}$.
However, we can easily enforce positivity by limiting the normalized 
displacement $U_i^k$, or the `upwind flux' $[Un]_i^k$, as
\begin{equation}\label{eq:PositivityConstraints.FD-form}
\begin{alignedat}{4}
  &\text{if $\alpha > 0$:} \quad& 
  0 \le\ &U_i^k \le 1 \quad && \text{or}\quad& 0\,\le\ &[Un]_i^k \le n_i^k,\\
  &\text{if $\alpha < 0$:} \quad& 
  -1\le\ &U_i^k \le 0 \quad && \text{or}\quad& -n_i^k\! \le\ &[Un]_i^k \le 0,
\end{alignedat}
\end{equation}
while the possibility of $\alpha = 0$ will always result in $U_i^k=0$.
Not surprisingly, these positivity constraints are equivalent to the 
accuracy constraint~\eqref{eq:PositivityConstraint.SL-form} on $\delta X$, 
that was proposed in the semi-Lagrangian formulation.

We choose to enforce the positivity 
constraints~\eqref{eq:PositivityConstraints.FD-form} by first computing the 
`nominal' high-order fluxes~$\Gamma_i$, as described in 
Section~\ref{sec:HighOrderCS.theory} for the quantity~$[Un]_i^k$, and then 
limiting their numerical values as follows:
\begin{equation}\label{eq:PositivityLimiter}
\begin{aligned}
  &\text{if $\alpha > 0$:} \qquad
    [Un]_i^k = \min\Par{\max\Par{\phantom{-}0\phantom{_i^k},\Gamma_i},n_i^k}, \\
  &\text{if $\alpha < 0$:} \qquad
    [Un]_i^k = \min\Par{\max\Par{-n_i^k,\Gamma_i},0\phantom{_i^k}}.
\end{aligned}
\end{equation}
This simple limiting strategy yields excellent results in practice,
as will be shown in Section~\ref{sec:NumericalTests}.

\subsection{Semi-discrete theory}
\label{sec:HighOrderCS.theory}

The purpose of this section is to construct a high-order version of the CS for 
the solution of the 1D constant advection 
equation~\eqref{eq:ConstantAdvection}, that is 
$\Par{\partial/\partial t + u\,\partial/\partial x}n(t,x) = 0$.

Given the hypothesis of sufficiently smooth initial conditions, we will Taylor 
expand both the analytical and numerical solutions 
to~\eqref{eq:ConstantAdvection}, and obtain local representations of the exact 
and numerical `one-step updates'.
A careful comparison between the two expressions, and the selection of a 
proper Ansatz for the high-order corrections, will allow us to construct an 
$\N$-th order accurate algorithm, which requires the evaluation of the first 
$\N-2$ spatial derivatives of the solution.

Regardless of how those derivatives are estimated, we will obtain a finite 
difference form of the high-order CS that is mass-conservative by 
construction, and is made positivity-preserving by the use of a simple limiter 
as described in Section~\ref{sec:HighOrderCS.formulations}.
A careful discussion of numerical differentiation strategies will be given in 
Section~\ref{sec:HighOrderCS.implementation}, where fully-discrete algorithms 
are presented.

For given initial conditions, the exact solution 
to~\eqref{eq:ConstantAdvection} is constant along any characteristic 
trajectory $X(t)$ with constant velocity~$u$, so that the identity 
$n(t+\Delta t,x) \equiv n(t,x-u\Delta t)$ holds: the 
solution at time $t+\Delta t$ is equal to the solution at time~$t$, `shifted' 
by a distance $u\Delta t$ in space.
By making use of the decomposition~\eqref{eq:CourantNumber}, that is 
$C:=u\Delta t/\Delta x = S+\alpha$, the analytic solution is reformulated as
\begin{equation}\label{eq:ExactSolution-0}
    n(t+\Delta t,x+S\Delta x)\ =\ n(t,x-\alpha\Delta x).
\end{equation}
By Taylor expanding the right hand side of~\eqref{eq:ExactSolution-0} about 
the point~$(t,x)$, we obtain a (spatially) local representation of the exact 
one-step update: 
\begin{equation}\label{eq:ExactSolution}
    n(t+\Delta t,x+S\Delta x)\ =\ n(t,x) + 
    \Par{\sum_{p=1}^{\N-1} (-\alpha)^p\frac{(\Delta x)^p}{p!}
    \frac{\partial^p}{\partial x^p} } n(t,x) + O\Par{\Delta x^\N}.
\end{equation}

A similar formulation of the CS update is derived in the following discussion.
For the sake of simplicity, we will focus on the case $\alpha \ge 0$ only, 
showing the full procedure to derive an N-th order accurate scheme; 
afterwards, we will show the general result that also holds for negative 
$\alpha$, but we will omit an unnecessary repetition of the calculations.
For $\alpha \ge 0$, we can rewrite the finite difference form of the CS update 
equation~\eqref{eq:CS.FD-form} as 
\begin{equation}\label{eq:CS.FD-form.2}
    n_{i+S}^{k+1} \,=\, n_i^k - \Par{U_i^k n_i^k - U_{i-1}^k n_{i-1}^k},
\end{equation}
which we will apply between time $t$ and $t+\Delta t$, at a generic location 
$x$.
We will further assume the CS solution to take on the exact sampled values at 
time $t$, so that $U_i^k n_i^k = U(t,x)n(t,x)$ and 
$U_{i-1}^k n_{i-1}^k = U(t,x-\Delta x)n(t,x-\Delta x)$.
Finally, the latter term can be Taylor expanded in space about the point 
$(t,x)$ in order to get a local representation of the one-step CS update, 
which depends on the first $\N-1$ spatial derivatives of the analytical 
solution at time~$t$:
\begin{equation}\label{eq:CS.update.HighOrderEstimate}
    n_\textsc{cs}(t+\Delta t,x+S\Delta x) \,=\, n(t,x) + 
    \Par{\sum_{p=1}^{\N-1} (-1)^p \frac{(\Delta x)^p}{p!}
    \frac{\partial^p}{\partial x^p}} U(t,x)\,n(t,x) + O\Par{\Delta x^\N}.
\end{equation}
At this point, we recall that $U = \alpha + \tilde{\alpha}$ as 
in~\eqref{eq:U(x,t)}, and that it is our intention to find the anti-diffusive 
Courant parameter $\tilde{\alpha}(t,x)$ that guarantees a local truncation 
error (LTE) of the order of~$(\Delta x)^\N$, i.e.
\begin{equation*}
    \mathcal{E}(t,x,\Delta t) \,:=\, 
    n(t+\Delta t,x+S\Delta x) - n_\textsc{cs}(t+\Delta t,x+S\Delta x) \,=\, 
    O(\Delta x^\N).
\end{equation*}
By direct comparison of \eqref{eq:ExactSolution} 
and~\eqref{eq:CS.update.HighOrderEstimate}, the last equation provides us with 
a necessary requirement on~$U(t,x)$, in the form
\begin{equation}\label{eq:CS.U-constraint}
    \sum_{p=1}^{\N-1} (-\alpha)^p \frac{(\Delta x)^p}{p!}
    \frac{\partial^p n}{\partial x^p} \,-\,
    \sum_{p=1}^{\N-1} (-1)^p \frac{(\Delta x)^p}{p!}
    \frac{\partial^p (U n)}{\partial x^p} \,=\, O(\Delta x^\N).
\end{equation}
It is useful to verify that for $\N=2$ the last equation simply yields
\begin{equation*}
    \alpha \frac{\partial n}{\partial x} - 
    \frac{\partial (U n)}{\partial x}\,=\, O(\Delta x),
\end{equation*}
which is satisfied by the standard version of the CS, which 
has~$U\equiv \alpha$.
(In fact, the standard CS is second order accurate in space.)
In a general situation with~$\N \ge 2$, equation~\eqref{eq:CS.U-constraint} 
unequivocally determines the first $\N-1$ terms of the spatial Taylor series 
of~$U(t,x)$.
Therefore, in order to simplify the following discussion, it is natural to 
choose a polynomial ansatz for the product~$Un$, such as
\begin{equation}\label{eq:CS.Ansatz}
    U(t,x)\,n(t,x) = \sum_{q=0}^{\N-2} (-1)^q\, \beta_q(\alpha)\, (\Delta x)^q
    \frac{\partial^q n(t,x)}{\partial x^q},
\end{equation}
where the $\N-1$ coefficients $\beta_q(\alpha)$ are unknown functions to be 
determined from~\eqref{eq:CS.U-constraint}.
One can easily verify that the high order polynomial corrections that were 
derived in~\citep{Guclu2012} are in fact consistent with the 
ansatz~\eqref{eq:CS.Ansatz} for~$\N=4$, when a uniform and constant velocity 
field $u$ is considered.

Making use of~\eqref{eq:CS.Ansatz}, the second term on the left hand side 
of~\eqref{eq:CS.U-constraint} becomes
\begin{equation*}
\begin{split}
&\sum_{p=1}^{\N-1} (-1)^p \frac{(\Delta x)^p}{p!}
    \frac{\partial^p (U n)}{\partial x^p}\ = \\
&=\ \sum_{p=1}^{\N-1} (-1)^p \frac{(\Delta x)^p}{p!}
    \frac{\partial^p}{\partial x^p} \Par{
    \sum_{q=0}^{\N-2} (-1)^q\, \beta_q(\alpha)\, (\Delta x)^q
    \frac{\partial^q n(t,x)}{\partial x^q} }\ = \\
&=\ \sum_{p=1}^{\N-1} \sum_{q=0}^{\N-2} (-1)^{p+q} \frac{\beta_q(\alpha)}{p!}
    (\Delta x)^{p+q} \frac{\partial^{p+q} n(t,x)}{\partial x^{p+q}}\ = \\
&=\ \sum_{r=1}^{\N-1} (-1)^r 
    \left[\sum_{q=0}^{r-1} \frac{\beta_q(\alpha)}{(r-q)!}\right]
    (\Delta x)^r \frac{\partial^r n(t,x)}{\partial x^r}\ +\ O(\Delta x^\N) ,
\end{split}
\end{equation*}
where the dummy index $r=p+q$ was introduced, and the term $O(\Delta x^\N)$ 
collects those terms in the sum having~$r\ge \N$.
After replacing the index~$r$ with the original~$p$, the substitution of the 
expression above into~\eqref{eq:CS.U-constraint} yields a lower triangular 
system of $\N-1$ linear equations,
\begin{equation}\label{eq:CS.LinearSystem}
\sum_{q=0}^{p-1} \frac{\beta_q(\alpha)}{(p-q)!}\,=\,\frac{\alpha^p}{p!}
\qquad (p = 1,2,\dots, \N-1),
\end{equation}
which can be solved for the $\N-1$ functions $\beta_q(\alpha)$ to be used 
in~\eqref{eq:CS.Ansatz} for any given~$\alpha$, for a specified order of 
accuracy~$\N$.
The application of forward-substitution to~\eqref{eq:CS.LinearSystem} gives 
the recursive relation
\begin{equation}\label{eq:CS.RecursiveRelation}
\begin{split}
\beta_0(\alpha) &= \alpha, \\
\beta_p(\alpha) &= \frac{\alpha^{p+1}}{(p+1)!} - 
                   \sum_{q=0}^{p-1}\frac{\beta_q(\alpha)}{(p+1-q)!}
\qquad (p \ge 1),
\end{split}
\end{equation}
which is sufficient for most practical purposes.
Explicit expressions for the coefficients $\beta_p(\alpha)$ can be 
obtained by noticing that the linear system~\eqref{eq:CS.LinearSystem} is 
in the form (here shown for~$\N=6$)
\begin{equation*}
\begin{bmatrix}[1.3]
\frac{1}{1!} & 0 & 0 & 0 & 0 \\
\frac{1}{2!} & \frac{1}{1!} & 0 & 0 & 0 \\
\frac{1}{3!} & \frac{1}{2!} & \frac{1}{1!} & 0 & 0 \\
\frac{1}{4!} & \frac{1}{3!} & \frac{1}{2!} & \frac{1}{1!} & 0 \\
\frac{1}{5!} & \frac{1}{4!} & \frac{1}{3!} & \frac{1}{2!} & \frac{1}{1!} \\
\end{bmatrix}
\cdot
\begin{bmatrix}[1.3]
\beta_0 \\ \beta_1 \\ \beta_2 \\ \beta_3 \\ \beta_4 \\
\end{bmatrix}
=
\begin{bmatrix}[1.3]
\frac{\alpha^1}{1!} \\
\frac{\alpha^2}{2!} \\
\frac{\alpha^3}{3!} \\
\frac{\alpha^4}{4!} \\
\frac{\alpha^5}{5!} \\
\end{bmatrix},
\end{equation*}
which can be inverted to give
\begin{equation}\label{eq:CS.MatrixVectorProduct}
\begin{bmatrix}[1.3]
\beta_0 \\ \beta_1 \\ \beta_2 \\ \beta_3 \\ \beta_4 \\
\end{bmatrix}
=
\begin{bmatrix}[1.3]
\frac{B_0}{0!} & 0 & 0 & 0 & 0 \\
\frac{B_1}{1!} & \frac{B_0}{0!} & 0 & 0 & 0 \\
\frac{B_2}{2!} & \frac{B_1}{1!} & \frac{B_0}{0!} & 0 & 0 \\
\frac{B_3}{3!} & \frac{B_2}{2!} & \frac{B_1}{1!} & \frac{B_0}{0!} & 0 \\
\frac{B_4}{4!} & \frac{B_3}{3!} & \frac{B_2}{2!} & \frac{B_1}{1!} & 
    \frac{B_0}{0!} \\
\end{bmatrix}
\cdot
\begin{bmatrix}[1.3]
\frac{\alpha^1}{1!} \\
\frac{\alpha^2}{2!} \\
\frac{\alpha^3}{3!} \\
\frac{\alpha^4}{4!} \\
\frac{\alpha^5}{5!} \\
\end{bmatrix},
\end{equation}
where
\begin{equation*}
B_0 = 1 \qquad
B_1 = -\frac{1}{2} \qquad
B_2 =  \frac{1}{6} \qquad
B_3 = 0 \qquad
B_4 = -\frac{1}{30}
\end{equation*}
and so on are the Bernoulli numbers of the first kind~\cite{Bernoulli1713}.
According to~\eqref{eq:CS.MatrixVectorProduct}, the polynomials 
$\beta_p(\alpha)$ can be compactly written as
\begin{equation}\label{eq:CS.Polynomials.Explicit}
\beta_p(\alpha) = \sum_{q=0}^p \frac{B_q}{q!} \frac{\alpha^{p+1-q}}{(p+1-q)!}
\qquad (p \ge 0).
\end{equation}

So far we have presented three different ways to compute the 
polynomials~$\beta_p(\alpha)$, and we may easily find arguments in favor of 
each of those options.
For instance, the recursive relation~\eqref{eq:CS.RecursiveRelation} does not 
require a table of Bernoulli numbers, the matrix-vector 
product~\eqref{eq:CS.MatrixVectorProduct} is easily amenable to algorithmic 
optimization, and the explicit expressions~\eqref{eq:CS.Polynomials.Explicit} 
can be hard-coded in the computation routines. 

Regarding the use of~\eqref{eq:CS.MatrixVectorProduct}, it is worth pointing 
out that very little storage is required, because the lower triangular 
Toeplitz matrix is completely defined by the first $\N-1$ Bernoulli numbers.
Further, it is likely that the matrix-vector product 
in~\eqref{eq:CS.MatrixVectorProduct} may be performed in fewer than the 
standard $\N(\N-1)$ operations, thanks to the peculiar structure of the 
vector on the right hand side.

Nevertheless, in the authors' experience there is very little to be gained 
from employing $\N > 20$ for calculations in double precision.
For modest~$\N$, which is therefore all that is needed here, the use of `fast' 
algorithms (possibly parallel) for the matrix-vector product is not optimal, 
because they incur substantial overhead that is only justified when $\N$ is 
large.

If we also take into account the fact that all the Bernoulli numbers with odd 
indexes are equal to zero (with the exception of $B_1 = -1/2$), then the 
simplest and most effective strategy appears to be hard-coding the symbolic
polynomials from~\eqref{eq:CS.Polynomials.Explicit} in the source code, as
\begin{equation*}
\begin{split}
\beta_0(\alpha)\ &=\ \alpha \\[0.2em]
\beta_1(\alpha)\ &=\ \frac{1}{2}\Par{\alpha^2-\alpha} \\
\beta_p(\alpha)\ &=\ \frac{\alpha^{p+1}}{(p+1)!}\, -\,
    \frac{1}{2} \frac{\alpha^p}{p!}\, +\, 
    \sum_{s=1}^{p/2} \frac{B_{(2s)}}{(2s)!} \frac{\alpha^{p+1-2s}}{(p+1-2s)!}
    \qquad (p \ge 2).
\end{split}
\end{equation*}
It is also possible to minimize the number of calculations, as well as the 
round-off errors, by rewriting the polynomials above in their Horner form.

It is worth pointing out that the expression~\eqref{eq:CS.Polynomials.Explicit}
is a rescaled version of the power summation formula
\begin{equation}\label{eq:CS.PowerSummationFormula}
\sum_{k=0}^{m-1} k^n\ =\
\sum_{k=0}^{n} \frac{B_k}{k!} \frac{n!}{(n+1-k)!} m^{n+1-k}\ =\
\frac{B_{n+1}(m)-B_{n+1}(0)}{n+1},
\end{equation}
derived by Jacob Bernoulli~\cite{Bernoulli1713} in 1690 for integer and 
positive values of~$m$, and which represents the oldest definition for the 
Bernoulli polynomials~$B_n(x)$.
When evaluating such polynomials at $x=0$ and~$x=1$, we recover the Bernoulli 
numbers of the first and second kind, respectively.
Since the second equality in~\eqref{eq:CS.PowerSummationFormula} holds for any 
value of~$m \in \mathbb{R}$, we can use it to rewrite the correction 
polynomials in~\eqref{eq:CS.Polynomials.Explicit} as 
\begin{equation}\label{eq:CS.Polynomials.Bernoulli}
\beta_p(\alpha) = \frac{B_{p+1}(\alpha)-B_{p+1}(0)}{(p+1)!}.
\end{equation}

We are now interested in extending the results so far obtained to the case 
of~$\alpha < 0$.
Starting from the second equation in~\eqref{eq:CS.FD-form} and using the same 
ansatz~\eqref{eq:CS.Ansatz}, one can repeat the calculations and derive an 
expression similar to~\eqref{eq:CS.Polynomials.Bernoulli}.
Finally, a general formula that holds for any value of~$\alpha \in \mathbb{R}$ 
is
\begin{equation}\label{eq:CS.Polynomials.Bernoulli-anyAlpha}
\begin{split}
\beta_0(\alpha)\ &=\ \alpha \,, \\
\beta_p(\alpha)\ &=\ \frac{B_{p+1}(\langle\alpha\rangle)-B_{p+1}(0)}{(p+1)!}
\qquad \Par{p \ge 1},
\end{split}
\end{equation}
where $\langle\alpha\rangle := \alpha\!\mod 1$, which means that 
$\langle\alpha\rangle = \alpha$ for~$0\le\alpha<1$, and 
$\langle\alpha\rangle = 1+\alpha$ for~$-1<\alpha<0$.
The functions $P_n(x) \equiv B_n(\langle x \rangle)$ are called `periodic 
Bernoulli polynomials', as they are the periodic continuation of the standard 
Bernoulli polynomials, given that the latter are first restricted to the 
range~$[0,1]$.

Finally, in Algorithm \ref{algo:semi-discrete} we present the general 
procedure for implementing the high order CS. 
Optimized versions of such an algorithm will be designed in the 
fully-discrete case, after one defines how to estimate the required spatial 
derivatives (see Section~\ref{sec:HighOrderCS.implementation}).
Without loss of generality, we will assume that~$\alpha \ge 0$.
\begin{algorithm}[!htb]
\caption{General form of the high order Convected Scheme.}
\label{algo:semi-discrete}
\vspace{1mm}
\begin{enumerate}
  \item Given the Courant parameter $C := u\Delta t/\Delta x$, decompose 
        it into integer and fractional parts as
        \[
          C = S + \alpha, \qquad
          \text{$S\in\mathbb{Z}$, $\alpha \in[0,1)\subset\mathbb{R}$};
        \]
  \item Given~$\alpha$, compute the $\N-2$ correction polynomials 
        $\beta_p(\alpha)$ needed to achieve a nominal order of accuracy~$\N$,
        according to \eqref{eq:CS.RecursiveRelation}, 
        \eqref{eq:CS.Polynomials.Explicit} 
        or~\eqref{eq:CS.Polynomials.Bernoulli}, and store the power series 
        coefficients:
        \[
          c_p = (-1)^p \beta_p(\alpha),
          \qquad p \in \{1,\dots,\N-2\};
        \]
  \item Compute the required $\N-2$ normalized spatial derivatives of 
        $n(t_k,x)$ at the cell centers~$x_i$, using some approximation
        \[
          d_{pi} \approx
          (\Delta x)^p \left.\frac{\partial^p n}{\partial x^p}\right|_i^k,
          \qquad p \in \{1,\dots,\N-2\};
        \]
  \item Compute the high order `upwind fluxes' according to the 
        ansatz~\eqref{eq:CS.Ansatz}, and apply the positivity  
        limiter~\eqref{eq:PositivityLimiter}:
        \[
          \Gamma_i =\, n_i^k\alpha + \sum_{p=1}^{\N-2} c_p\,d_{pi},
          \qquad [Un]_i^k = \min\Par{\max\Par{0,\Gamma_i}\!,n_i^k},
          \qquad \forall i;
        \]
  \item Obtain the solution $\{n_i^{k+1}\}$ at time $t_{k+1} = t_k +\Delta t$, 
        according to the update
        \[
          n_i^{k+1} = n_{i-S}^k + [Un]_{i-S-1}^k - [Un]_{i-S}^k ,
          \qquad \forall i.
        \]
\end{enumerate}
\end{algorithm}

\subsection{Fully discrete implementation}
\label{sec:HighOrderCS.implementation}

In this section we derive arbitrarily high order implementations of the CS, 
which basically differ as to how step 3 of Algorithm \ref{algo:semi-discrete} 
is carried out.
For achieving a truncation error $O(\Delta x^{\N})$, we need a proper means of 
evaluating the required $\N-2$ spatial derivatives of $n(t,x)$;
this is equivalent to replacing the ansatz~\eqref{eq:CS.Ansatz} with an 
algebraic (e.g.~finite difference) approximation, which must be accurate 
enough to ensure the local truncation error be~$O\!\Par{\Delta x^\N}$.
Since in~\eqref{eq:CS.U-constraint} the leading error in our approximation to 
$Un$ is multiplied by powers of~$\Delta x$ with exponents~$\ge 1$, it is 
sufficient to estimate the products
\begin{equation*}
(\Delta x)^q \left.\frac{\partial^q n(t,x)}{\partial x^q}\right|_i^k
\end{equation*}
with an error no larger than~$O(\Delta x^{\N-1})$.
In order to approximate the aforementioned quantities with the required 
accuracy, in this section we follow the simple strategy of differentiating a 
(polynomial or trigonometric) function that interpolates a set of points 
containing~$x_i$.

In Section~\ref{sec:HighOrderCS.implementation.polynomial} we explore the 
use of a symmetric (local) polynomial interpolation, which leads to centered 
finite difference approximations to the required derivatives.
We provide a detailed description of the algorithm, complete with explicit 
formulas for any even order of accuracy.

In Section~\ref{sec:HighOrderCS.implementation.trigonometric} we construct a 
(global) trigonometric interpolant, which is appropriate when periodic 
boundary conditions are employed.
Thanks to the linearity of the discrete Fourier transform, we efficiently 
compute the first 20 derivatives in Fourier space; accordingly, the 
calculation of the corrected quantity $[Un]_i^k$ at all locations $\{x_i\}$ 
only requires one application of the fast Fourier transform (FFT) algorithm, 
and one application of its inverse (IFFT).
We also discuss the use of filtering in order to reduce spurious oscillations 
at the Nyquist frequency.

\begin{remark*}[Order of accuracy]
We recall that in the Convected Scheme (CS), as in most semi-Lagrangian 
schemes, the discretization parameters $\Delta x$ and $\Delta t$ are chosen 
independently of each other.
Therefore, we identify the accuracy of our 1D advection solvers based on their 
local truncation error (LTE).
In other words, we refer to a CS with a LTE = $O(\Delta x^{\N})$ as an `N-th 
order accurate CS'.
For ease of comparison, in the numerics section we present 1D refinement 
studies at a fixed Courant number; in such cases we observe an order of 
convergence of $\N-1$.
\end{remark*}

\subsubsection{Fully discrete implementation: Polynomial interpolation}
\label{sec:HighOrderCS.implementation.polynomial}

In this section we reconstruct $n(x)$ from its $\N-1$ point values at 
time~$t_k$,
using Lagrangian interpolation of order~$\N-2$; the resulting polynomial 
$L(x)$ is then differentiated $\N-2$ times at the required location~$x_i$.
Since the leading term in the truncation error of the scheme is 
proportional to the $\N$-th derivative of the solution, we choose $\N$ even 
to obtain a scheme that has a high-order diffusive behavior, not dispersive, 
which is ideal for suppressing under-resolved modes in a Vlasov simulation.
Moreover, we use a centered stencil $\{i+r\}$, where 
$r \in\left\{-R,\ \dots,\ -1,\ 0,\ 1,\ \dots,\ R \right\}$ and $R:=\N/2-1$, 
because this choice minimizes the error constant for a given~$\N$.

There is a simple way to compute the finite difference coefficients on the 
given stencil, for all the required derivatives.
In fact, it is sufficient to rewrite the interpolating polynomial $L(x)$ in 
Taylor form to observe its $\N-2$ derivatives at the point~$x_i$:
\begin{equation*}
  L(x) = \sum_{q=0}^{N-2} a_q \Par{x-x_i}^q = \sum_{q=0}^{N-2} 
  \left.\frac{\partial^q L}{\partial x^q}\right|_{x_i}\!\!\frac{(x-x_i)^q}{q!}.
\end{equation*}
Since $L(x)$ interpolates the values $\{n_{i+r}^k\}$ at the locations 
$\{x_{i+r}\}$, the following system of $\N-1$ scalar equations holds:
\begin{equation*}
  n_{i-R+p} = \sum_{q=0}^{N-2} \frac{1}{q!} \mat{V}_{pq} d_q
  \qquad \text{with} \quad
  \begin{cases}
    d_q := (\Delta x)^q\left.\dfrac{\partial^q L}{\partial x^q}\right|_{x_i}\\
    \mat{V}_{pq} := (-R+p)^q
  \end{cases}
  \qquad (p = 0,1,\cdots,\N-2) ,
\end{equation*}
where $d$ is the vector of the normalized derivatives of $L(x)$ at $x=x_i$, 
and~$\mat{V}$ is a classical Vandermonde matrix.
Solving for the vector~$d$ yields
\begin{equation*}
  d_p = \sum_{q=0}^{N-2} \mat{D}_{pq}\, n_{i-R+q}, \qquad
  \text{with $\mat{D}_{pq} = p! \left[\mat{V}^{-1}\right]_{pq}$},
\end{equation*}
where each row $p$ of the matrix $\mat{D}$ contains the required finite 
difference coefficients for the $p$-th derivative of~$L(x)$.
If the exact solution $n(t,x)$ is sufficiently smooth, the quantities~$d_p$ 
so obtained approximate the normalized derivatives of $n(t_k,x)$ at the 
location $x_i$ with an error no larger than~$O(\Delta x^{\N-1})$.
Specifically we have
\begin{equation*} 
  \Par{\Delta x}^p \left.\dfrac{\partial^p n}{\partial x^p}\right|_i^k =\
  d_p +  
  \begin{cases}
    O\Par{\Delta x^{\N}}   & \text{if $p$ is even},\\
    O\Par{\Delta x^{\N-1}} & \text{if $p$ is odd}.
  \end{cases}
\end{equation*}

\subsubsection*{Example: 6th order finite difference scheme}
As an example, we provide the finite difference coefficients for constructing 
a scheme with a 6th-order truncation error.
By choosing a 5-point centered stencil $\{i+r\}$ with $r\in\{-2,1,0,1,2\}$ we 
can approximate the first four spatial derivatives of $n(t,x)$ at the point 
$(t_k,x_i)$ by means of the finite difference schemes
\begin{align*}
\Delta x\ \left.\frac{\partial n}{\partial x}\right|_i^k &=\ 
    \frac{n_{i-2}^k -8\,n_{i-1}^k +8\,n_{i+1}^k -n_{i+2}^k}{12} + 
    O\!\Par{\Delta x^5},\\
\Par{\Delta x}^2 \left.\frac{\partial^2 n}{\partial x^2}\right|_i^k &=\
    \frac{-n_{i-2}^k+16\,n_{i-1}^k-30\,n_i^k+16\,n_{i+1}^k-n_{i+2}^k}{12} +
    O\!\Par{\Delta x^6},\\
\Par{\Delta x}^3 \left.\frac{\partial^3 n}{\partial x^3}\right|_i^k &=\
    \frac{-n_{i-2}^k+2\,n_{i-1}^k-2\,n_{i+1}^k-n_{i+2}^k}{2} +
    O\!\Par{\Delta x^5},\\
\Par{\Delta x}^4 \left.\frac{\partial^4 n}{\partial x^4}\right|_i^k &=\
    n_{i-2}^k-4\,n_{i-1}^k+6\,n_i^k-4\,n_{i+1}^k+n_{i+2}^k +
    O\!\Par{\Delta x^6}.
\end{align*}
which are obtained by differentiating in space the unique quartic polynomial 
that interpolates the five point values~$\{n_{i+r}^k\}$.

\subsubsection*{Algorithm}
We are now ready to describe the implementation details in Algorithm
\ref{algo:polynomial}.
As usual, without loss of generality we will assume that the fractional part 
of the Courant number is non-negative ($\alpha\ge 0$).
\begin{algorithm}[!htb]
\caption{High order Convected Scheme with Lagrange interpolation.}
\label{algo:polynomial}
\vspace{1mm}

\begin{enumerate}
  \item {[\bf{Preprocessing}]} Given the (even) order of accuracy 
        $\N = 2(R+1)$ with $R\in\mathbb{N}$, compute the $(\N-1)\times(\N-1)$ 
        matrix $\mat{D}$ of finite difference coefficients:
        \begin{equation*}
          \begin{split}
            \mat{V}_{pq} &= (-R+p)^q \\
            \mat{D}_{pq} &= p! \left[\mat{V}^{-1}\right]_{pq}
          \end{split}
          \qquad p,q \in \{0,1,\dots,2R\};
        \end{equation*}
  \item Given the Courant parameter $C := u\Delta t/\Delta x$, decompose 
        it into integer and fractional parts as
        \[
          C = S + \alpha, \qquad
          \text{$S\in\mathbb{Z}$, $\alpha \in[0,1)\subset\mathbb{R}$};
        \]
  \item Given $\alpha$, compute the correction polynomials $\beta_p(\alpha)$ 
        according to \eqref{eq:CS.RecursiveRelation}, 
        \eqref{eq:CS.Polynomials.Explicit} 
        or~\eqref{eq:CS.Polynomials.Bernoulli}, and store the power series 
        coefficients
        \[
          c_p = (-1)^p \beta_p(\alpha),
          \qquad p \in \{0,1,\dots,2R\};
        \]
  \item Combine the finite difference coefficients $\mat{D}_{pq}$ and the 
        power series coefficients~$c_p$ to obtain the stencil weights
        \[
          w_q = \sum_{p=0}^{2R} c_p \mat{D}_{pq},
          \qquad q\in \{0,1,\dots,2R\};
        \]
  \item Compute the fluxes by convolving the solution $\{n_i^k\}$ at time 
        $t_k$ with the stencil weights~$\{w_q\}$, and enforce positivity 
        using the limiter~\eqref{eq:PositivityLimiter}:
        \[
          \Gamma_i = \sum_{q=0}^{2R} w_q\, n_{i-R+q}^k ,
          \qquad [Un]_i^k = \min\Par{\max\Par{0,\Gamma_i}\!,n_i^k},
          \qquad \forall i;
        \]
  \item Obtain the solution $\{n_i^{k+1}\}$ at time $t_{k+1} = t_k +\Delta t$, 
        according to the update
        \[
          n_i^{k+1} = n_{i-S}^k + [Un]_{i-S-1}^k - [Un]_{i-S}^k ,
          \qquad \forall i.
        \]
\end{enumerate}
\end{algorithm}

\subsubsection{Fully discrete implementation: Trigonometric interpolation}
\label{sec:HighOrderCS.implementation.trigonometric}

When the computational domain is periodic, a more accurate way of computing 
the derivatives is to use the discrete Fourier transform.
Further, when a large order of accuracy $\N$ is required, a Convected Scheme 
(CS) based on fast Fourier transform (FFT) algorithms may be faster than an 
equivalent CS that uses high order polynomial interpolation.
In fact, thanks to the linearity of the ansatz~\eqref{eq:CS.Ansatz}, we can 
avoid evaluating the spatial derivatives of $n(t,x)$ altogether: the 
successive application of the Fourier transform operator 
$\mathcal{F}[\,\cdot\,](\xi)$ and its inverse $\mathcal{F}^{-1}[\,\cdot\,](x)$ 
yields
\begin{equation}\label{eq:CS.spectral}
\begin{split}
Un(x) &=\ \mathcal{F}^{-1}\left\{ \mathcal{F}\left[
    	      \sum_{q=0}^{\N-2} (-1)^q\, \beta_q(\alpha)\, (\Delta x)^q
    	      \frac{\partial^q n}{\partial x^q} \right] \right\} = \\
      &=\ \mathcal{F}^{-1}\left\{
          \sum_{q=0}^{\N-2} (-1)^q\, \beta_q(\alpha)\, (\iu \xi \Delta x)^q
    	      \cdot \mathcal{F}[n] \right\},
\end{split}
\end{equation}
where $\iu:=\sqrt{-1}$ is the imaginary unit.
Since the product $\left[Un\right]_i^k$ has to be computed at all points~$x_i$ 
on a uniform mesh, considerable speedup is obtained by using a fast discrete 
Fourier transform (FFT) and its inverse (IFFT) in place of the operators 
$\mathcal{F}$ and $\mathcal{F}^{-1}$, respectively.
Given a uniform grid with $\N_x$ cells of size $\Delta x$, the normalized 
wave-number $\xi\Delta x$ assumes equally spaced values in the interval 
$[-\pi,\pi]$; hence when computing its powers, numerical overflow does not 
occur in double precision unless $\N > 622$.

For smooth solutions ($C^\infty$) and sufficiently large order of 
accuracy~$\N$, the procedure 
just outlined permits us to pursue machine precision, because the $\N-2$ 
computed derivatives of $n(x)$ are spectrally accurate and the remaining 
truncation error $O(\Delta x^\N)$ quickly falls below machine precision as 
the mesh is refined.
We have observed in practice that the error introduced by the FFT/IFFT process 
becomes dominant for~$\N\ge\N_{sp}$, where $\N_{sp}\approx 20$ in double 
precision.

Because of round-off error, the numerically computed discrete Fourier 
transform of $n(x)$ is corrupted with white noise, which has approximately 
constant amplitude across Fourier space.
Since the relative error will be greater for those modes that have smaller 
magnitude, we zero-out any Fourier coefficient that falls below a certain 
threshold.
Nevertheless, any residual numerical noise will be amplified at the higher 
frequencies, because the Fourier coefficients of these are multiplied by the 
higher values of $\Par{\xi\Delta x}^q$ in~\eqref{eq:CS.spectral}.
In practice this limits the order of accuracy to approximately 
$\N_\text{max}\approx 25$ in double precision.

For the reasons above, we will refer to the scheme with an `optimal' order of 
accuracy $\N_{sp} \le \N \le \N_\text{max}$ as a \emph{spectrally accurate} 
Convected Scheme.
Indeed, a nominal order of accuracy $\N=22$ is chosen for the numerical tests 
in Section~\ref{sec:NumericalTests}, and the resulting scheme shows spectral 
convergence instead of algebraic, until it reaches machine precision.

\subsubsection*{Filtering}
We point out that, when applied to the Vlasov equation, the `Spectral-CS' just 
described is not able to dissipate those features that naturally fall below 
the mesh size~$\Delta x$.
Because of the `filamentation' phenomenon, energy is transferred toward higher 
modes, and eventually reaches the Nyquist frequency 
$\xi_\text{max} = \pm\, \pi/\Delta x$.
In the absence of a dissipation mechanism, the Nyquist mode will quickly grow 
in time and pollute the solution everywhere in the domain.
Multiple strategies can be envisioned to deal with this problem; a detailed 
investigation, comparison and optimization of those is a demanding task that 
falls beyond of the scope of this work.

A rigorous approach to cope with filamentation in Vlasov-Poisson and 
Vlasov-Maxwell simulations was envisioned by Klimas in the middle 
1980s~\cite{Klimas1987}: this consists of evolving a `filtered' distribution 
function (i.e.\ the original distribution convolved with a Gaussian filter in 
velocity space), according to a modified Vlasov equation.
On the one hand, such a formulation is particularly interesting because the 
evolution of the electric field is identical to the original system; on the 
other hand, the modified system is not Hamiltonian (and hence a symplectic 
time-integrator cannot be used), it does not lend itself to efficient $(x,v)$ 
splitting~\cite{Klimas1994}, and extension to higher dimensions is unclear.

In this work we prefer to follow the original Vlasov-Poisson formulation, and 
we propose to use a simple non-adaptive filter in Fourier space, which we 
describe in detail in \ref{sec:Appendix-A}.
Our approach is a simplified version of the windowing strategy by Sun, Zhou, 
Li and Wei~\cite{Sun2006}, who investigated the use of (time) adaptive 
filtering to overcome similar difficulties, arising in pseudo-spectral 
schemes with Runge-Kutta (RK) time-stepping:
\begin{enumerate}
  \item at each time-step, a tentative RK step is taken everywhere in the 
        domain without using filtering;
  \item the increase in the total variation (TV) of the solution is tested 
        against a prescribed tolerance;
  \item if the TV test is successful the updated solution is accepted, 
        otherwise the RK step is repeated with the filter turned on.
\end{enumerate}
Such a procedure, with a CS update in place of the RK step, can provide 
extremely accurate results if the tunable parameters (i.e.\ the TV tolerance 
and the filter strength) are carefully optimized for the specific profile at 
hand, and for a given number of mesh cells.

In the authors' experience, performing such an optimization for the 
Vlasov-Poisson system is too complicated and expensive, as the `optimal' 
parameters vary greatly between different test-cases, and even within the 
same simulation.
Instead, the use of a filter with moderate strength, constant in time, has 
proven to be sufficient to stabilize all the Vlasov-Poisson test-cases that 
will be shown in Section~\ref{sec:NumericalTests.VlasovPoisson}.
The resulting Spectral-CS is not able to capture shocks, but it effectively 
`localizes' spurious oscillations to a relatively small neighborhood of the 
discontinuity.

\subsubsection*{Algorithm}
We are now ready to present the detailed implementation of the `Spectral-CS' 
in Algorithm \ref{algo:spectral}.
As usual, without loss of generality we will assume that the fractional part 
of the Courant number is non-negative ($\alpha \ge 0$).
In the following description, both the spatial index $i$ and frequency 
index~$r$ run from $0$ to~$\N_x\!-\!1$.
Moreover, the scalar quantity $\omega := \exp\Par{2\pi\iu/\N_x}$ represents 
the $\N_x$-th primitive root of unity, with $\iu := \sqrt{-1}$ the imaginary 
unit.
The construction of the filter $\hat{K}(\cdot)$ is described in 
\ref{sec:Appendix-A}.

\begin{algorithm}[!htb]
\caption{High order Convected Scheme with filtered 
         trigonometric interpolation.}
\label{algo:spectral}
\vspace{1mm}

\begin{enumerate}
  \item {[\bf{Preprocessing}]} Given the number of mesh subdivisions $\N_x$, 
        compute the normalized wave-numbers $\xi_r\Delta x \in [\pi,\pi]$ 
        supported by the mesh, and sample the filter $\hat{K}(\cdot)$ at 
        those locations:
        \[
          \xi_r\Delta x = 
          \begin{cases}
            2\pi r/\N_x       &\text{for $r \le \N_x/2$},\\
            2\pi(r-\N_x)/\N_x &\text{otherwise},
          \end{cases}
          \qquad
          \hat{K}_r = \hat{K}\!\Par{\xi_r\Delta x}, \qquad \forall r;
        \]
  \item Given the Courant parameter $C := u\Delta t/\Delta x$, decompose 
        it into integer and fractional parts as
        \[
          C = S + \alpha, \qquad
          \text{$S\in\mathbb{Z}$, $\alpha \in[0,1)\subset\mathbb{R}$};
        \]
  \item Given $\alpha$, compute the correction polynomials $\beta_q(\alpha)$ 
        according to \eqref{eq:CS.RecursiveRelation}, 
        \eqref{eq:CS.Polynomials.Explicit} 
        or~\eqref{eq:CS.Polynomials.Bernoulli}, and store the power series 
        coefficients needed to achieve a nominal order of accuracy $\N$:
        \[
          c_q = (-1)^q \beta_q(\alpha),
          \qquad q \in \{1,2,\dots,\N\!-\!2\};
        \]
  \item Compute the discrete Fourier transform of the solution $\{n_i^k\}$ at 
        time $t_k$, using an FFT algorithm
        \[ \hat{n}_r = \sum_{i=0}^{\N_x-1} n_i^k\,\omega^{-ir},
           \qquad \forall r;
        \]
  \item Reduce roundoff noise by suppressing the modes below a threshold 
        $\eps$ (in double precision we choose $\eps = 2\cdot10^{-15}$):
        \[
          A = \max_r{|\hat{n}_r|},
          \qquad \text{if $|\hat{n}_r| < A\eps$ set $\hat{n}_r = 0$},
          \qquad \forall r;
        \]
  \item Compute the Fourier coefficients of the filtered high-order flux 
        corrections 
        \[
          \hat{H}_r = 
          \Par{\sum_{q=1}^{\N-2}c_q\cdot\Par{\,\iu\,\xi_r\Delta x}^q} 
          \hat{K}_r\, \hat{n}_r,
          \qquad \forall r;
        \]
  \item Compute the inverse discrete Fourier transform of $\{\hat{H}_r\}$ 
        using an IFFT algorithm, add the high-order correction to the 
        low-order flux, and enforce positivity using the 
        limiter~\eqref{eq:PositivityLimiter}
        \[
          H_i = \frac{1}{\N_x}\!\sum_{r=0}^{\N_x-1} \hat{H}_r\,\omega^{ir}
          \qquad  \Gamma_i =\, \alpha\,n_i^k + H_i
          \qquad [Un]_i^k = \min\Par{\max\Par{0,\Gamma_i}\!,n_i^k}
          \qquad \forall i;
        \]
  \item Obtain the solution $\{n_i^{k+1}\}$ at time $t_{k+1} = t_k +\Delta t$, 
        according to the update
        \[
          n_i^{k+1} = n_{i-S}^k + [Un]_{i-S-1}^k - [Un]_{i-S}^k
          \qquad \forall i.
        \]
\end{enumerate}

\end{algorithm}

\section{Numerical tests}
    \label{sec:NumericalTests}
In this section we assess the performance of the proposed numerical methods on 
increasingly more complicated problems.
Section~\ref{sec:NumericalTests.Advection1D} tests the basic 1D constant 
advection solvers.
Section~\ref{sec:NumericalTests.Advection2D} investigates the interaction 
between the 1D solver and several operator splitting schemes, on a 2D rotating 
advection test-case.
In Section~\ref{sec:NumericalTests.LinearVlasov} the 1D-1V linear Vlasov 
equation is solved, with a given electric field (constant in time) that 
indefinitely confines the electrons in the domain.
Finally, Section~\ref{sec:NumericalTests.VlasovPoisson} focuses on the 1D-1V 
Vlasov-Poisson system, where the nonlinear electron dynamics is solved 
self-consistently with the electric field: the solver accurately captures the 
formation and evolution of periodic electronic structures in a uniform ion 
background.

\subsection{1D constant advection}
\label{sec:NumericalTests.Advection1D}
In this section we will assess the numerical properties of different 
high-order implementations of the CS: a 4th-order scheme that 
uses parabolic interpolation over a 3-point stencil, which we refer to as `P4',
a 6th-order scheme that uses quartic interpolation over a 5-point stencil 
(`P6'), and a 22nd-order CS based on the fast Fourier transform (`F22'). 
The aforementioned schemes are applied to the solution of the 1D constant 
advection equation
\begin{equation}\label{eq:tests.advection1D.eq}
\frac{\partial n}{\partial t} + \frac{\partial n}{\partial x} = 0,
\quad x \in [-0.5, 0.5],
\quad t \in [0, T],
\end{equation}
with the periodic boundary condition $n(t,-0.5) \equiv n(t,0.5)$.
(For clarity, in this section a constant advection flow field $u \equiv 1$ is 
chosen.)
The final time $T$ and the initial condition $n(0,x) = n_0(x)$ will depend on 
the test case at hand.

Section~\ref{sec:NumericalTests.Advection1D.refinement} presents a refinement 
study on smooth initial conditions.
Section~\ref{sec:NumericalTests.Advection1D.LongTimeInt} shows the long-time 
effect of error accumulation, for initial conditions with various degrees of 
smoothness.
Section~\ref{sec:NumericalTests.Advection1D.oscillations} analyzes the 
behavior of the F22-CS for discontinuous initial conditions.

Overall, we can conclude that higher-order versions of the CS have advantages 
over lower-order versions, as they allow coarser meshes while achieving better 
long-time accuracy.
The computational savings are very considerable for problems of high 
dimensionality $d$, because increasing the mesh size by a factor of 2 reduces 
the memory requirement by a factor of $2^d$.
Moreover, increasing the order of the scheme tends to increase the number of 
floating-point operations over the number of memory access events: this 
feature enables better scaling on modern parallel architectures.

\subsubsection{Refinement study}
\label{sec:NumericalTests.Advection1D.refinement}

The purpose of this first test-case is to verify mass conservation, positivity 
preservation, and the expected order of accuracy (with respect to $\Delta x$) 
of the different high-order implementations of the CS.
We solve the 1D constant advection equation~\eqref{eq:tests.advection1D.eq} 
until the final time $T=1$, which corresponds to exactly one pass through the 
periodic domain.
The initial condition $n(0,x) = n_0(x)$ is given by the symmetric 
superposition of three Gaussian bells:
\begin{equation}\label{eq:tests.advection1D.ICs}
n_0(x) \,=\, 
0.5\,e^{-\Par{\frac{x+0.2}{0.03}}^2} \,+\,
     e^{-\Par{\frac{x    }{0.06}}^2} \,+\,
0.5\,e^{-\Par{\frac{x-0.2}{0.03}}^2}.
\end{equation}
Although the odd derivatives of $n_0(x)$ are discontinuous at the 
boundaries of the domain, our numerical calculations in double precision 
cannot resolve such singularities, and for all practical purposes we can 
consider $n_0(x)$ to be a smooth function.

In these simulations we choose a constant CFL parameter $C = \Delta t / 
\Delta x = 0.32$ and we subdivide the spatial domain into a number of cells 
which is a multiple of 32 (i.e. $\N_x = 32 k$ with $k \in \mathbb{N}^+$); 
accordingly, the number of time-steps during each simulation is a multiple of 
100 (i.e. $\N_t = 100 k$).

Since the exact solution is known in the form $n(t,x) = n_0(x-t)$, we can 
compute the numerical error at all (discrete) time instants and mesh 
locations.
A scalar measure of the error is obtained by first taking its 
$L^2$-norm in space, and then the $L^\infty$-norm in time,
\begin{equation}\label{eq:tests.advection1D.ErrNorm}
E \,=\, \max_k \left\{
\sqrt{\sum_i \Bigl[ n(t_k,x_i)-n_\textsc{cs}(t_k,x_i) \Bigr]^2 \Delta x
     } \right\},
\end{equation}
which is to say that we compute the \emph{maximum in time} of the spatial 
$L^2$-norm of the error.

For each scheme (P4, P6 and F22), we run the same simulation with an 
increasing number $\N_x$ of spatial cells (a refinement study) and we compute 
the (incremental) algebraic order of convergence of the scheme based on the 
norm~\eqref{eq:tests.advection1D.ErrNorm}.
Further, we record the minimum value of the numerical solution 
$n_{\min}=\min_{(i,k)}\{n_i^k\}$ during each simulation (the exact minimum 
being approximately~$6.93\times 10^{-31}$), and we verify that this is 
positive or equal to zero.
The results of such a convergence analysis are summarized in 
Tables~\ref{table:Convergence1D-P4}, \ref{table:Convergence1D-P6} 
and~\ref{table:Convergence1D-F22}, where we also report the results obtained 
\emph{without} using the positivity limiter~\eqref{eq:PositivityLimiter}.

The P4 and P6 schemes achieve their expected order of convergence of 3 and 5, 
respectively, which equals the order of their local truncation error (LTE) 
minus 1.
The F22 scheme reaches machine precision very soon, for $\N_x = 256$, and its
expected order of convergence cannot be observed.
We point out that higher-order implementations appear to be more efficient 
than lower-order ones even when low accuracy is required: for instance, the 
errors of the P4, P6 and F22 schemes with $\N_x=128$, 64 and 32, respectively, 
are all of comparable magnitude.

In all simulations the total mass $M(t_k) = \sum_i n_i^k \Delta x$ is 
conserved to machine precision.
Further, whenever the positivity limiter~\eqref{eq:PositivityLimiter} is 
employed, the solution remains positive or equal to zero everywhere in the 
domain and for all times, i.e.~$n_i^k \ge 0$ $\forall\,i,k$, while the error 
level and the order of convergence are substantially unaffected.
These results constitute strong evidence of the effectiveness of our simple 
limiting strategy.

We notice that a reported zero minimum value, $n_{\min}=0$, is a clear signal 
that the positivity limiter was active during the simulation.
In the case of the polynomial schemes P4 and P6, the limiter is not triggered 
in those simulations that use a large number of mesh cells $\N_x$, where the 
nominal order of convergence is achieved.
In the case of the spectral scheme F22, the limiter is active 
in all simulations, even beyond the point where machine precision is achieved;
in fact, the global FFT/IFFT procedure employed pollutes the solution 
everywhere in the domain, with an error level of the order of the machine 
epsilon ($2.22\times 10^{-16}$ in double precision), which is larger in 
magnitude than the minimum value of the exact solution.

\begin{table}[!htb]
\newcommand{\pz}{\phantom{0}}
\newcommand{\pd}{\phantom{-}}
\begin{center}
\setlength{\bigstrutjot}{1.5pt}
\begin{tabular}{|r|ccc|ccc|}
	\hline
	\bigstrut[t]
	 & \multicolumn{3}{c|}{\bf{P4}} & 
	   \multicolumn{3}{c|}{\bf{P4 -- no limiter}}\\
	\cline{2-7}
	\bigstrut[t]
	\bf{N$_x$} & $\bm{L^2}$ \bf{error} & \bf{Order} & \bf{Min value}
	           & $\bm{L^2}$ \bf{error} & \bf{Order} & \bf{Min value}\\
	\hline
	\hline
	\bigstrut[t]
$  32$ & $1.41\times 10^{-1}$ &   ---  & $    0 $
       & $1.41\times 10^{-1}$ &   ---  & $   -1.83\times 10^{-2\pz}$\\
$  64$ & $5.99\times 10^{-2}$ & $1.24$ & $    0 $
       & $6.05\times 10^{-2}$ & $1.22$ & $   -2.23\times 10^{-2\pz}$\\
$ 128$ & $2.28\times 10^{-2}$ & $1.39$ & $    0 $
       & $2.42\times 10^{-2}$ & $1.32$ & $   -1.35\times 10^{-2\pz}$\\
$ 256$ & $5.44\times 10^{-3}$ & $2.07$ & $    0 $
       & $5.42\times 10^{-3}$ & $2.16$ & $   -1.10\times 10^{-3\pz}$\\
$ 512$ & $7.94\times 10^{-4}$ & $2.78$ & $    0 $
       & $7.94\times 10^{-4}$ & $2.77$ & $   -2.98\times 10^{-8\pz}$\\
$1024$ & $1.02\times 10^{-4}$ & $2.96$ & $\pd 6.22\times 10^{-31}$
       & $1.02\times 10^{-4}$ & $2.96$ & $\pd 6.22\times 10^{-31}$\\
$2048$ & $1.28\times 10^{-5}$ & $2.99$ & $\pd 8.59\times 10^{-31}$
       & $1.28\times 10^{-5}$ & $2.99$ & $\pd 8.59\times 10^{-31}$\\
\hline
\end{tabular}
\caption{
Convergence analysis for the high-order Convected Scheme applied to the 1D 
constant advection equation~\eqref{eq:tests.advection1D.eq}, with periodic 
boundary conditions and `smooth' initial 
condition~\eqref{eq:tests.advection1D.ICs}.
The number of spatial cells in the domain is progressively increased from 
$\N_x=32$ to $\N_x=2048$. 
We compare numerical solutions obtained with the `P4' scheme, a 4th-order CS 
that uses parabolic interpolation over a 3-point stencil, with and without the 
use of the positivity limiter~\eqref{eq:PositivityLimiter}.
The `$L^2$ error' columns report the maximum value in time of the $L^2$-norm 
of the error, according to~\eqref{eq:tests.advection1D.ErrNorm}.
The `Order' columns refer to the algebraic order of convergence, computed as 
the base-2 logarithm of the ratio of two successive error norms.
The `Min value' columns contain the minimum value in time and space of the 
numerical solution (the exact minimum being 
approximately~$6.93\times 10^{-31}$); a reported $0$ value corresponds exactly 
to the double-precision zero.
All simulations employ a constant Courant parameter 
$C = \Delta t / \Delta x = 0.32$.
\label{table:Convergence1D-P4}
}
\end{center}
\end{table}

\begin{table}[!htb]
\newcommand{\pz}{\phantom{0}}
\newcommand{\pd}{\phantom{-}}
\begin{center}
\setlength{\bigstrutjot}{1.5pt}
\begin{tabular}{|r|ccc|ccc|}
	\hline
	\bigstrut[t]
	 & \multicolumn{3}{c|}{\bf{P6}} & 
	   \multicolumn{3}{c|}{\bf{P6 -- no limiter}}\\
	\cline{2-7}
	\bigstrut[t]
	\bf{N$_x$} & $\bm{L^2}$ \bf{error} & \bf{Order} & \bf{Min value}
	           & $\bm{L^2}$ \bf{error} & \bf{Order} & \bf{Min value}\\
	\hline
	\hline
	\bigstrut[t]
$  32$ & $7.68\times 10^{-2}$ &   ---  & $    0 $
       & $7.45\times 10^{-2}$ &   ---  & $   -3.86\times 10^{-2\pz}$\\
$  64$ & $2.55\times 10^{-2}$ & $1.59$ & $    0 $
       & $2.97\times 10^{-2}$ & $1.33$ & $   -2.48\times 10^{-2\pz}$\\
$ 128$ & $4.45\times 10^{-3}$ & $2.52$ & $    0 $
       & $4.45\times 10^{-3}$ & $2.74$ & $   -3.54\times 10^{-4\pz}$\\
$ 256$ & $2.14\times 10^{-4}$ & $4.37$ & $    0 $
       & $2.14\times 10^{-4}$ & $4.37$ & $   -7.40\times 10^{-16}$\\
$ 512$ & $7.03\times 10^{-6}$ & $4.93$ & $\pd 1.30\times 10^{-30}$
       & $7.03\times 10^{-6}$ & $4.93$ & $\pd 1.30\times 10^{-30}$\\
$1024$ & $2.21\times 10^{-7}$ & $4.99$ & $\pd 9.10\times 10^{-31}$
       & $2.21\times 10^{-7}$ & $4.99$ & $\pd 9.10\times 10^{-31}$\\
$2048$ & $6.93\times 10^{-9}$ & $5.00$ & $\pd 8.13\times 10^{-31}$
       & $6.93\times 10^{-9}$ & $5.00$ & $\pd 8.13\times 10^{-31}$\\
\hline
\end{tabular}
\caption{
Convergence analysis as in Table \ref{table:Convergence1D-P4}, 
but for the `P6' scheme, a 6th-order CS that uses quartic interpolation 
over a 5-point stencil.
\label{table:Convergence1D-P6}
}
\end{center}
\end{table}

\begin{table}[!htb]
\newcommand{\pz}{\phantom{0}}
\newcommand{\pd}{\phantom{-}}
\begin{center}
\setlength{\bigstrutjot}{1.5pt}
\begin{tabular}{|r|ccc|ccc|}
	\hline
	\bigstrut[t]
	 & \multicolumn{3}{c|}{\bf{F22}} & 
	   \multicolumn{3}{c|}{\bf{F22 -- no limiter}}\\
	\cline{2-7}
	\bigstrut[t]
	\bf{N$_x$} & $\bm{L^2}$ \bf{error} & \bf{Order} & \bf{Min value}
	           & $\bm{L^2}$ \bf{error} & \bf{Order} & \bf{Min value}\\
	\hline
	\hline
	\bigstrut[t]
$  32$ & $2.47\times 10^{-2\pz}$ &       ---  & $ 0 $
       & $1.95\times 10^{-2\pz}$ &       ---  & $-2.82\times 10^{-2\pz}$\\
$  64$ & $1.84\times 10^{-4\pz}$ & $\pz 7.07$ & $ 0 $
       & $1.96\times 10^{-4\pz}$ & $\pz 6.64$ & $-1.53\times 10^{-4\pz}$\\
$ 128$ & $7.55\times 10^{-11  }$ & $   21.22$ & $ 0 $
       & $7.55\times 10^{-11  }$ & $   21.30$ & $-2.18\times 10^{-12}$\\
$ 256$ & $1.02\times 10^{-13  }$ & $\pz 9.54$ & $ 0 $
       & $1.02\times 10^{-13  }$ & $\pz 9.54$ & $-4.39\times 10^{-16}$\\
$ 512$ & $1.20\times 10^{-13  }$ & (\emph{m.p.}) & $ 0 $
       & $1.20\times 10^{-13  }$ & (\emph{m.p.}) & $-3.59\times 10^{-16}$\\
$1024$ & $4.44\times 10^{-13  }$ & (\emph{m.p.}) & $ 0 $
       & $4.44\times 10^{-13  }$ & (\emph{m.p.}) & $-2.77\times 10^{-16}$\\
$2048$ & $4.20\times 10^{-13  }$ & (\emph{m.p.}) & $ 0 $
       & $4.20\times 10^{-13  }$ & (\emph{m.p.}) & $-3.03\times 10^{-16}$\\
\hline
\end{tabular}
\caption{
Convergence analysis as in Tables \ref{table:Convergence1D-P4} and 
\ref{table:Convergence1D-P6}, but for the `F22' scheme, a 22nd-order CS based 
on fast Fourier transforms.
Since the F22 scheme reaches machine precision (\emph{m.p.}) for $\N_x = 256$, 
the error does not decrease further beyond that point.
\label{table:Convergence1D-F22}
}
\end{center}
\end{table}

\subsubsection{Long-time integration}
\label{sec:NumericalTests.Advection1D.LongTimeInt}

We now test the performance of the proposed numerical schemes at low 
resolution levels (i.e.\ coarse grid spacing), for very long-time 
integration.
For this purpose, we solve the 1D constant advection 
equation~\eqref{eq:tests.advection1D.eq} until the final time $T=100$, and we 
use initial conditions $n(0,x) = n_0(x)$ with compact support and of different 
degree of smoothness: first we test a 6th-order cosine bell
\begin{equation}\label{eq:tests.advection1D.LongTime.Cos6Bell}
n_0(x) \,=\, 0.1 + 
  \begin{cases}
     \cos(2\pi x)^6 & \text{if $|x|< 0.25$},\\
     0              & \text{otherwise},
  \end{cases}
\end{equation}
which is of class $\mathcal{C}^5$, and then we test a triangular profile
\begin{equation}\label{eq:tests.advection1D.LongTime.Triangle}
n_0(x) \,=\, 0.1 +
  \begin{cases}
    1-|4x| & \text{if $|x|< 0.25$},\\
    0     & \text{otherwise},
  \end{cases}
\end{equation}
which is of class $\mathcal{C}^1$.

\begin{figure}[htb!]
   \centering
   \includegraphics[width=\textwidth]{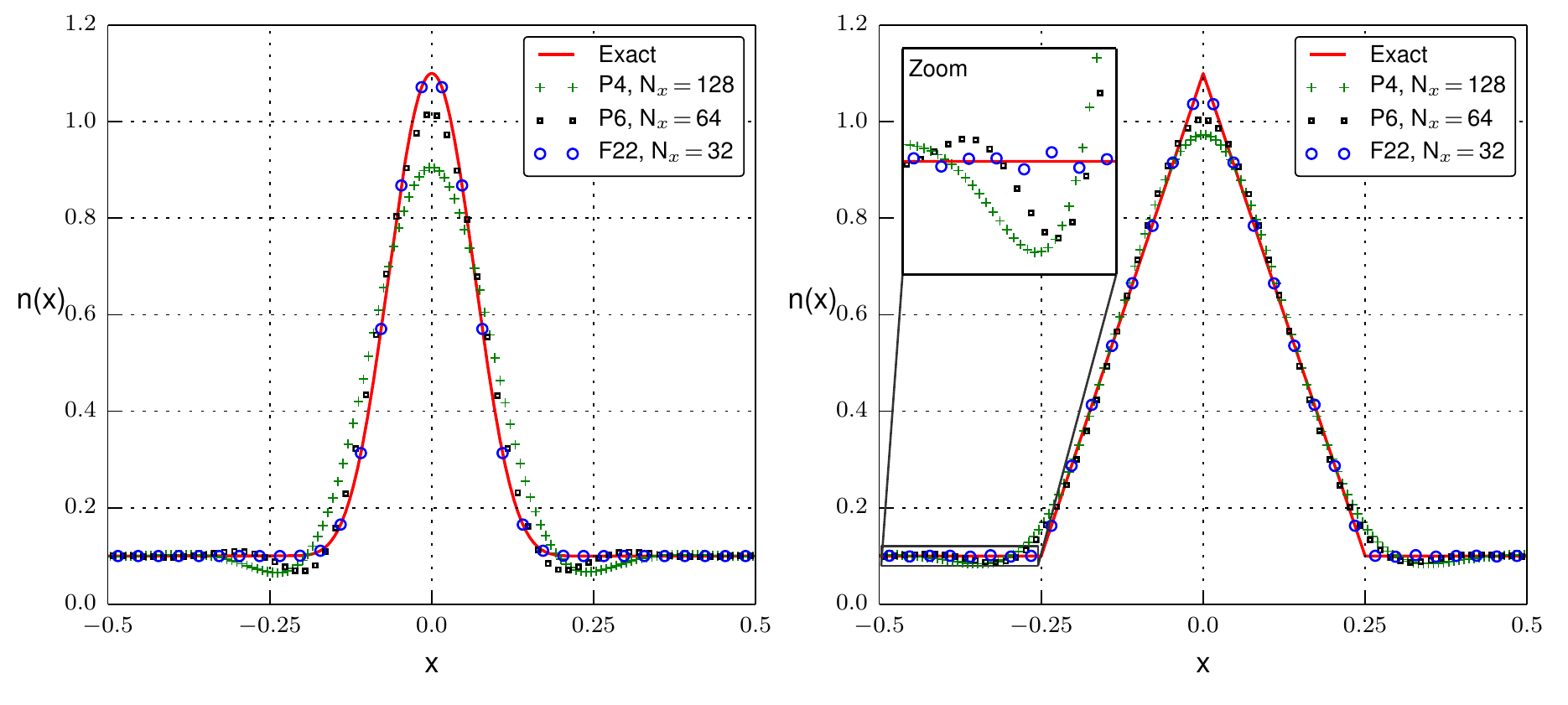}
   \caption{Long time integration of the 1D constant advection 
   equation~\eqref{eq:tests.advection1D.eq} with periodic boundary conditions,
   and initial conditions \eqref{eq:tests.advection1D.LongTime.Cos6Bell} (left 
   plot) and~\eqref{eq:tests.advection1D.LongTime.Triangle} (right plot).
   The exact solutions at time $t=100$ are compared to the numerical solutions 
   obtained with high-order versions of the CS: `P4' uses quadratic polynomial 
   interpolation over a three-point stencil, `P6' uses quartic polynomial 
   interpolation over a five-point stencil, and `F22' is based on FFTs.}
   \label{fig:test1d.cos6&triH.final}
\end{figure}

For all of these simulations we use a constant Courant parameter 
$C = \Delta t / \Delta x = 0.32$, but different mesh spacings depending 
on the nominal order of accuracy: the P4 scheme uses $\N_x = 128$ cells in the 
domain, the P6 scheme uses 64, and the F22 scheme uses only 32.
Figure~\ref{fig:test1d.cos6&triH.final} compares the 
numerical solutions at the final time $T=100$ with the exact solutions.
In both test-cases, F22 with 32 cells outperforms P6 with 64 cells, which in 
turn outperforms P4 with 128 cells.

The right plot of Figure~\ref{fig:test1d.cos6&triH.final} includes an 
enlargement of the nominally flat 
region to the left of the triangle, where all three schemes show spurious 
oscillations: while the long-wavelength oscillations of the P4 and P6 schemes 
are characteristic of their high-order diffusive behavior, the F22 scheme 
introduces ringing artifacts at very short wavelengths (close to the Nyquist 
value of $2\Delta x$) which are caused by the FFT/IFFT process.
In other words, the error in the P4 and P6 schemes comes from the truncation 
error, while the error in the F22 scheme is due to the Gibbs phenomenon; for 
this reason we sometimes refer to the latter scheme as a `Spectral-CS'.

The amplitude of Gibbs' oscillations for the F22 scheme depends on the 
asymptotic behavior of the Fourier spectrum of the initial conditions $n_0(x)$, 
which in turn can be related to its `degree of smoothness'.
In general, if $n_0(x)$ is of class $\mathcal{C}^a$, then the spurious 
oscillations have amplitude $O(\Delta x^a)$; this implies that for 
discontinuous profiles the scheme is not uniformly convergent.
In the case of the 6th-order cosine bell initial 
conditions~\eqref{eq:tests.advection1D.LongTime.Cos6Bell}, Gibbs' oscillations 
have amplitude below $3\times 10^{-4}$.

\begin{figure}[htb!]
   \centering
   \includegraphics[width=\textwidth]{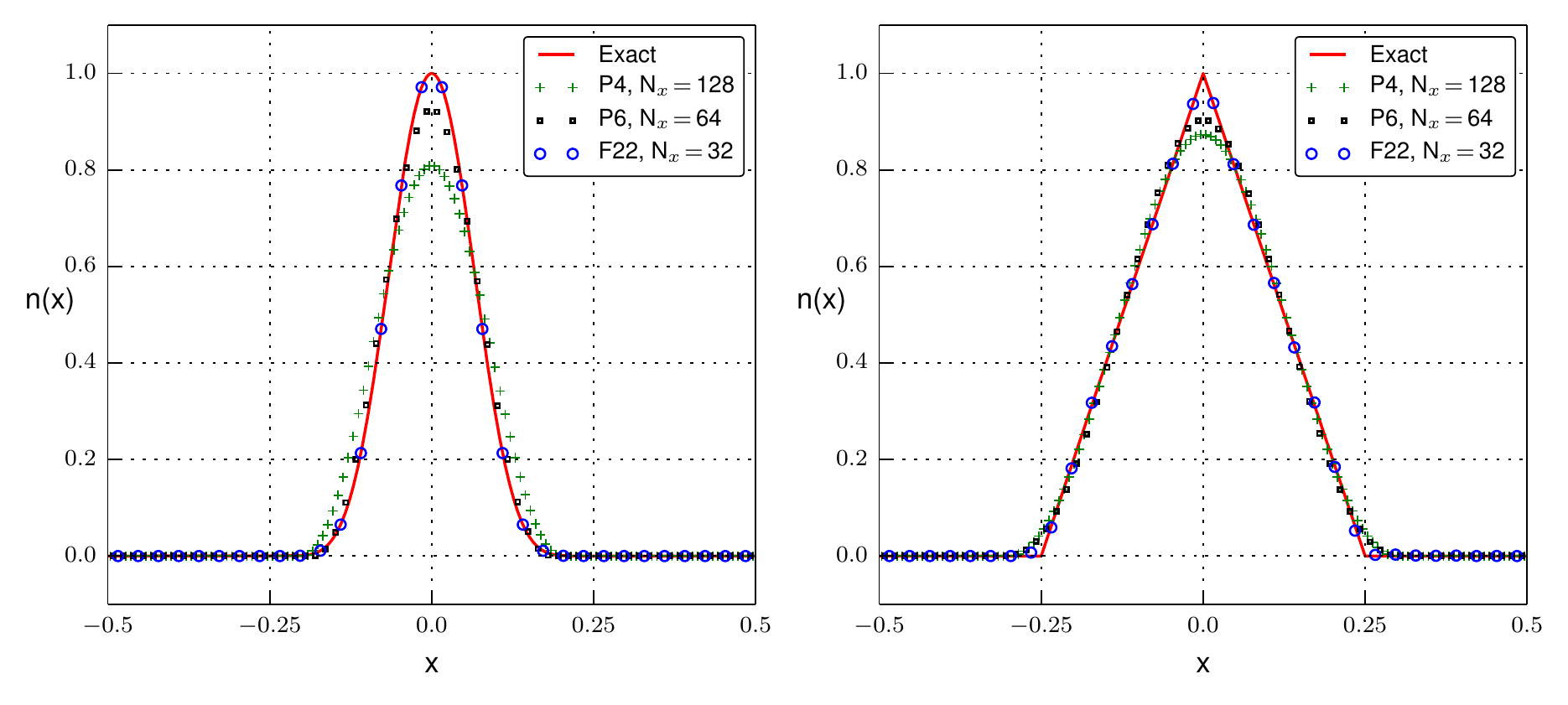}
   \caption{Long time integration of the 1D constant advection 
   equation~\eqref{eq:tests.advection1D.eq} with periodic boundary conditions,
   and initial condition given by the profiles 
   \eqref{eq:tests.advection1D.LongTime.Cos6Bell} (left) 
   and~\eqref{eq:tests.advection1D.LongTime.Triangle} (right), but with 
   baselines of $0$ instead of $0.1$.
   The exact solutions at time $t=100$ are compared to the numerical solutions 
   obtained with high-order versions of the CS: `P4' uses quadratic polynomial 
   interpolation over a three-point stencil, `P6' uses quartic polynomial 
   interpolation over a five-point stencil, and `F22' is based on FFTs.}
   \label{fig:test1d.cos6&triL.final}
\end{figure}

\begin{figure}[htb!]
   \centering
   \includegraphics[width=\textwidth]{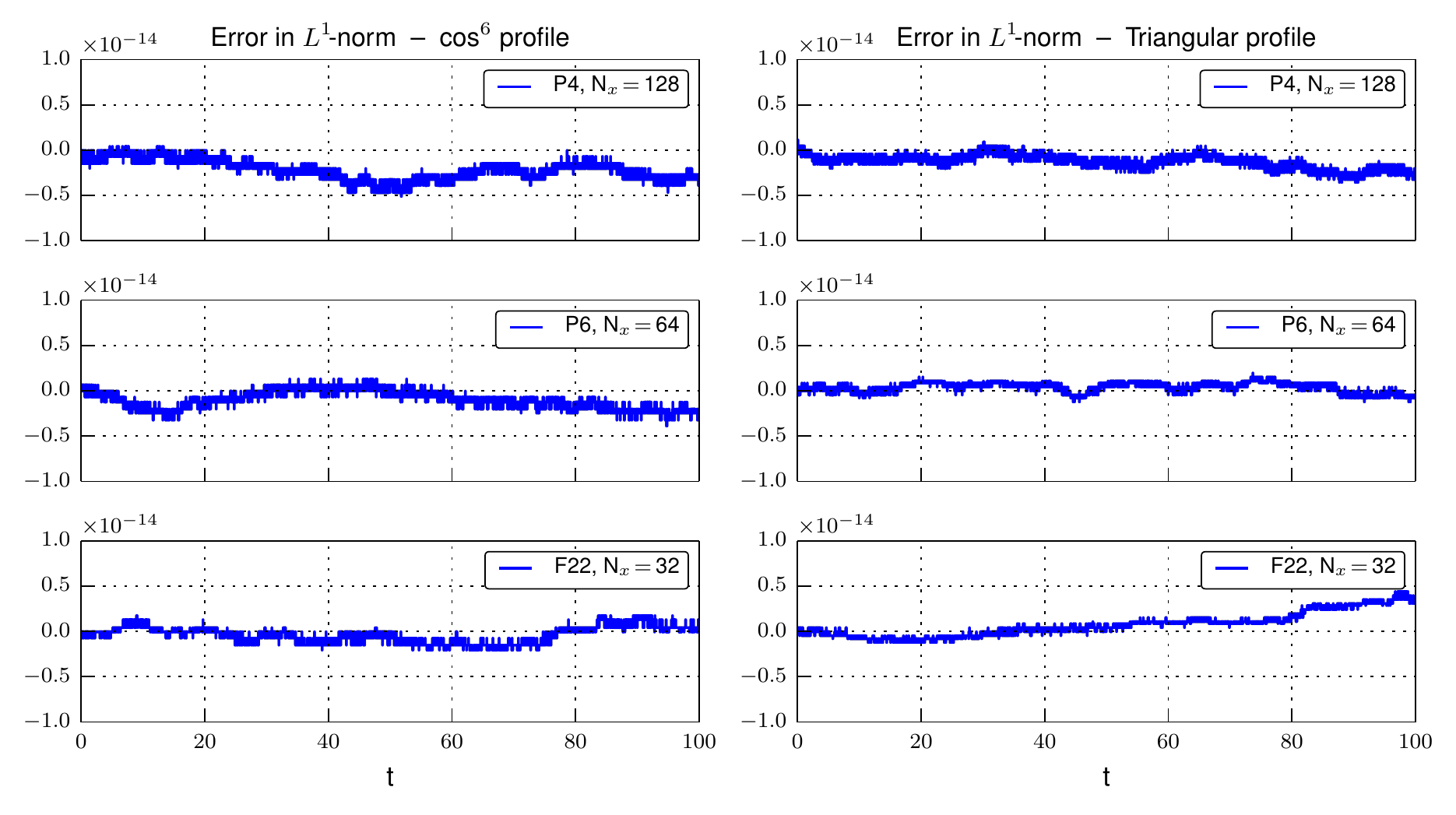}
   \caption{Long time integration of the 1D constant advection 
   equation~\eqref{eq:tests.advection1D.eq} with periodic boundary conditions,
   and initial condition given by the profiles 
   \eqref{eq:tests.advection1D.LongTime.Cos6Bell} (left) 
   and~\eqref{eq:tests.advection1D.LongTime.Triangle} (right), but with 
   baselines of $0$ instead of $0.1$.
   We plot the relative error in the $L^1$-norm of the numerical solutions
   obtained with high-order versions of the CS: `P4' uses quadratic polynomial 
   interpolation over a three-point stencil, `P6' uses quartic polynomial 
   interpolation over a five-point stencil, and `F22' is based on FFTs.}
   \label{fig:test1d.cos6&triL.invariants}
\end{figure}

As an aside, we point out that all the presented implementations of the CS 
strictly preserve the positivity of the solution if initial conditions 
$n_0(x) \ge 0$ are given; accordingly, the spurious oscillations here observed 
cannot possibly cause the numerical solution to become negative.
Such a desirable property is easily tested by progressively lowering the 
baseline of the initial profiles \eqref{eq:tests.advection1D.LongTime.Cos6Bell} 
or~\eqref{eq:tests.advection1D.LongTime.Triangle} from $0.1$ to $0$; 
in so doing, we have observed the following behavior:
\begin{itemize}
  \item the `super-diffusive' oscillations of the P4 and P6 schemes decrease 
        in amplitude until they disappear;
  \item the Gibbs oscillations of the F22 scheme get shifted upward, so that 
        their local minima remain positive while their amplitude is 
        substantially unchanged.
\end{itemize}
The final solutions for the two `lowered' profiles are reported side by side 
in Figure~\ref{fig:test1d.cos6&triL.final}.
Since the scheme is mass-conservative and positivity-preserving, in all our 
simulations the $L^1$-norm is only affected by round-off error in 
double-precision, and it would be exactly conserved if exact arithmetic were 
used.
In Figure~\ref{fig:test1d.cos6&triL.invariants} we show the relative error in 
the $L^1$-norm, which remains below $10^{-14}$ for the whole length of the 
simulations.

\subsubsection{Discontinuous solution: propagation of oscillations}
\label{sec:NumericalTests.Advection1D.oscillations}

We now intend to analyze the behavior of the F22 scheme under the most 
challenging conditions, namely a discontinuous profile.
Accordingly, we solve the 1D constant advection 
equation~\eqref{eq:tests.advection1D.eq} until the final time $T=1$, with 
initial condition $n(0,x) = n_0(x)$ given by the sum of a rectangular profile 
and a Gaussian bell:
\begin{equation}\label{eq:tests.advection1D.rect+bell}
  n_0(x) \,=\, 0.1 + 
  \begin{cases}
	1.0                             & \text{if $-0.4 \le x \le -0.2$}, \\
	e^{-\Par{\frac{x-0.2}{0.04}}^2} & \text{if $-0.1 \le x \le  0.5$}, \\
	0                               & \text{otherwise}.
  \end{cases}
\end{equation}
Because of the rectangular wave, an FFT-based numerical scheme will naturally 
introduce large spurious oscillations (the Gibbs phenomenon), which inevitably 
deteriorate even the smooth regions of the solution.
This issue can be mitigated (and eventually suppressed) by increasing the 
strength of the filter, but at the cost of introducing excessive numerical 
diffusion in the smooth regions.
As a trade-off, we use a relatively weak filter, with the purpose of 
confining the Gibbs oscillations to a finite interval around the 
discontinuities.

\begin{figure}[htb!]
   \centering
   \includegraphics[width=\textwidth]{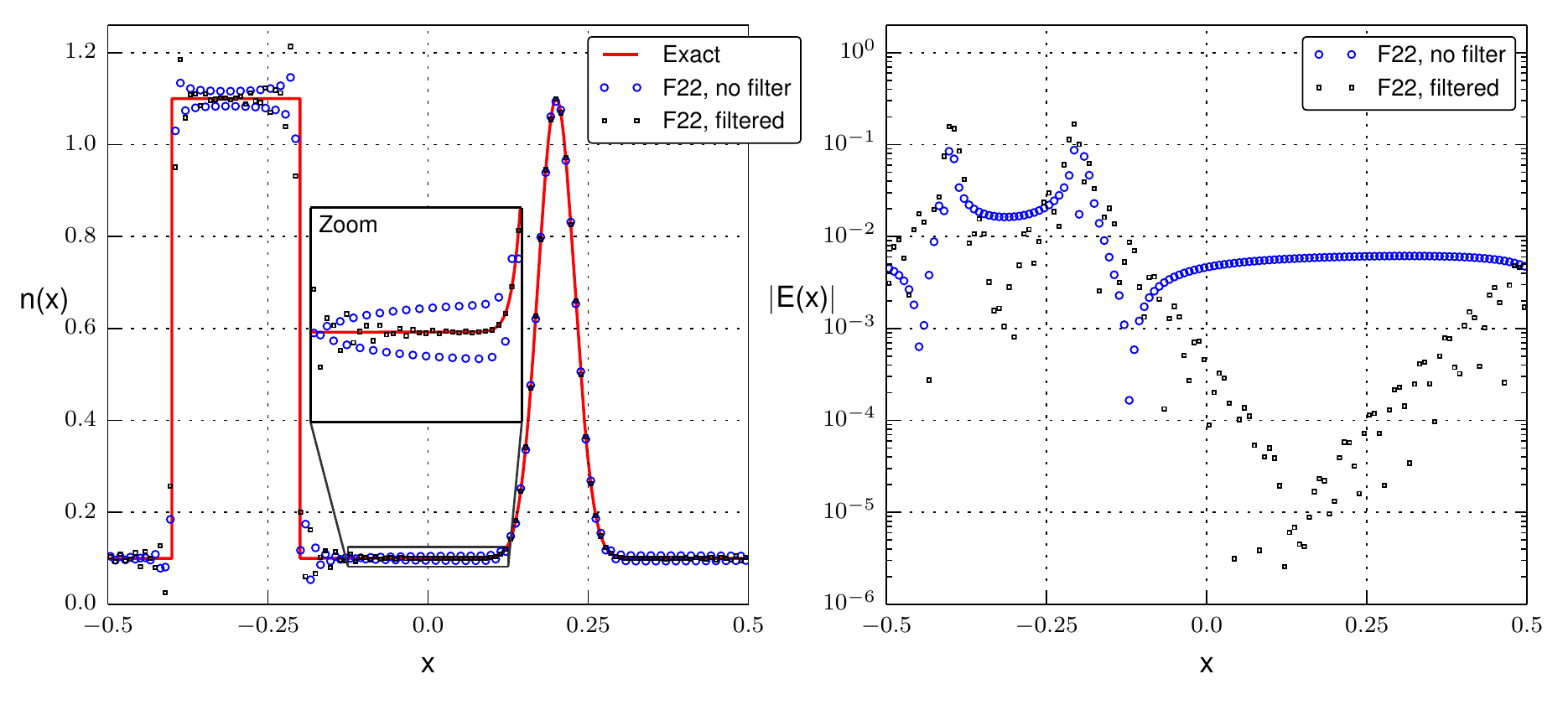}
   \caption{Time integration of the 1D constant advection 
   equation~\eqref{eq:tests.advection1D.eq} with periodic boundary conditions,
   and initial condition given by~\eqref{eq:tests.advection1D.rect+bell}.
   The exact solution at time $t=1$ is compared to the 
   numerical solutions obtained with the F22 scheme (FFT-based, 
   with nominally 22nd order truncation error), with and without using 
   filtering.
   A constant CFL parameter of $C = \Delta t / \Delta x = 0.32$ is used, 
   and the computational domain is subdivided into $\N_x = 128$ cells.   
   The left plot shows the solution itself, with a zoom on the central region, 
   while the right plot shows the absolute value of the local error 
   $E(x) := n_\text{ex}(t\!=\!1,x)-n_\text{CS}(t\!=\!1,x)$ on a logarithmic scale.   
   By virtue of the filtering process, the amplitude of the spurious 
   oscillations introduced by the rectangular profile decreases exponentially 
   fast in space from the jump discontinuities, allowing the scheme to 
   accurately resolve the Gaussian bell on the right-hand side of the domain.
   }
   \label{fig:test1d.rect+bell}
\end{figure}

The efficacy of this approach is shown in Figure~\ref{fig:test1d.rect+bell}. 
On the left plot, the numerical solutions obtained using the F22 scheme, with 
and without filtering, are compared to the exact solution at the final time 
$T=1$.
In addition, the right plot shows the absolute 
value of the local error in the final solution, on a logarithmic scale.
The simulations were run with a constant CFL parameter of 
$C = \Delta t / \Delta x = 0.32$, and the $x$ axis was subdivided into 
$\N_x = 128$ cells.

Both numerical schemes appear very noisy in the vicinity of the rectangular 
profile, but while the unfiltered solution is corrupted in the whole domain, 
filtering successfully suppresses the spurious oscillations within a finite 
distance (here approximately 30 cells) from the discontinuities; this permits 
us to accurately resolve the Gaussian bell.

\subsection{2D rotating advection}
\label{sec:NumericalTests.Advection2D}

We now focus on a simple two-dimensional advection problem, which we will 
solve by means of a time-splitting procedure similar in all regards to the one 
described in Section~\ref{sec:OperatorSplitting} for the Vlasov-Poisson 
equation: this will permit us to investigate the interaction between the 
time-integration algorithm and the basic 1D advection solver.
We consider the 2D continuity equation 
\begin{equation*}
  \frac{\partial n}{\partial t} + 
  \frac{\partial \Par{u n}}{\partial x} + 
  \frac{\partial \Par{v n}}{\partial y} \,=\, 0,
\end{equation*}
where $n(t,x,y)$ is the scalar density, while $u(t,x,y)$ and $v(t,x,y)$ are the 
$x$ and $y$ components, respectively, of the vector velocity $\vect{u}(t,x,y)$.
If we assume $\vect{u}$ to be a rotating flow-field with constant angular 
velocity $\omega$ and center of rotation $(x_0,y_0)$, so that
\begin{equation*}
  u(y) \,=\, -\Par{y-y_0}\omega, \qquad
  v(x) \,=\,  \Par{x-x_0}\omega,
\end{equation*}
then the 2D continuity equation can be dimensionally split into two 2D 
`transverse advection equations'
\begin{equation*}
  \frac{\partial n}{\partial t} + u(y)\frac{\partial n}{\partial x} = 0,\qquad
  \frac{\partial n}{\partial t} + v(x)\frac{\partial n}{\partial y} = 0.
\end{equation*}
After discretization of the spatial variable $y$, the first of the equations 
above can be interpreted as a family of 1D constant advection equations along 
the direction $x$, parametric in the velocity $u_j \equiv u(y_j)$.
Similarly, after discretization of $x$ in the second equation, we obtain a 
family of 1D constant advection equations along the direction $y$, parametric 
in the velocity $v_i \equiv v(x_i)$.

Given the solution at time $t$, we will obtain the solution at time 
$t+\Delta t$ by alternately evolving the two families of equations over 
prescribed substeps: the full algorithm satisfies appropriate order conditions,
which ensure that the splitting error is of the required order of accuracy.
The 1D constant advection equations will be accurately solved using the CS, 
which guarantees that the mass is conserved and the solution remains positive.

In the following tests, we choose $\omega = 2\pi$ in order to complete a full 
rotation at time $t = 1$, and for simplicity we choose $(x_0,y_0) = (0,0)$.
By computing the solution on the square domain $[-1,1] \times [-1,1]$, and 
giving an initial condition with compact support within the circle $x^2+y^2 
\le 1$, the non-zero portion of the exact solution will always be contained 
in the domain.
The resulting \emph{2D rotating advection equation} is
\begin{equation}\label{eq:tests.advection2D.eq}
  \frac{\partial n}{\partial t} - 
  \Par{2\pi y}\frac{\partial n}{\partial x} + 
  \Par{2\pi x}\frac{\partial n}{\partial y} = 0, \qquad
  (x,y) \in [-1,1] \times [-1,1], \qquad
  t \in [0,T],
\end{equation}
where the final time $T>0$ depends on the test-case.
The initial condition $n(0,x,y) = n_0(x,y)$ is given by the superposition of 
two identical 22nd-order cosine bells with elliptical cross-section, having 
the major axes aligned along the $x$ and $y$ axes, respectively.
The resulting `22nd-order cosine-cross' is
\begin{subequations}\label{eq:tests.advection2D.cosine-cross}
\begin{align}
  n_0(x,y) \,&=\, 0.5\,B\bigl(r_1(x,y)\bigr) \,+\, 0.5\,B\bigl(r_2(x,y)\bigr), \\
  B(r) &= \begin{cases}
  	        \cos\Par{\frac{\pi r}{2a}}^{22} & \text{if $r \le a$}, \\
	        0                & \text{otherwise},
          \end{cases} \\
  r_1(x,y) &= \sqrt{  (x-x_c)^2 + 8 (y-y_c)^2}, \\
  r_2(x,y) &= \sqrt{8 (x-x_c)^2 +   (y-y_c)^2},
\end{align}
\end{subequations}
where the length of the major axes is $2a=1$, and the center of the profile is 
$(x_c,y_c) = (0.5,0)$.
The initial condition~\eqref{eq:tests.advection2D.cosine-cross} is of class 
$\mathcal{C}^{21}\!\Par{\mathbb{R}^2}$, which ensures the truncation error of 
the F22 algorithm is nominally $O(\Delta x^{22},\Delta y^{22})$.

\begin{table}[!ht]
\begin{center}
{\setlength{\bigstrutjot}{1pt}
\begin{tabular}{|c|l|c|c|c|}
\hline
\bigstrut[t]
\bf{Label} & \bf{Description} & \bf{Order} & \bf{Stages} & \bf{Refs.} \\
\hline
\hline
\bigstrut[t]
LF2   & Leap-frog / Strang / Störmer-Verlet & 2 & 1 & \cite{Strang1968} \\
Y4    & Triple-jump composition of LF2
      & 4 & 3 & \cite{ForestRuth1990,Yoshida1990,CandyRozmus1991} \\
O6-4  & 4th-order RKN, optimized & 4 & 6  & \cite{BlanesMoan2002} \\
O11-6 & 6th-order RKN, optimized & 6 & 11 & \cite{BlanesMoan2002} \\
O14-6 & 6th-order RKN, optimized & 6 & 14 & \cite{BlanesMoan2002} \\
\hline
\end{tabular}}
\caption{
Symmetric splitting methods employed in the simulations.
The `label' column refers to the names used in the figures; for an easier 
comparison, we borrow our labels from~\cite{BlanesMoan2002}.
RKN is an acronym for `Runge-Kutta-Nyström'.
All integrators are symplectic.
\label{table:TimeSplittingIntegrators}}
\end{center}
\end{table}

\subsubsection{Time convergence and efficiency}
\label{sec:NumericalTests.Advection2D.refinement}

We now solve~\eqref{eq:tests.advection2D.eq} with initial 
condition~\eqref{eq:tests.advection2D.cosine-cross} until the final time $T=1$, 
using five different symplectic integrators, of order from 2 to 6.
For each of the five integrators we run a series of simulations with a 
progressively smaller time-step, and we verify that the numerical solution 
converges to the exact analytical profile.
The non-dimensional quantity that we will monitor is the number of time-steps 
per rotation, $\N_t := (2\pi/\omega) / \Delta t = 1/ \Delta t$.

We want all simulations to rely on `well resolved' 1D solutions, with a 
spatial error below machine precision: for this purpose we use a computational 
grid of $(\N_x,\N_y) = (256,256)$ cells, and we employ the F22 scheme 
(the 22nd-order `Spectral' CS) as our base 1D constant advection solver.

Since the exact solution is rigidly rotating with constant angular velocity, 
we can compute the numerical error at all (discrete) time instants $t_k$ and 
mesh locations $(x_i,y_j)$.
A scalar measure of the error is obtained by first taking its 
$L^2$-norm in space, and then the $L^\infty$-norm in time,
\begin{equation}\label{eq:tests.advection2D.ErrNorm}
E \,=\, \max_k \left\{
\sqrt{\sum_{i,j} \Bigl[ n(t_k,x_i,y_j)-n_\textsc{cs}(t_k,x_i,y_j) \Bigr]^2 
      \Delta x\, \Delta y
     } \right\},
\end{equation}
which is to say that we compute the \emph{maximum in time} of the spatial 
$L^2$-norm of the error.

\begin{figure}[htb!]
   \centering
   \includegraphics[width=0.75\textwidth]{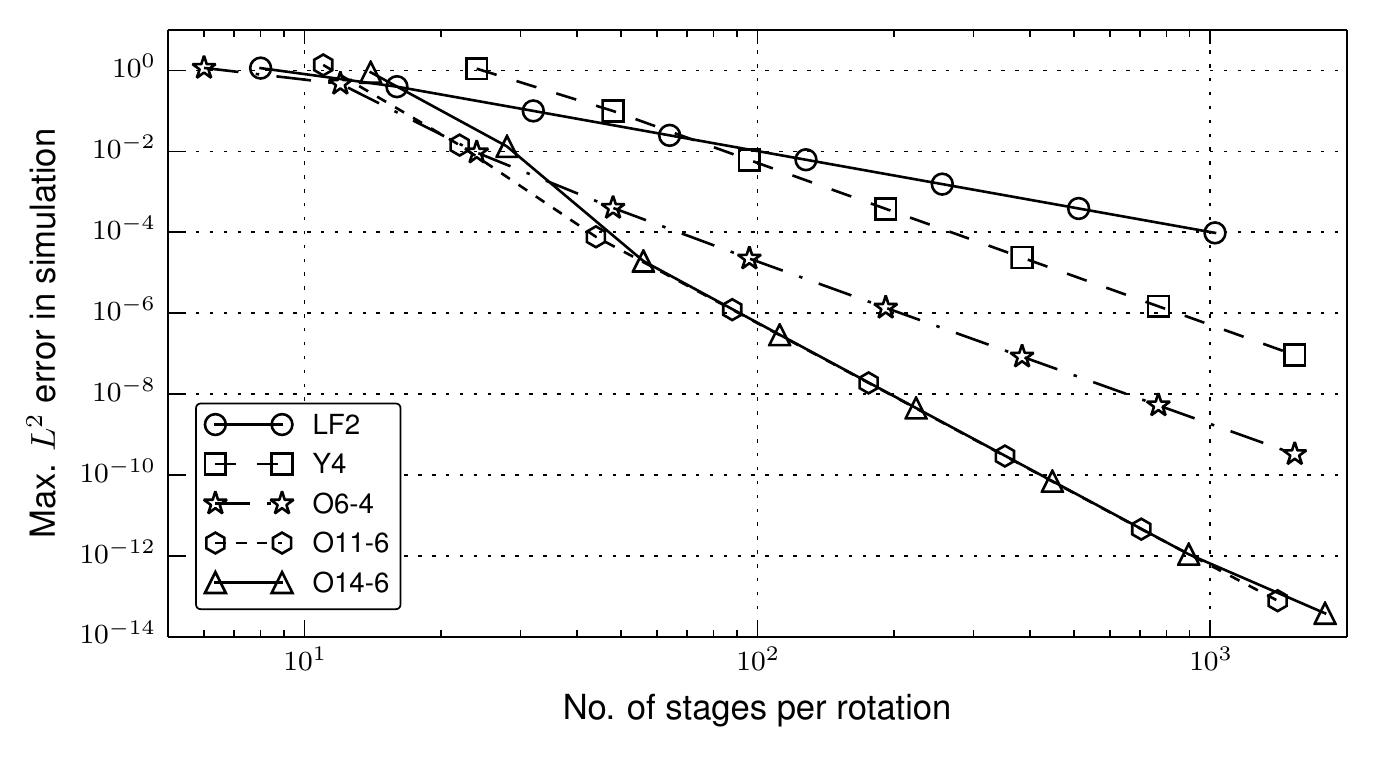}
   \caption{2D rotating advection test-case: comparison of symmetric splitting 
   time integration schemes.
   Equation~\eqref{eq:tests.advection2D.eq} is given initial 
   condition~\eqref{eq:tests.advection2D.cosine-cross}, and it is integrated 
   in time until $t=1$.
   All simulations employ the F22 scheme as base 1D solver, with a spatial 
   discretization of $(\N_x,\N_y) = (256,256)$ cells.
   The $y$ axis reports the maximum $L^2$-norm of the error during the 
   simulation, according to~\eqref{eq:tests.advection2D.ErrNorm};
   the $x$ axis reports the total number of stage evaluations in one full 
   rotation, to which the computational cost of the scheme is linearly 
   proportional.
   A brief description of the different integrators, complete with references, 
   is given in Table~\ref{table:TimeSplittingIntegrators}.
   }
   \label{fig:test2d.cos22cross.efficiency}
\end{figure}

In order to assess the computational efficiency of the different 
time-splitting algorithms as $\Delta t$ is decreased, we must consider that 
the computational cost for evolving the solution over one time-step is 
linearly proportional to the number of stages $S$ of the scheme.
Accordingly, Figure~\ref{fig:test2d.cos22cross.efficiency} reports an accurate 
comparison based on the total number of stages in one rotation, $S\,\N_t$.
Thanks to the logarithmic plot we can easily see that the error 
measure~\eqref{eq:tests.advection2D.ErrNorm} decreases with the expected 
algebraic rates in the interval $S\,\N_t \in [100,1000]$: 2nd-order for LF2, 
4th-order for Y4 and O6-4, 6th-order for O11-6 and O14-6.
In line with what was reported in~\cite{BlanesMoan2002}, we observe that O6-4 
performs consistently better than Y4: for the same amount of work, O6-4 
achieves an error that is approximately 250 times smaller; and for the same 
accuracy, it requires about 4 times fewer computations.
With regard to the 6th-order schemes (O11-6 and O14-6), they appear to be 
completely equivalent when $S\,\N_t > 50$, but O11-6 performs slightly better 
at lower resolutions.

\subsubsection{Long-time integration and conserved quantities}
\label{sec:NumericalTests.Advection2D.LongTimeInt}

Again, we solve~\eqref{eq:tests.advection2D.eq} with initial 
condition~\eqref{eq:tests.advection2D.cosine-cross}, using the F22 CS as the 
base 1D advection solver on a mesh of $(\N_x,\N_y) = (256,256)$ cells; but 
differently than before, we now perform a longer integration, until the final
time $T=100$, and we analyze the conservation properties of the scheme when 
comparatively large time steps are employed.
The numerical solution is advanced using three splitting schemes from 
Table~\ref{table:TimeSplittingIntegrators}: LF2, O6-4 and O11-6.
The time-step is chosen so that these have identical computational cost: LF2 
uses $\N_t = 1/\Delta t = 33$ time-steps per rotation, O6-4 has $\N_t = 5.5$ 
and O11-6 has $\N_t = 3$.

\begin{figure}[htb!]
   \centering
   \includegraphics[width=0.75\textwidth]{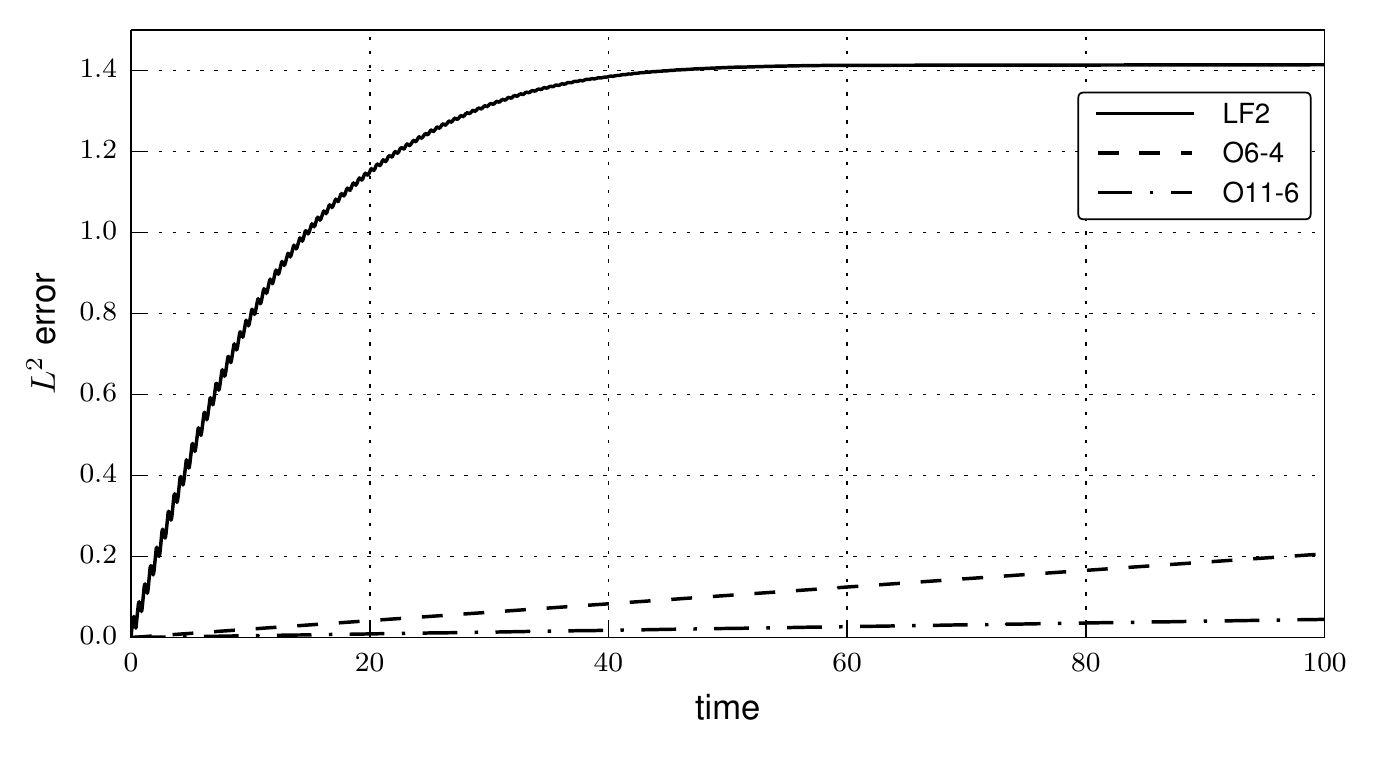}
   \caption{2D rotating advection: long-time growth of the error in the 
   numerical solutions.
   The spatial $L^2$-norm of the error is computed at each time step and 
   plotted versus time.
   Equation~\eqref{eq:tests.advection2D.eq} is given initial 
   condition~\eqref{eq:tests.advection2D.cosine-cross}, and it is integrated 
   in time until $t=100$, using three different symplectic schemes:
   LF2 is the common 2nd-order Strang splitting algorithm~\cite{Strang1968}, 
   O6-4 is a 6-stage 4th-order Runge-Kutta-Nyström (RKN) method, and O11-6 is 
   an 11-stage 6th-order RKN method~\cite{BlanesMoan2002}.
   All simulations employ the F22 scheme as their base 1D solver, and a 
   spatial discretization of $(\N_x,\N_y) = (256,256)$ cells.
   The time step size is chosen so that the total number of stage evaluations, 
   and hence the computational cost, is the same for the three simulations: 
   LF2 uses $\N_t = 1/\Delta t = 33$ time steps per rotation, O6-4 uses 
   $\N_t = 5.5$, and O11-6 uses $\N_t = 3$.
   }
   \label{fig:test2d.cos22cross.error_growth}
\end{figure}

\begin{figure}[htb!]
   \centering
   \includegraphics[width=\textwidth]{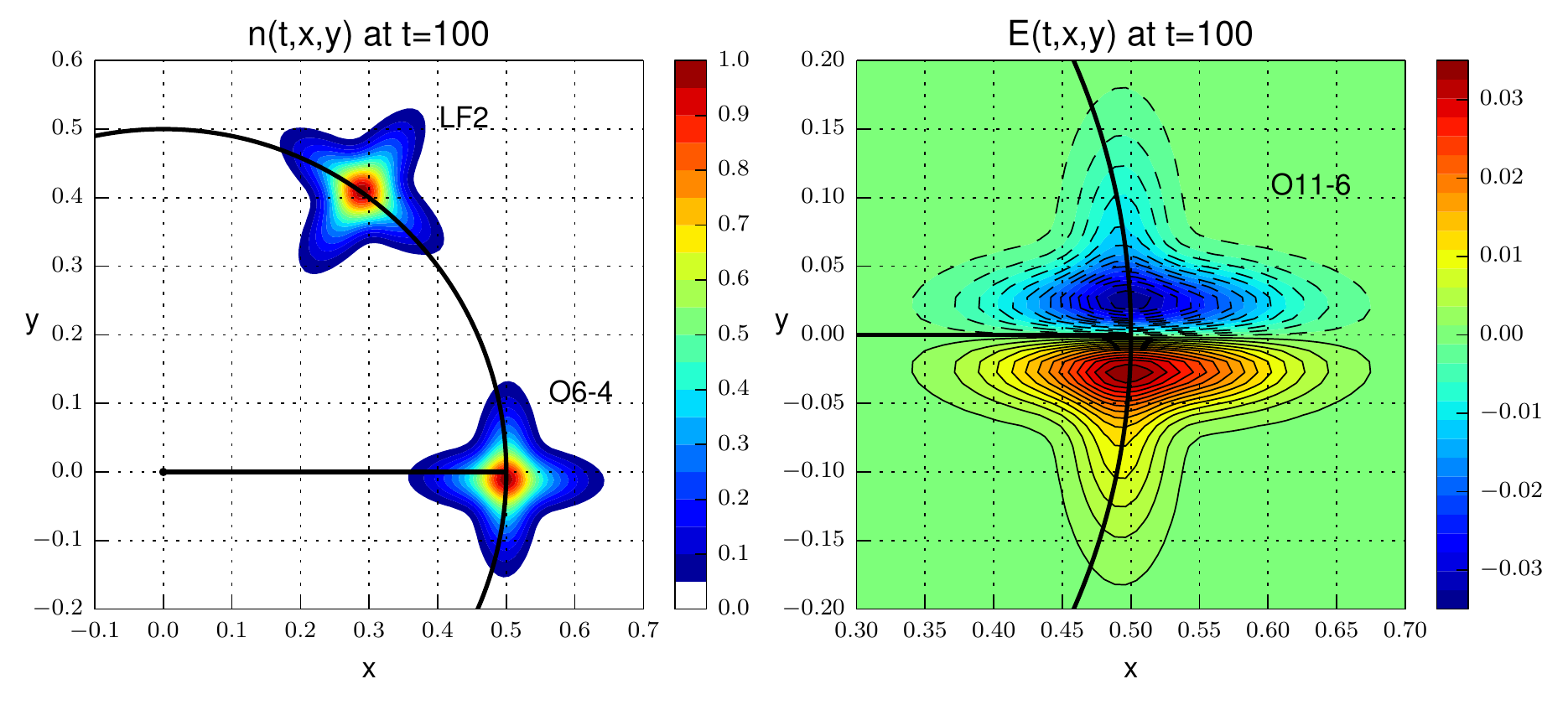}
   \caption{2D rotating advection: long-time drift of the numerical solution 
   along the exact circular trajectory.
   We plot the solution (left) and the error (right) after 100 rotations.
   Equation~\eqref{eq:tests.advection2D.eq} is given initial 
   condition~\eqref{eq:tests.advection2D.cosine-cross}, and it is integrated 
   in time until $t=100$, using three different symplectic schemes from 
   Table~\ref{table:TimeSplittingIntegrators}: LF2, O6-4 and O11-6.
   All simulations employ the F22 scheme as their base 1D solver, and a 
   spatial discretization of $(\N_x,\N_y) = (256,256)$ cells.
   Moreover, the LF2 integrator uses $\N_t = 1/\Delta t = 33$ time steps per 
   rotation, while O6-4 uses $\N_t = 5.5$ and O11-6 uses $\N_t=3$, so that 
   they have identical computational cost.
   For reference, in both plots the solid black curve represents the circular 
   path described by the peak of the exact solution, while the straight black 
   line connects the center of rotation to the instantaneous location of the 
   peak.
   The left-hand plot depicts the final numerical solutions for LF2 and O6-4 
   (for the sake of clarity, the first contour level is at 0.05):
   both numerical solutions have correct radial location and tangential 
   alignment, but the LF2 scheme introduces substantial displacement along the 
   exact trajectory (phase error).
   The right-hand plot shows the final error for O11-6: the spatial 
   distribution reveals a small phase-shift, but no appreciable diffusion or 
   dispersion.
   }
   \label{fig:test2d.cos22cross.final_solutions&error}
\end{figure}

Figure~\ref{fig:test2d.cos22cross.error_growth} illustrates the time evolution 
of the spatial $L^2$-norm of the error.
While the error of O6-4 and O11-6 grows linearly in time for the full length 
of the simulation, the error of LF2 saturates at a maximum value of 
approximately 1.4, which is just about twice the $L^2$-norm of the analytical 
solution itself; this suggests that the largest contribution to the error is 
due to the rigid displacement of the numerical solution with respect to the 
exact one.
This hypothesis is confirmed by the contour plot on the left-hand side of 
Figure~\ref{fig:test2d.cos22cross.final_solutions&error}, which depicts the 
final solution of LF2 and O6-4, enlarged on a portion of the computational 
domain: while the shape of the two profiles looks identical to the exact 
solution, to the eye, both solutions appear to be misplaced along the exact 
trajectory (phase error).
As expected, the LF2 scheme introduces a much larger phase error than O6-4 
does.

We do not plot the final solution computed using the O11-6 scheme, because it 
would be hardly distinguishable from the exact solution.
Instead, we provide a contour plot of the spatial error of O11-6 on the 
right-hand side of Figure~\ref{fig:test2d.cos22cross.final_solutions&error}.
The vertical antisymmetry of the error profile, and the absence of ringing 
phenomena, strongly suggest that the primary source of error is a small shift 
of the numerical solution toward negative values of $y$.

\begin{figure}[htb!]
   \centering
   \includegraphics[width=0.75\textwidth]{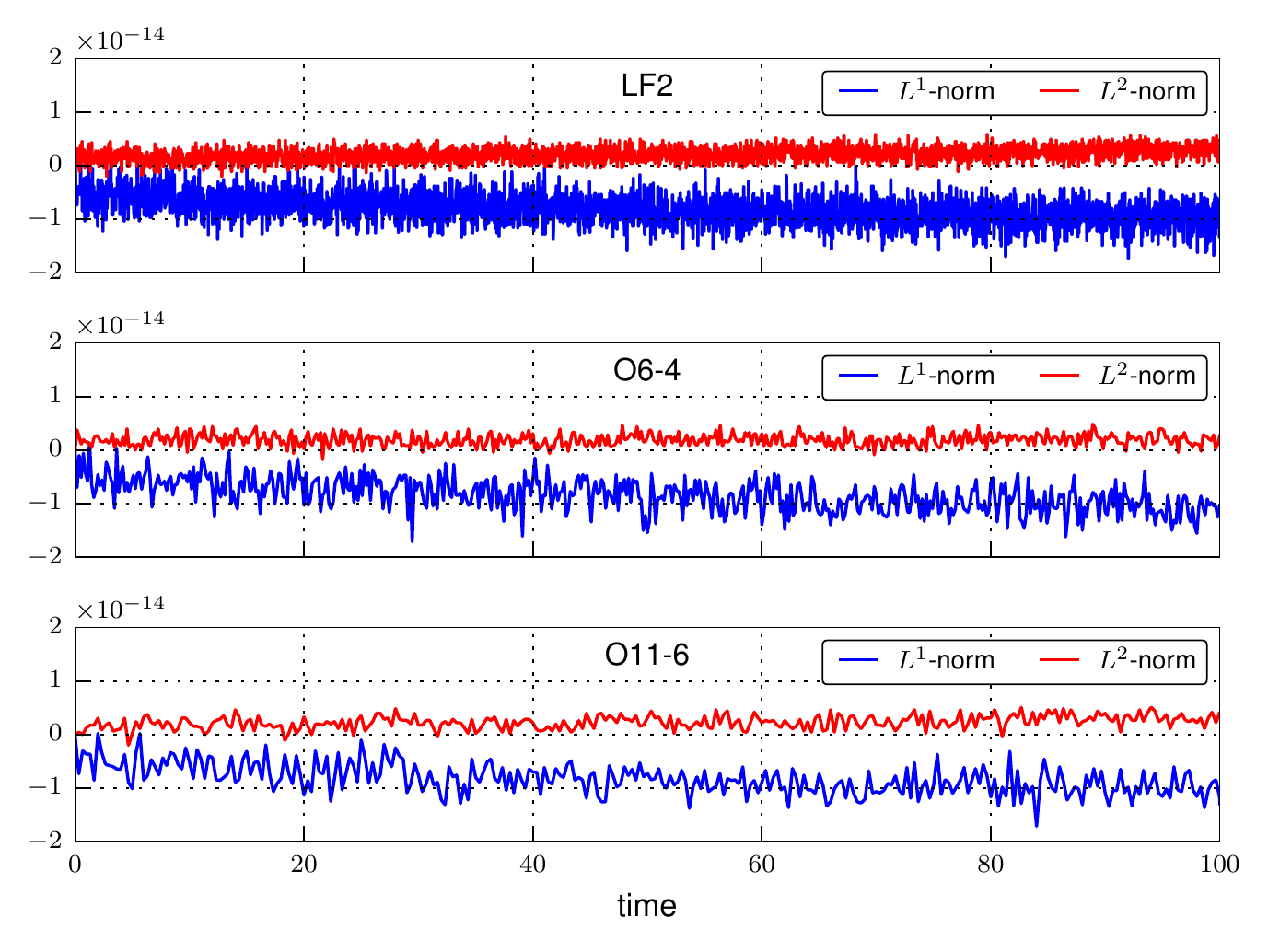}
   \caption{2D rotating advection: conservation of the invariants.
   We plot the relative deviation of the $L^1$ and $L^2$-norms of the 
   numerical solution from the initial values, versus time.
   Equation~\eqref{eq:tests.advection2D.eq} is given initial 
   condition~\eqref{eq:tests.advection2D.cosine-cross}, and it is integrated 
   in time until $t=100$, using three different symplectic schemes from 
   Table~\ref{table:TimeSplittingIntegrators}: LF2, O6-4 and O11-6.
   All simulations employ the F22 scheme as their base 1D solver, and a 
   spatial discretization of $(\N_x,\N_y) = (256,256)$ cells.
   Moreover, the LF2 integrator uses $\N_t = 1/\Delta t = 33$ time steps per 
   rotation, while O6-4 uses $\N_t = 5.5$ and O11-6 uses $\N_t=3$, so that 
   they have identical computational cost.
   In all three cases the invariants are conserved up to round-off error.
   }
   \label{fig:test2d.cos22cross.invariants}
\end{figure}

Finally, Figure~\ref{fig:test2d.cos22cross.invariants} reports the relative 
errors in two conserved quantities, the $L^1$- and $L^2$-norms of the solution, 
for the three simulations.
The base 1D solver is mass-conservative and positivity-preserving, and hence 
it exactly conserves the $L^1$-norm of the solution; any deviation from exact 
conservation is due to accumulation of round-off errors in a finite precision 
calculation.
Conservation of the $L^2$-norm of the solution was also achieved in all 
simulations, thanks to the combination of a spectrally-accurate 1D solver with 
a volume-preserving time integrator; because of the filtering process in 
Fourier space, the $L^2$-norm cannot be strictly conserved when the solution 
is under-resolved.

\subsection{1D-1V linear Vlasov equation with stationary field}
\label{sec:NumericalTests.LinearVlasov}
As an intermediate test between the 2D rotating advection equation and the 
1D-1V Vlasov-Poisson system, in this section we combine operator splitting and 
our base 1D spectral CS for solving the 1D-1V linear Vlasov equation, which 
takes the form 
\begin{equation}\label{eq:tests.LinearVlasov.GeneralEq}
  \Par{ \frac{\partial}{\partial t} + v \frac{\partial}{\partial x} + 
        \frac{q}{m} E(t,x) \frac{\partial}{\partial v} } f(t,x,v) = 0,
\end{equation}
where the electric field $E(t,x)$ is a prescribed function of time $t$ and 
space $x$.
As usual, $q$ and $m$ are the particle charge and mass of the species 
under consideration.
Since $E(t,x)$ does not depend on the distribution function $f(t,x,v)$, the 
above equation is indeed linear.

We will use~\eqref{eq:tests.LinearVlasov.GeneralEq} to describe the dynamics 
of electrons trapped in an electrostatic field, a situation that is ubiquitous 
in plasma physics.
Particularly important is the condition with a stationary electrostatic field 
$E(x) = -d\phi(x)/dx$, where the electrons remain indefinitely confined in 
the absence of collisions: as we will see, characteristic trajectories in 
phase-space describe closed stationary curves, which coincide with the level 
sets of a total energy function $W_\text{tot}(x,v)$.

\begin{figure}[htb!]
   \centering
   \includegraphics[width=\textwidth]{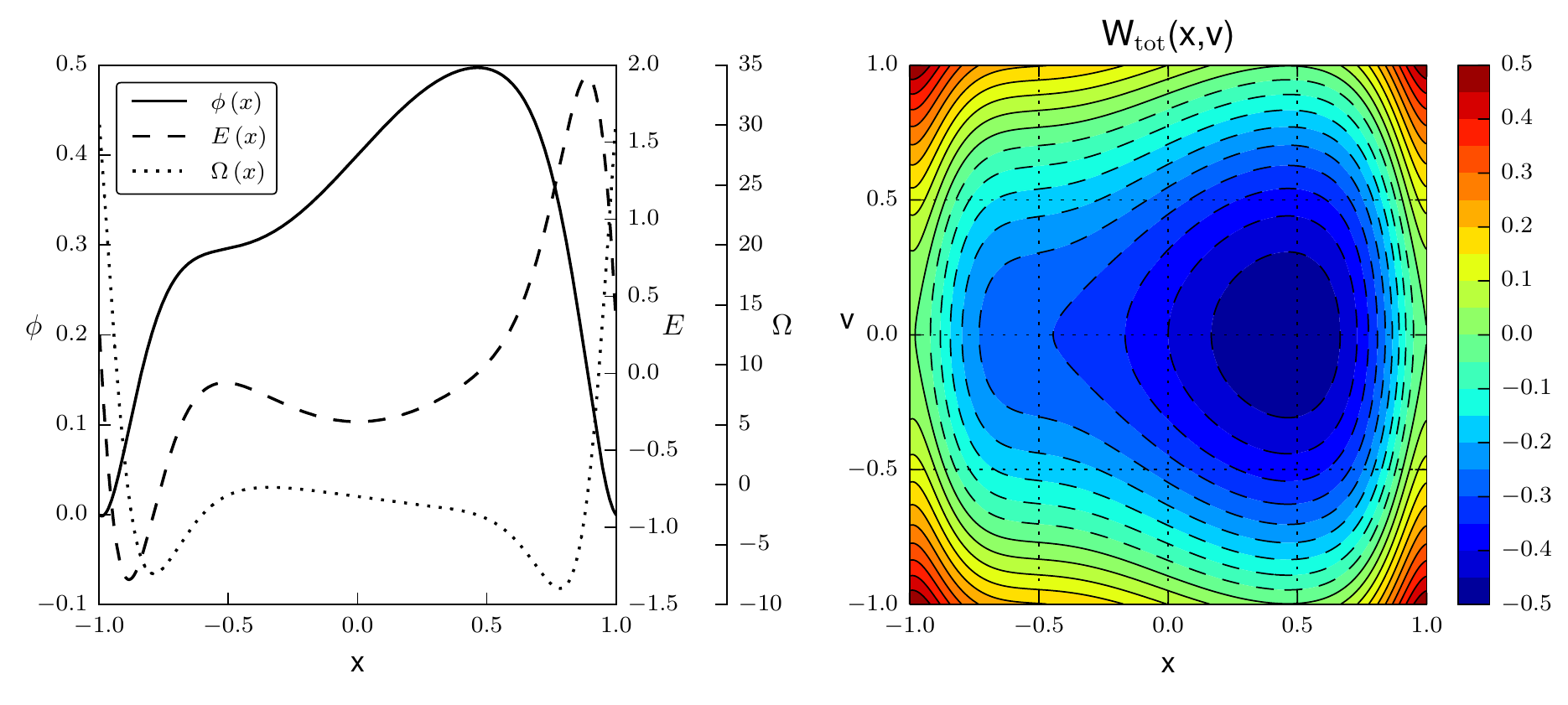}
   \caption{1D-1V linear Vlasov equation. 
   The left-hand plot shows the electrostatic potential $\phi(x)$, 
   electric field $E(x) = -d\phi/dx$, and phase-space vorticity 
   $\Omega(x) = d^2\phi/dx^2-1$ 
   (see equations \eqref{eq:tests.LinearVlasov.phi} 
   and~\eqref{eq:tests.LinearVlasov.vorticity}).
   The right-hand plot depicts the total energy 
   $W_\text{tot}(x,v) = v^2/2-\phi(x)$ 
   (sum of kinetic and potential energy) in phase-space.
   Contour levels with $W_\text{tot} < 0$ (dashed lines) identify closed 
   orbits and hence trapped electrons.
   $W_\text{tot}(x,v)$ is a stream function of the phase-space flow: the 
   volume flow rate of phase-space fluid between any two streamlines is equal 
   to the difference in total energy between them.
   }
   \label{fig:testLV.fields&total_energy}
\end{figure}

After non-dimensionalization of~\eqref{eq:tests.LinearVlasov.GeneralEq}, our 
model equation for electrons in a stationary electrostatic field is simply
\begin{equation}\label{eq:tests.LinearVlasov.eq}
  \Par{ \frac{\partial}{\partial t} + v \frac{\partial}{\partial x} +
        \frac{d\phi}{dx} \frac{\partial}{\partial v} } f(t,x,v) = 0,
  \qquad (x,v) \in [-1,1] \times [-1,1], \qquad  t\in [0,T],
\end{equation}
where the final time $T$ and the initial condition $f(0,x,v)=f_0(x,v)$ will 
depend on the test-case.
We impose a strongly asymmetric potential profile 
\begin{equation}\label{eq:tests.LinearVlasov.phi}
  \phi(x) = 0.2 + 0.2\cos(\pi x^4) + 0.1\sin(\pi x),
\end{equation}
which we plot on the left-hand side of 
Figure~\ref{fig:testLV.fields&total_energy}, together with the electric 
field $E(x)$ and the \emph{phase-space vorticity} $\Omega(x)$.
The latter quantity is explained in the next paragraph.

We notice that the 1D-1V linear Vlasov 
equation~\eqref{eq:tests.LinearVlasov.eq} may be interpreted as a 2D continuity 
equation in phase-space, with a prescribed divergence-free velocity field 
$\vect{u}(x,v)=[v,a]^T$, where $a(x) = d\phi/dx$ is the normalized acceleration.
Further, we recall that in the 2D rotating advection test case 
(Section~\ref{sec:NumericalTests.Advection2D}) the physical time-scale to be 
resolved was determined by the angular velocity $\omega$, so that in general 
$\Delta t \ll 2\pi/\omega$ for accuracy reasons.
In that case $\omega$ was proportional to the (uniform) 
vorticity field $\Omega(t,x,y) \equiv 2\omega$; by analogy, we expect that the 
smaller time scale to be resolved here is indeed determined by the phase-space 
vorticity
\begin{equation}\label{eq:tests.LinearVlasov.vorticity}
  \Omega(t,x,v) = \frac{\partial a}{\partial x} -
                  \frac{\partial v}{\partial v}
                = \frac{d^2\phi}{dx^2}-1 = \Omega(x).
\end{equation}
The Jacobian matrix of the phase-space flow field is~\cite{LandauLifshitz1987}
\begin{equation*}
  \nabla \vect{u} \,=\,
  \begin{bmatrix} 
    \frac{\partial v}{\partial x} & \frac{\partial v}{\partial v} \\[0.5em]
    \frac{\partial a}{\partial x} & \frac{\partial a}{\partial v}
  \end{bmatrix} \,=\,
  \begin{bmatrix} 
    0            & 1 \\[0.5em]
    d^2\phi/dx^2 & 0
  \end{bmatrix} \,=\,
  \begin{bmatrix} 
    D & 0 \\[0.5em]
    0 & D
  \end{bmatrix} \,+\, 
  \mat{R}_{\theta}
  \cdot\!
  \begin{bmatrix} 
    S & 0 \\[0.5em]
    0 & \!\!-S
  \end{bmatrix}
  \!\cdot
  \mat{R}_{\,-\theta} \,+\,
  \begin{bmatrix} 
    0      &\!\!-\omega \\[0.5em]
    \omega & 0
  \end{bmatrix},
\end{equation*}
where $\mat{R}_{\theta}$ is the rotation matrix corresponding to an angle of 
rotation $\theta$, from which we obtain
\begin{itemize}
  \setlength{\itemsep}{0pt}
  \item The expansion rate $D(x,v)\equiv 0$,
  \item The shear rate $S(x,v) = (d^2\phi/dx^2+1)/2 = \Omega(x)/2 + 1$,
  \item The angular velocity $\omega(x,v) = (d^2\phi/dx^2-1)/2 = \Omega(x)/2$,
  \item $\theta = 45^\circ$, hence the principal strain-rate directions are 
        $v=\pm x$.
\end{itemize}
Of course $D=0$ because the phase-space flow is incompressible.
As opposed to the 2D rotating advection, we now have a shear rate $S \neq 0$, 
which leads to \emph{filamentation}: a `blob' of fluid stretches along 
the direction of the flow, diminishing its cross-section.
Both the shear rate and the angular velocity are linearly proportional to the 
vorticity; the time step is thus chosen so that $|\Omega| \Delta t \lesssim 1$ 
everywhere in the domain.

We know that $f(t,x,v)$ is propagated along the characteristic trajectories 
$(X(t),V(t))$ in phase-space, which obey the equations of motion $\dot{X} = V$ 
and $\dot{V} = E(X)$ with initial conditions $(X(0),V(0)) = (X_0,V_0)$.
Since the electric field is conservative, $E(x)=-d\phi/dx$, the sum of the 
kinetic and potential energies along any such trajectory is a constant of 
motion, which we call `total energy' and define as
\begin{equation*}
  W_\text{tot}(X_0,V_0) := \frac{1}{2} V_0^2 - \phi(X_0) 
  \equiv \frac{1}{2} V(t)^2 - \phi(X(t)),
\end{equation*}
where the dependence on the initial conditions was made explicit.
Since $(X_0,V_0)$ are just Eulerian coordinates, a contour plot of the 
function $W_\text{tot}(x,v)$ in phase-space (see the right-hand side of 
Figure~\ref{fig:testLV.fields&total_energy}) provides us with the path 
followed by the characteristic trajectories, which are in fact orbits with 
constant total energy.

In analogy with the 2D advection equation, the total energy 
$W_\text{tot}(x,v)$ is a \emph{stream function} of the phase-space flow: 
the phase-space velocity can be obtained as the curl of a vector field 
orthogonal to the phase-plane having magnitude $W_\text{tot}$.
Moreover, since the flow is incompressible, the difference in the values of 
the total energy on two different streamlines is equal to the volume flow rate 
of fluid between them; accordingly, the flow velocity is inversely 
proportional to the local distance between the streamlines.

We notice that the constant in the potential 
profile~\eqref{eq:tests.LinearVlasov.phi} was chosen so that trajectories with 
$W_\text{tot} < 0$ describe closed orbits in phase-space, which correspond to 
`trapped' or `confined' electrons: if we give initial conditions 
to~\eqref{eq:tests.LinearVlasov.eq} that have compact support in the region of 
phase-space with $W_\text{tot} < 0$, then the exact solution will be 
indefinitely confined within the same region.
For our numerical scheme to correctly reproduce this behavior, two important 
ingredients are needed:
\begin{enumerate}
  \item The scheme should be \emph{energy stable}, i.e.\ any numerical 
        streamline should remain bounded in a finite interval 
        $[W_e-\eps,W_e+\eps]$ about its exact total energy $W_e$;
  \item The scheme should have minimal numerical diffusion in the direction 
        perpendicular to the streamlines.
\end{enumerate}
Failure to meet these criteria may lead to a secular drift in the total energy 
of our solution, and ultimately to the artificial `escape' of electrons from 
their confinement region.
We will show that our spectrally accurate 1D solver, combined with a 
high-order symplectic time-splitting, has very good performance in this regard.

In Section~\ref{sec:NumericalTests.LinearVlasov.filamentation} we will test 
the ability of our scheme to resolve the aforementioned `filamentation' 
phenomenon.
This can be a limiting factor for semi-Lagrangian solvers, which are 
required to resolve ever-shrinking features on a fixed mesh, while avoiding 
numerical diffusion in the direction perpendicular to the streamlines.
The same physical process may be less of a problem for fully Lagrangian 
(particle) schemes, in that it is perhaps easier to add extra particles in 
regions of filamentation~\cite{Christlieb2009} than to adaptively refine an 
Eulerian mesh.

In Section~\ref{sec:NumericalTests.LinearVlasov.stationary} we test the 
capacity of our scheme to preserve a steady-state solution for a very long 
period of time.
Equation~\eqref{eq:tests.LinearVlasov.eq} admits as steady-state solution any 
distribution function that depends on the phase space coordinates only through 
the total energy, i.e.\ $f(t,x,v) \equiv f_0(x,v) = g(W_\text{tot}(x,v))$.
Here the solution will be well resolved by the computational mesh; the 
challenge is keeping numerical diffusion to minimal levels.

\subsubsection{Filamentation of solution: refinement study}
\label{sec:NumericalTests.LinearVlasov.filamentation}

We now present a test-case that illustrates well the phenomenon of 
`filamentation' of the distribution function of electrons confined in an 
electrostatic field. 
With this intention, we solve the linear Vlasov 
equation~\eqref{eq:tests.LinearVlasov.eq} with the stationary electrostatic 
potential~\eqref{eq:tests.LinearVlasov.phi}, imposing an initial condition with 
compact support within the region of phase-space having negative total energy.
We use the 22nd-order cosine bell 
\begin{equation}\label{eq:tests.LinearVlasov.filamentation.ICs}
  f_0(x,v) = 
  \begin{cases}
  \cos\Par{\dfrac{\pi r}{2R}}^{22} & \text{if $r\le R$},\\
  0                               & \text{otherwise},
  \end{cases}
  \qquad
  \text{with $r = \sqrt{(x-x_c)^2 + (v-v_c)^2}$},
\end{equation}
where the radius of the bell is $R=0.75$, and the center of the profile 
is $(x_c,v_c) = (-0.2,0)$.
For this test-case, the final time is $T=3.2$.

As usual, we use the F22-CS as our base 1D advection solver, combined with 
a symmetric splitting method from Table~\ref{table:TimeSplittingIntegrators}.
Since the initial condition~\eqref{eq:tests.LinearVlasov.filamentation.ICs} 
is a function of class $\mathcal{C}^{21}\Par{\mathbb{R}^2}$, the truncation 
error of the F22 algorithm is nominally $O\Par{\dx^{22},\dv^{22}}$.
We employ a mesh of $(\N_x,\N_v) = (1024,1024)$ cells, which guarantees that 
the (exact) final solution is still fully resolved on the grid.

\begin{figure}[htb!]
   \centering
   \includegraphics[width=\textwidth]
     {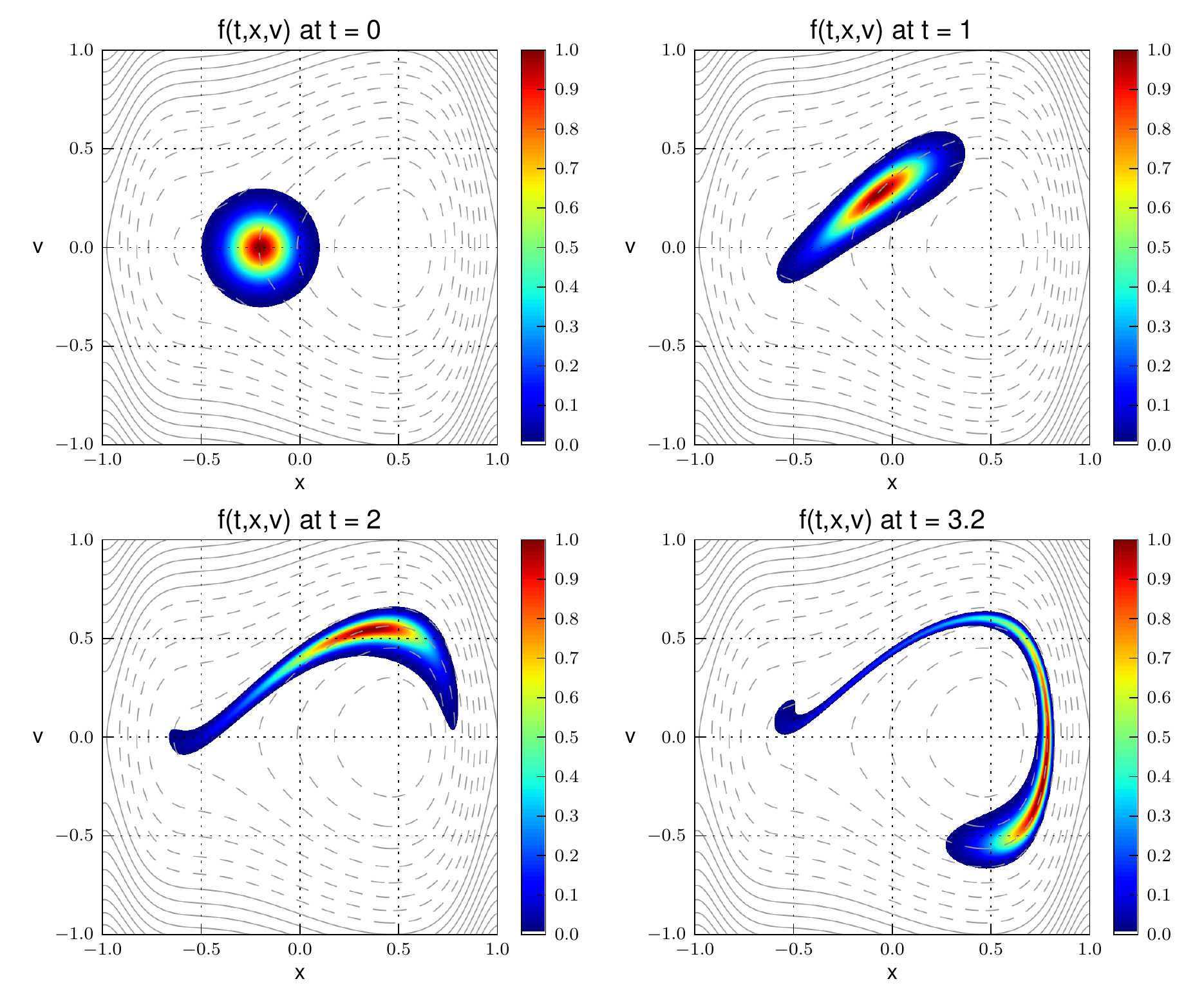}
   \caption{1D-1V linear Vlasov equation: filamentation of the initial bell 
   profile by the highly inhomogeneous phase-space flow field.
   We solve the initial value problem~\eqref{eq:tests.LinearVlasov.eq}, with 
   potential profile~\eqref{eq:tests.LinearVlasov.phi} and initial 
   condition~\eqref{eq:tests.LinearVlasov.filamentation.ICs}, until the final 
   time $T=3.2$.
   We show here the reference solution at the time instants 
   $t_k\in\{0,1,2,3.2\}$, obtained by the method of characteristics on a mesh 
   of $(\N_x,\N_v)=(1024,1024)$ cells.
   The Lagrangian trajectories are integrated backward in time using the 
   Runge-Kutta-Nyström method O11-6 from 
   Table~\ref{table:TimeSplittingIntegrators} with a time-step $\dt = 0.025$.
   The first contour level is at $f(t_k,x,v)=0.01$.
   }
   \label{fig:testLV.filamentation.snapshots}
\end{figure}

In order to accurately estimate the error in our numerical simulations, we 
first compute a reference solution by using the method of characteristics: 
since the electric field is given, we can trace each individual trajectory 
ending on a grid point back in time from $t=3.2$ to $t=0$, and then sample the 
initial condition.
Here the sampling process introduces no error, because the initial 
condition~\eqref{eq:tests.LinearVlasov.filamentation.ICs} 
is known analytically at any point in phase-space.
Further, very accurate time integration of the trajectories (within machine 
accuracy) can be easily achieved by taking very small time-steps with a high 
order symplectic integrator: for this purpose we use the O11-6 scheme from 
Table~\ref{table:TimeSplittingIntegrators} with $\dt = 0.025$.
Figure~\ref{fig:testLV.filamentation.snapshots} illustrates 4 successive 
snapshots of the reference solution: the initial condition at time $t=0$, 
intermediate solutions at $t=1$ and $t=2$, and the final result at $t=3.2$.

\begin{figure}[htb!]
   \centering
   \includegraphics[width=0.75\textwidth]
     {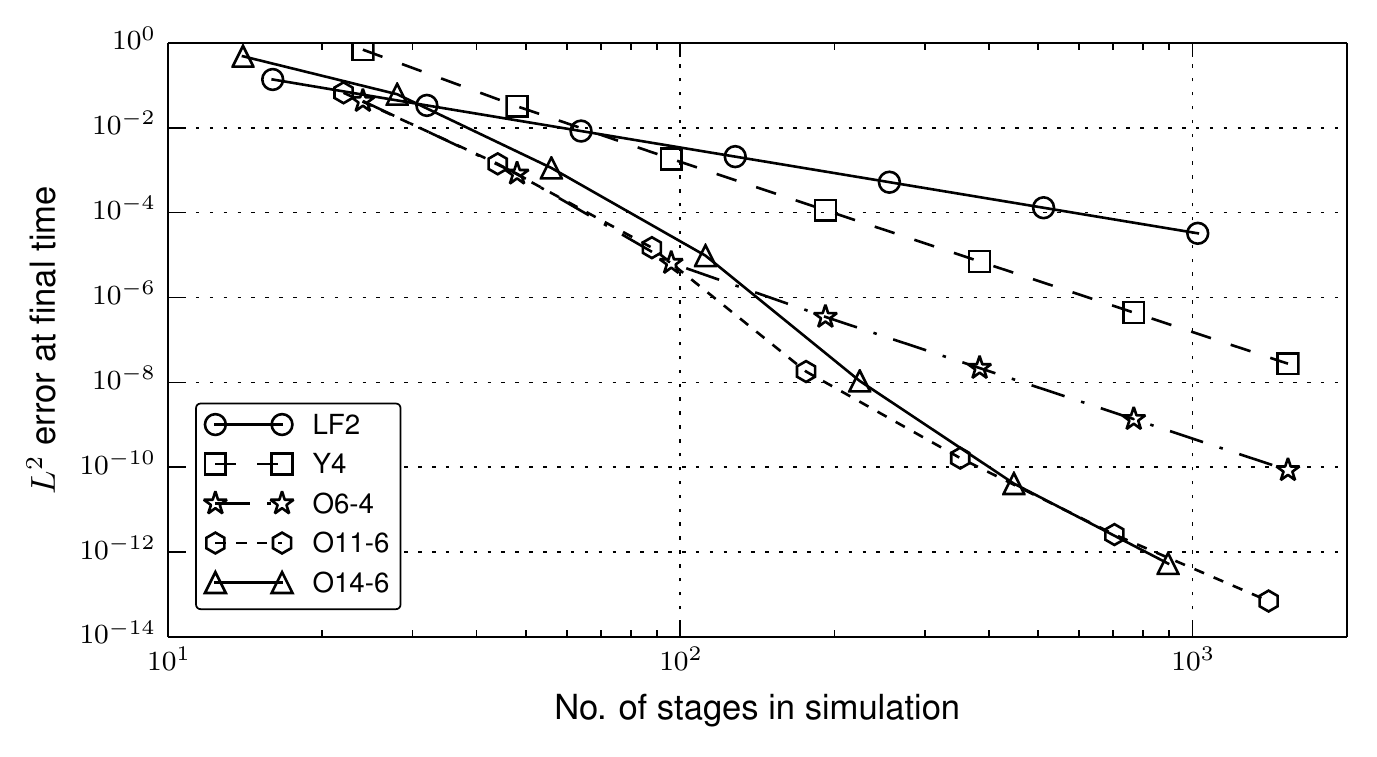}
   \caption{1D-1V linear Vlasov equation: comparison of symmetric splitting 
   time integration schemes.
   We solve the initial value problem~\eqref{eq:tests.LinearVlasov.eq}, with 
   potential profile~\eqref{eq:tests.LinearVlasov.phi} and initial 
   condition~\eqref{eq:tests.LinearVlasov.filamentation.ICs}, until the final 
   time $T=3.2$.
   All simulations employ the F22 scheme as base 1D solver, and a phase-space 
   discretization of $(\N_x,\N_y) = (1024,1024)$ cells.
   The $y$ axis reports the $L^2$-norm of the error at the final time, with 
   respect to the reference solution in 
   Figure~\ref{fig:testLV.filamentation.snapshots}; the $x$ axis reports the 
   total number of stage evaluations during the simulation, to which the 
   computational cost of the scheme is linearly proportional.
   A brief description of the different integrators, complete with references, 
   is given in Table~\ref{table:TimeSplittingIntegrators}.
   }
   \label{fig:testLV.filamentation.efficiency}
\end{figure}

Once the reference solution is computed, we perform a refinement study in the 
$\Delta t$ parameter for the five splitting methods from 
Table~\ref{table:TimeSplittingIntegrators}.
The efficiency of these methods is compared through the error-work diagram 
in Figure~\ref{fig:testLV.filamentation.efficiency}: as observed in the 2D 
rotating advection test case (see Section~\ref{sec:NumericalTests.Advection2D} 
and Figure~\ref{fig:test2d.cos22cross.efficiency}), the most efficient scheme 
overall is O11-6, an optimized Runge-Kutta-Nyström method of order 6 by Blanes 
and Moan~\cite{BlanesMoan2002}.
The outstanding efficiency of the O11-6 scheme is matched by O6-4 for errors 
above $10^{-5}$, and by O14-6 for errors below $10^{-10}$.
Moreover, in Table~\ref{table:testLV.convergence} we report the $L^2$-norm of 
the errors at the final time, and the convergence rates based on those, for 
LF2 (Strang splitting), O6-4 and O11-6.

\begin{table}[htbp!]
   \centering
		{\setlength{\bigstrutjot}{1.5pt}
		\begin{tabular}{|c|cc|cc|cc|}
		\hline
		\bigstrut[t]
		 & \multicolumn{2}{c|}{\bf{LF2}} & 
		   \multicolumn{2}{c|}{\bf{O6-4}} & 
		   \multicolumn{2}{c|}{\bf{O11-6}}\\
		\cline{2-7}
		\bigstrut[t]
		$\bm{\Delta t}$ & $\bm{L^2}$ \bf{error} & \bf{Order} 
		                & $\bm{L^2}$ \bf{error} & \bf{Order}
		                & $\bm{L^2}$ \bf{error} & \bf{Order} \\
		\hline
		\hline
		\bigstrut[t]
			$1.60\times 10^{0\phantom{-}}$  & & & & & 
			$6.67\times 10^{-2\phantom{0}}$ & ---
		\\
			$8.00\times 10^{-1}$ & & & 
			$4.26\times 10^{-2\phantom{0}}$ & --- & 
			$1.41\times 10^{-3\phantom{0}}$ & $5.56$
		\\
			$4.00\times 10^{-1}$ & & & 
			$8.24\times 10^{-4\phantom{0}}$ & $5.69$ & 
			$1.49\times 10^{-5\phantom{0}}$ & $6.56$
		\\
			$2.00\times 10^{-1}$ & 
			$1.37\times 10^{-1}$ & --- & 
			$6.50\times 10^{-6\phantom{0}}$ & $6.99$ & 
			$1.80\times 10^{-8\phantom{0}}$ & $9.70$
		\\
			$1.00\times 10^{-1}$ & 
			$3.36\times 10^{-2}$ & $2.03$ & 
			$3.49\times 10^{-7\phantom{0}}$ & $4.22$ & 
			$1.65\times 10^{-10}$ & $6.77$
		\\
			$5.00\times 10^{-2}$ & 
			$8.36\times 10^{-3}$ & $2.01$ & 
			$2.18\times 10^{-8\phantom{0}}$ & $4.00$ & 
			$2.60\times 10^{-12}$ & $5.99$
		\\
			$2.50\times 10^{-2}$ & 
			$2.09\times 10^{-3}$ & $2.00$ & 
			$1.36\times 10^{-9\phantom{0}}$ & $4.00$ & 
			$7.04\times 10^{-14}$ & $5.20$
		\\
			$1.25\times 10^{-2}$ & 
			$5.21\times 10^{-4}$ & $2.00$ & 
			$8.53\times 10^{-11}$ & $4.00$ & &
		\\
			$6.25\times 10^{-3}$ & 
			$1.30\times 10^{-4}$ & $2.00$ & & & &
		\\
			$3.125\times 10^{-3}$ & 
			$3.26\times 10^{-5}$ & $2.00$ & & & & \\
		\hline
		\end{tabular}}
   \caption{1D-1V linear Vlasov equation: refinement analysis for three 
   different time-splitting integrators.
   We solve the initial value problem~\eqref{eq:tests.LinearVlasov.eq}, with 
   potential profile~\eqref{eq:tests.LinearVlasov.phi} and initial 
   condition~\eqref{eq:tests.LinearVlasov.filamentation.ICs}, until the final 
   time $T=3.2$.
   All simulations employ the F22 scheme as base 1D solver, and a phase-space 
   discretization of $(\N_x,\N_y) = (1024,1024)$ cells.
   The table reports the $L^2$-norm of the error at the final time, for 
   progressively smaller time-step size $\dt$.
   The `Order' columns refer to the algebraic order of convergence, computed 
   as the base-2 logarithm of the ratio of two successive error norms.
   A brief description of the different integrators, complete with references, 
   is given in Table~\ref{table:TimeSplittingIntegrators}.
   }
   \label{table:testLV.convergence}
\end{table}

\subsubsection{Steady-state solution: conservation of the invariants}
\label{sec:NumericalTests.LinearVlasov.stationary}

We now focus on another challenging test-case for mesh-based solvers: 
minimizing the secular deviation from a steady-state condition caused by 
numerical diffusion across the streamlines.
Again, we use the linear Vlasov equation~\eqref{eq:tests.LinearVlasov.eq} 
subject to a stationary electrostatic 
potential~\eqref{eq:tests.LinearVlasov.phi} to describe the dynamics of 
`trapped' electrons.
Differently than the previous test-case, we now give an initial condition that 
is consistent with a steady-state solution, and we perform a very long time 
integration to verify the capacity of our numerical scheme to maintain the 
distribution function within the `trapping region'.

As previously discussed, the characteristic trajectories have constant total 
energy; hence, the contour levels of the total energy function 
$W_\text{tot}(x,v) = v^2/2 - \phi(x)$ are stationary streamlines of the 
phase-space flow.
Since a steady-state solution to~\eqref{eq:tests.LinearVlasov.eq} is constant 
along each streamline, such a solution must explicitly depend on the total 
energy itself, i.e.\ $f(t,x,v) \equiv f_0(x,v) = g\Par{W_\text{tot}(x,v)}$.
Accordingly, we give the initial condition in the form of a 22nd-order cosine 
bell that is a function of the total energy only, as
\begin{equation}\label{eq:tests.LinearVlasov.stationary.ICs}
  f_0(x,v) = 
  \begin{cases}
  \cos\Par{\dfrac{\pi r}{2R}}^{22} & \text{if $r\le R$},\\
  0                               & \text{otherwise},
  \end{cases}
  \qquad
  \text{with $r(x,v) = 1-\frac{W_\text{tot}(x,v)}{W_\text{tot}^\text{(min)}}$},
\end{equation}
where the radius of the bell is $R=0.9$ and the minimum total energy in the 
domain is $W_\text{tot}^\text{(min)} \approx -0.4972446622065941$ within 16 
digits of accuracy.
For this test-case, the final time is $T=1000$.

As usual, we use the F22-CS as our base 1D advection solver.
We employ a phase-space mesh of $(\N_x,\N_v) = (256,256)$ cells, and we evolve 
the solution with the O6-4 time-splitting scheme from 
Table~\ref{table:TimeSplittingIntegrators} and a relatively large time-step 
$\dt = 0.2$.

\begin{figure}[htb!]
   \centering
   \includegraphics[width=\textwidth]{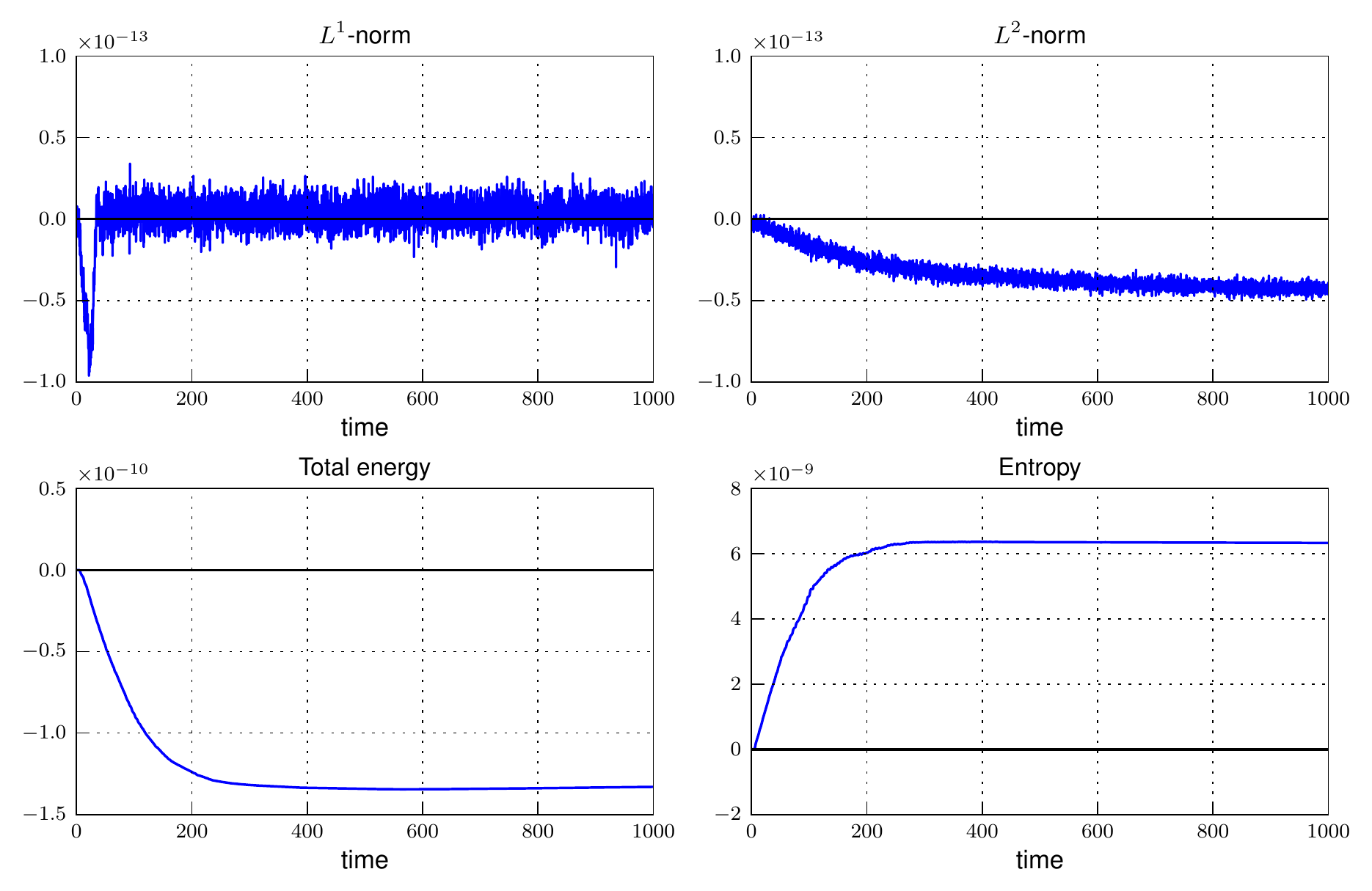}
   \caption{1D-1V linear Vlasov equation: long time preservation of a 
   stationary solution.
   We solve the initial value problem~\eqref{eq:tests.LinearVlasov.eq}, with 
   potential profile~\eqref{eq:tests.LinearVlasov.phi} and initial 
   condition~\eqref{eq:tests.LinearVlasov.stationary.ICs}, until the final 
   time $T=1000$.
   The F22 scheme is used as the base 1D solver, with a phase-space 
   discretization of $(\N_x,\N_v) = (256,256)$ cells.
   The solution is evolved in time by the O6-4 time-splitting scheme from 
   Table~\ref{table:TimeSplittingIntegrators} with a time-step $\dt = 0.2$.
   }
   \label{fig:testLV.steady-state.invariants}
\end{figure}

For this specific problem, the exact solution is stationary and coincides with 
the initial condition~\eqref{eq:tests.LinearVlasov.stationary.ICs}.
In general, an exact solution to the linear Vlasov 
equation~\eqref{eq:tests.LinearVlasov.eq} is time-dependent but has 
infinitely-many invariants, e.g.\ all the $L^p$-norms in phase-space, the total 
energy, and the entropy.
Therefore, we assess the error in the numerical simulation by monitoring the 
following four \emph{discrete} invariants,
\begin{subequations}\label{eq:tests.LinearVlasov.discrete-invariants}
\begin{align}
	\label{eq:tests.LinearVlasov.L1-norm}
	I_1(t_k) &= \sum_{i,j} |f_{i,j}^k| \dx\dv & \text{$L^1$-norm}, \\
	\label{eq:tests.LinearVlasov.L2-norm}
	I_2(t_k) &= 
       \biggl[\sum_{i,j} \Par{f_{i,j}^k}^2 \dx\dv \biggr]^{\frac{1}{2}}
       & \text{$L^2$-norm}, \\
    \label{eq:tests.LinearVlasov.TotEn}
    I_W(t_k) &= 
       \sum_{i,j} f_{i,j}^k \left[v_j^2/2-\phi(x_i)\right] \dx\dv
       & \text{Total energy}, \\
    \label{eq:tests.LinearVlasov.Entropy}
    I_\eta(t_k) &= \sum_{i,j} f_{i,j}^k \ln\Par{f_{i,j}^k+\eps} \dx\dv
       & \text{Entropy},
\end{align}
\end{subequations}
where $f_{i,j}^k \approx f\Par{t_k,x_i,v_j}$ is the numerical solution at time 
$t_k$ and location $\Par{x_i,v_j}$.
In~\eqref{eq:tests.LinearVlasov.Entropy} we employ the small tolerance 
$\eps \approx 2.23\times 10^{-308}$ (the smallest floating-point real positive 
number that can be used in double precision operations) to avoid taking the 
natural logarithm of zero.
In Figure~\ref{fig:testLV.steady-state.invariants} we report the relative 
errors in the above invariants during the simulation, computed as
\begin{equation}
   E[I](t_k) \,=\, \frac{I(t_k) - I(0)}{I(0)}.
\end{equation}
Accordingly, a positive (negative) value in the relative error corresponds to 
the invariant being larger (smaller) than it was at $t=0$.

\subsection{1D-1V Vlasov-Poisson system}
\label{sec:NumericalTests.VlasovPoisson}

We will now assess the behavior of the spectral-CS for the one-dimensional 
Vlasov-Poisson system.
In the following numerical tests we will evolve the distribution function of a 
single species, of negative charge; a constant homogeneous background of 
positive charge ensures global charge neutrality in the domain, and allows for 
the use of periodic boundary conditions.
Such a simple model is able to describe the self-consistent dynamics of 
electrons in an infinite, homogeneous and quasi-neutral plasma; the 
contribution of the slow and heavy positive ions to fluctuations in the 
electrostatic field is neglected.

The Vlasov-Poisson system~\eqref{eq:VlasovPoisson-e} for electrons is 
non-dimensionalized in \ref{sec:Appendix-B}, and it is reduced to 1D-1V 
(i.e.\ one dimension in configuration space and one in velocity space) in 
\ref{sec:Appendix-C}.
After truncation of the velocity domain, the model equations are
\begin{equation}\label{eq:tests.VlasovPoisson.Eq}
  \begin{alignedat}{2}
  \Par{ \frac{\partial}{\partial t} + v \frac{\partial}{\partial x} -
        E \frac{\partial}{\partial v} } f(t,x,v) = 0,
  &\qquad&
  (x,v) \in [-L,L] \times [-V,V], \\
  \frac{\partial E(t,x)}{\partial x} =\, n_0 -\int_{\!-V}^{\,V} f(t,x,v)\, dv,
  && t\in [0,T], \\
  \end{alignedat}
\end{equation}
with periodic boundary conditions in $(x,v)$ and initial condition 
$f(0,x,v) = f_0(x,v)$.
The domain size $(2L,2V)$, the final time $T$ and the initial condition $f_0$ 
will depend on the test-case.
In all numerical examples, the instantaneous electric field is calculated by 
means of a simple pseudo-spectral Fourier method, which we describe in 
\ref{sec:Appendix-D}.

Similarly to the linear Vlasov equation, the exact solution to the 
Vlasov-Poisson system~\eqref{eq:tests.VlasovPoisson.Eq} has infinitely many 
invariants.
Again, we will monitor the \emph{discrete} $L^1$-norm, $L^2$-norm, total 
energy, and entropy: the definitions are the same as 
in~\eqref{eq:tests.LinearVlasov.discrete-invariants}, but with a subtle 
difference in the total energy $I_W$, which now reads
\begin{equation}\label{eq:tests.VlasovPoisson.TotEn}
  I_W(t_k) = \sum_i 
  \Par{\frac{1}{2} \sum_j f_{i,j}^k v_j^2 \dv + \frac{1}{2}\Par{E_i^k}^2} \dx.
\end{equation}

As usual, the selection of the discretization parameters $\Delta x$, $\Delta v$ 
and $\Delta t$ in the numerical simulations is based on accuracy 
considerations.
Accordingly, the Courant number plays no role in this process.
Nevertheless, as a means of comparison to other mesh-based solvers, we report 
the maximum Courant numbers for advection along $x$ and $v$, defined as
\begin{equation}\label{eq:CxCv}
  C_x := \max_j |v_j| \frac{\Delta t}{\Delta x},
  \qquad
  C_v := \max_{i,k} \left| E_i^k \right| \frac{\Delta t}{\Delta v}.
\end{equation}

In Section~\ref{sec:NumericalTests.VlasovPoisson.linear-landau} we simulate 
the `linear' Landau damping, where the electrostatic energy decays 
exponentially in time according to well-known linear theory.
The spectral-CS accurately reproduces the analytical results, even using a 
coarse mesh and taking large time-steps.

In Section~\ref{sec:NumericalTests.VlasovPoisson.bump-on-tail} we simulate 
the `bump-on-tail' instability, which is a strongly non-linear test-case where 
the electrostatic energy increases in time.
For this challenging problem we present a time refinement study for a solution 
that is fully resolved in phase-space.

In Section~\ref{sec:NumericalTests.VlasovPoisson.bump-on-tail.long-time} we 
extend the simulation of the `bump-on-tail' instability to hundreds of plasma 
oscillation periods, and we observe the slow convergence toward a stable 
asymptotic solution, due to the high-order numerical diffusion intrinsic to 
our scheme.

\subsubsection{Linear Landau damping}
\label{sec:NumericalTests.VlasovPoisson.linear-landau}
Landau damping is the collisionless damping of electrostatic waves in a 
plasma, where electrostatic energy is converted into electron kinetic energy 
through phase-mixing.
In apparent contrast to the Hamiltonian nature of the Vlasov-Poisson system, 
Landau damping drives the system from a perturbed \emph{stable} equilibrium 
($f(t,x,v) = f_\text{eq}(v) + \delta f(t,x,v)$) 
towards a spatially homogenous asymptotic state (as $t\to \infty$).
This fundamental process was mathematically predicted by Landau in 
1946~\cite{Landau1946}, and it was confirmed experimentally in 1964 for both 
ions~\cite{Wong1964} and electrons~\cite{Malmberg1964}; it has been studied 
theoretically, experimentally and numerically ever since.
We refer the reader to the recent work by Mouhot and 
Villani~\cite{MouhotVillani2011} for an extensive review of the theory.

In his pioneering work~\cite{Landau1946}, Landau solved the Cauchy problem for 
the \emph{linearized} Vlasov-Poisson system in a periodic domain, with a 
perturbed Maxwellian distribution as initial condition, and he showed the 
electric field decays exponentially as $t\to \infty$.
More specifically, for a mode of wavenumber $k$, the electric field oscillates 
at a frequency $\omega_r(k) \in \mathbb{R}$ while its amplitude decays at an 
exponential rate $\gamma(k) \in \mathbb{R}^+$, i.e.\ 
$E_k(x,t)\propto \exp{(i\omega t)}$ with $\omega(k) = \omega_r(k)+i\gamma(k)$, 
where $i := \sqrt{-1}$.
The process described by this linear theory is properly called `linear Landau 
damping', and its most notable characteristic is that the complex frequency 
$\omega(k)$ depends on the equilibrium distribution $f_\text{eq}(v)$ but not 
on the initial perturbation $\delta f_0(x,v)$, as long as this is sufficiently 
smooth.

In the case of the fully \emph{nonlinear} Vlasov-Poisson system, the general 
existence of Landau damping was established in 2010 by Mouhot and 
Villani~\cite{MouhotVillani2011}, who concluded that Landau damping is indeed a linear 
phenomenon, which survives nonlinear perturbation due to the structure of the 
Vlasov–Poisson system. 
Specifically, the nonlinearity manifests itself by the presence of `plasma 
wave echoes'~\cite{Gould1967}.
On the one hand the asymptotic behavior is in general different from the limit 
predicted by the linear theory, and depends on the initial condition.
On the other hand the linearized system, or higher-order expansions, provide 
a good approximation, as they belong to a convergent Newton iteration that can 
be used to approximate the long-time limit with arbitrarily high precision.

Here we intend to test our (nonlinear) Vlasov-Poisson solver on initial 
conditions that consist of a small perturbation (1 \% in amplitude) over a 
Maxwellian distribution.
The resulting long-time dynamics is accurately described by the linear theory 
of Landau damping, to the results of which we compare our numerical simulation.
Accordingly, we solve the non-dimensional Vlasov-Poisson 
system~\eqref{eq:tests.VlasovPoisson.Eq} on the square domain $[-L,L] \times [-V,V]$ 
with $L = 2\pi$ and $V = 7$, and initial condition
\begin{equation}\label{eq:tests.VlasovPoisson.linear-landau}
  f_0(x,v) = \frac{1}{\sqrt{2\pi}} 
             \bigl( 1+\eps\cos(k\,x) \bigr) e^{-v^2\!/2},
  \qquad \text{with $\eps = 0.01$, $k = 0.5$},
\end{equation}
until the final time $T=60$.
The same test case has been used by several authors before 
(e.g.\ see~\cite{Filbet2001,Qiu2010,Rossmanith2011}), but we truncate the 
velocity domain at a higher $V$ in order to obtain a reference solution 
that is accurate within machine precision.
For the initial condition~\eqref{eq:tests.VlasovPoisson.linear-landau}, the 
linear theory predicts an oscillation frequency $\omega_r \approx 1.41566$ 
and a damping rate $\gamma \approx 0.153359$.
In the numerical tests we will monitor the decay of the (discrete) 
electrostatic energy in the domain,
\begin{equation}\label{eq:tests.VlasovPoisson.linear-landau.ElectricEnergy}
  W_e = \frac{\Delta x}{2} \sum_i \Par{E_i^k}^2, 
\end{equation}
as well as the conservation of the usual discrete invariants ($L^1$-norm, 
$L^2$-norm, total energy, and entropy).

We run two simulations: one we call `converged', meaning that its numerical 
error is dominated by machine round-off; the other we call `numerical', as it 
uses a coarser mesh and a much larger time-step, which are typical of 
real-world applications.
Both simulations use the F22-CS as their base 1D constant advection solver, 
but they employ different time integrators.
The numerical parameters for the two simulations, as well as the maximum 
Courant parameters observed, are reported in 
Table~\ref{table:testVP.linear-landau.parameters}.

\begin{table}[htb!]
   \centering
   {\setlength{\bigstrutjot}{1pt}
   \begin{tabular}{|c||c|c||c|c|c||c|c|}
   \hline
   \bigstrut[t]
   \bf{Label in figures} & \bf{Advection solver} & \bf{Time integrator} & 
   \bf{N}$\bm{_x}$ & \bf{N}$\bm{_v}$ & $\bm{\Delta t}$ & $\bm{C_x}$ & $\bm{C_v}$\\
   \hline
   \hline
   \bigstrut[t]
   `Numerical' & F22 & O6-4  &  8 & 256 & 1.0 & 4.4  & 0.34  \\
   `Converged' & F22 & O11-6 & 16 & 512 & 0.1 & 0.89 & 0.072 \\
   \hline
   \end{tabular}}
   \caption{1D-1V Vlasov-Poisson system, linear Landau damping: numerical 
   parameters employed in the simulations.
   We solve the initial value problem~\eqref{eq:tests.VlasovPoisson.Eq} with 
   $L=2\pi$ and $V=7$, with periodic boundary conditions along $x$ and $v$ 
   and initial condition~\eqref{eq:tests.VlasovPoisson.linear-landau}, until 
   the final time $T=60$.
   The 1D constant advection solver is described in 
   Section~\ref{sec:HighOrderCS.implementation.trigonometric}, and the time 
   integrators are described in Table~\ref{table:TimeSplittingIntegrators}.
   The $C_x$ and $C_v$ columns report the maximum Courant parameters along $x$ 
   and $v$, respectively, according to~\eqref{eq:CxCv}.
   }
   \label{table:testVP.linear-landau.parameters}
\end{table}

\begin{figure}[htb!]
   \centering
   \includegraphics[width=0.75\textwidth]
     {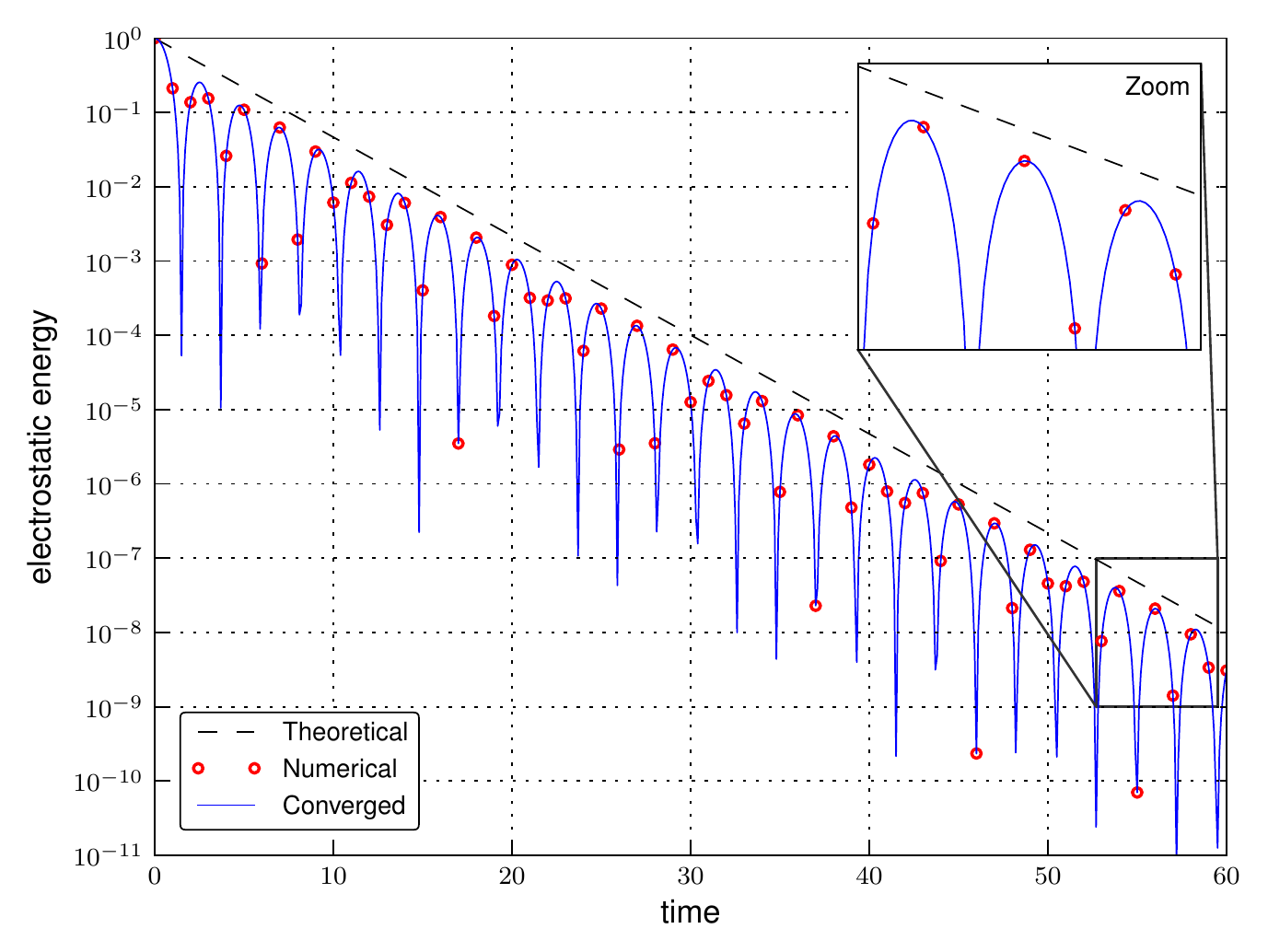}
   \caption{1D-1V Vlasov-Poisson system, linear Landau damping: exponential 
   time decay of the electrostatic energy in the domain.
   We solve the initial value problem~\eqref{eq:tests.VlasovPoisson.Eq} with 
   $L=2\pi$ and $V=7$, with periodic boundary conditions along $x$ and $v$ 
   and initial condition~\eqref{eq:tests.VlasovPoisson.linear-landau}, until 
   the final time $T=60$.
   We plot the electrostatic energy in the domain, computed according 
   to~\eqref{eq:tests.VlasovPoisson.linear-landau.ElectricEnergy}, for a fully 
   resolved (`Converged', solid blue) and a less demanding calculation 
   (`Numerical', red circles), which use the parameters given in 
   Table~\ref{table:testVP.linear-landau.parameters}.
   The `Theoretical' line (dashed black) corresponds to the exponential decay 
   in the amplitude of the oscillation ($\gamma = 0.153359$), as 
   predicted by Landau's linear theory~\cite{Landau1946}.
   }
   \label{fig:testVP.linear-landau.energy_decay}
\end{figure}

In Figure~\ref{fig:testVP.linear-landau.energy_decay} we plot the time 
evolution of the electrostatic energy in the domain.
As expected, the `converged' solution shows excellent agreement with the 
exponential decay predicted by the linear theory.
More surprisingly, the `numerical' solution is practically indistinguishable 
from the converged one, despite the very large time-step employed (the full 
simulation requires only 60 time-steps).
This makes a strong case for the use of optimized high-order time splitting 
integrators in conjunction with accurate semi-Lagrangian solvers.

\begin{figure}[htb!]
   \centering
   \includegraphics[width=\textwidth]{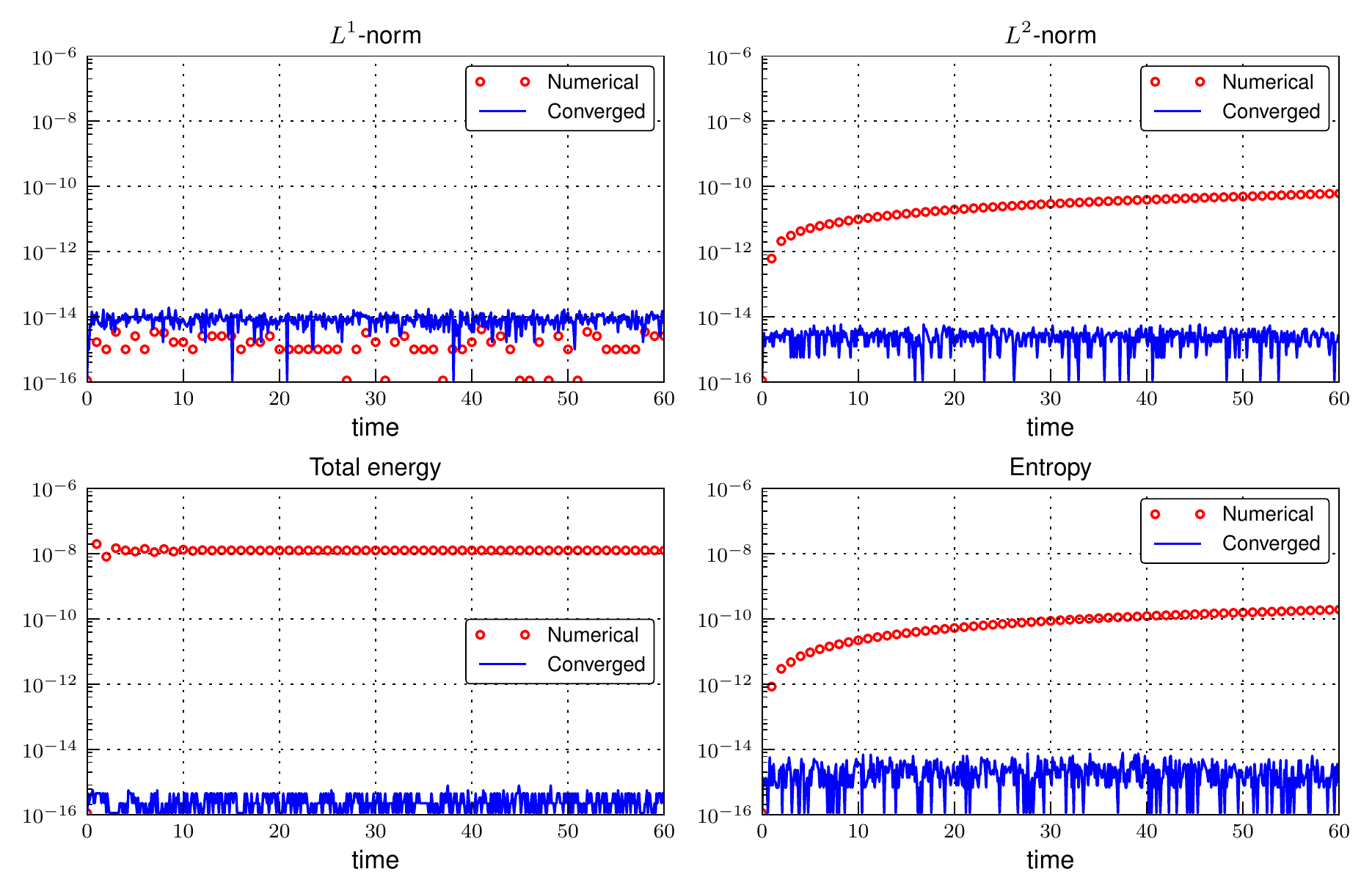}
   \caption{1D-1V Vlasov-Poisson system, linear Landau damping: relative 
   errors in the discrete conserved quantities.
   We solve the initial value problem~\eqref{eq:tests.VlasovPoisson.Eq} with 
   $L=2\pi$ and $V=7$, with periodic boundary conditions along $x$ and $v$ 
   and initial condition~\eqref{eq:tests.VlasovPoisson.linear-landau}, until 
   the final time $T=60$.
   We plot the results for a fully resolved (`Converged', solid blue) and a 
   less demanding calculation (`Numerical', red circles), which use the 
   parameters given in Table~\ref{table:testVP.linear-landau.parameters}.
   The conserved quantities are computed according to equations 
   \eqref{eq:tests.LinearVlasov.L1-norm},
   \eqref{eq:tests.LinearVlasov.L2-norm},
   \eqref{eq:tests.VlasovPoisson.TotEn} 
   and~\eqref{eq:tests.LinearVlasov.Entropy}, respectively.
   }
   \label{fig:testVP.linear-landau.conservations}
\end{figure}

In Figure~\ref{fig:testVP.linear-landau.conservations} we plot the relative 
errors in the conservation of the discrete invariants.
Since the CS is mass-conservative and positivity-preserving, the $L^1$-norm 
will be exactly conserved (so far as roundoff error permits in double 
precision), no matter what numerical parameters are employed; the other 
invariants, instead, will be approximate, and become exact only in the limit 
as $\Delta x,\Delta v,\Delta t \to 0$.
For the `converged' simulation, all invariants are conserved to roundoff error.
For the `numerical' simulation, the conservation of the $L^1$-norm is even 
better (the smaller number of operations makes roundoff less severe), while 
the truncation error shows up in the other invariants.
The $L^2$-norm and the entropy show a very small deviation, linear in time, 
from their initial values, and at the end of the simulation their relative 
errors are of the order of $10^{-10}$; this is achieved thanks to the spectral 
accuracy of the F22 advection solver, which has virtually no numerical 
diffusion for `fully-resolved' wavelengths $l \gtrsim 3\Delta x$.
The relative error in the total energy is approximately $10^{-8}$ throughout 
the whole simulation, and does not increase with time; this is achieved thanks 
to the energy-stability of the symplectic time integrator, combined with the 
spectral accuracy of the CS.

\subsubsection{Bump-on-tail instability: time refinement study}
\label{sec:NumericalTests.VlasovPoisson.bump-on-tail}
The Vlasov-Poisson system admits a class of homogeneous equilibria that are 
linearly unstable, in the sense that a small wave-like perturbation with will 
tend to grow in time, under certain conditions on its wavelength, until some 
kind of nonlinear saturation is obtained.
In fact, one of the most fundamental instabilities in plasma physics is the 
so-called `bump-on-tail' instability: a wave perturbation whose phase velocity 
lies along the positive slope of a bump on the tail of the distribution 
function becomes unstable.
As a consequence, electron kinetic energy is converted into electrostatic 
energy, but the instability eventually saturates: in the asymptotic state the 
phase-space shows a Bernstein-Greene-Kruskal (BGK) vortex structure traveling 
at the phase-velocity of the wave~\cite{Bernstein1957}.

Here we want to test the ability of our solver to correctly reproduce this 
linear instability and its nonlinear evolution to the BGK mode.
Accordingly, we solve the normalized Vlasov-Poisson 
system~\eqref{eq:tests.VlasovPoisson.Eq} on the square domain 
$[-L,L] \times [-V,V]$ with dimensions $L=2\pi/0.3$ and $V=8$, 
and initial condition
\begin{equation}\label{eq:tests.VlasovPoisson.bump-on-tail}
  f_0(x,v) = \frac{1}{\sqrt{2\pi}} 
             \Bigl( 1+0.04\cos(0.3\,x) \Bigr)
             \Bigl( 0.9\,e^{-v^2/2} + 0.2\,e^{-4(v-4.5)^2} \Bigr).
\end{equation}
As usual, we apply periodic boundary conditions along $x$ and $v$.
This is the same test-case setup used by Arber and Vann~\cite{Arber2002}, 
Banks and Hittinger~\cite{Banks2010,Hittinger2013}, and 
Seal~\cite{Seal2012_Thesis}.
Other authors have used initial conditions similar 
to~\eqref{eq:tests.VlasovPoisson.bump-on-tail}, but including multiple 
wavelengths of the initial density perturbation; this was achieved by means 
of a larger domain and/or a smaller wavelength 
(e.g.\ see~\cite{Nakamura1999,Crouseilles2010}). 
We notice that in~\cite{Hittinger2013} the maximum velocity was increased 
from 8 to 10 to allow for the asymmetric `bump'; while this was probably 
necessary for a long time simulation with $T=100$, it did not appear to effect 
the numerical error in our shorter simulations.

\begin{figure}[htb!]
   \centering
   \includegraphics[width=\textwidth]{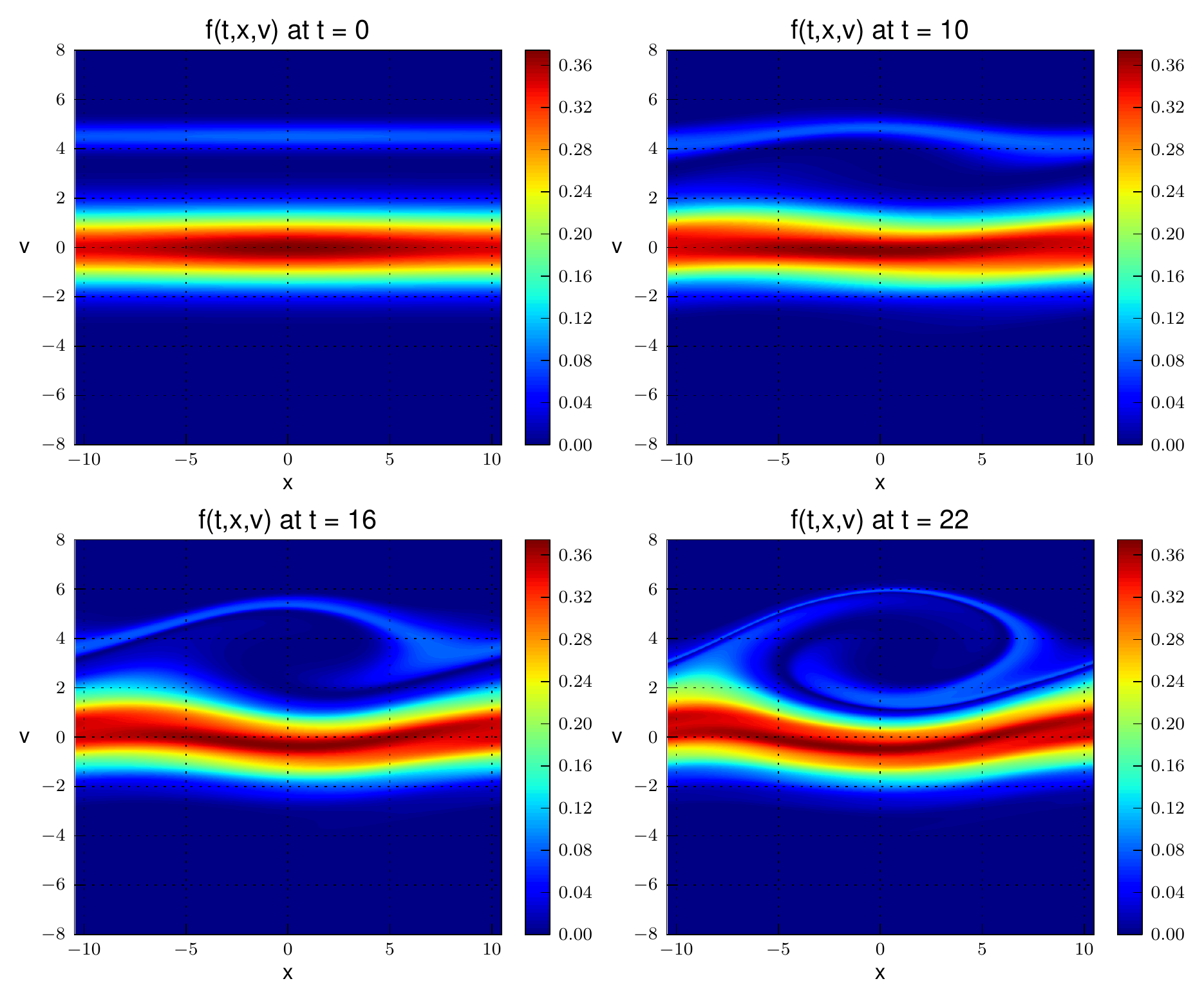}
   \caption{1D-1V Vlasov-Poisson system, `bump-on-tail': numerical solution at 
   successive time instants.
   We solve the initial value problem~\eqref{eq:tests.VlasovPoisson.Eq} with 
   $L=10\pi/3$ and $V=8$, with periodic boundary conditions along $x$ and $v$ 
   and initial condition~\eqref{eq:tests.VlasovPoisson.bump-on-tail}, until 
   the final time $T=22$.
   We employ the F22 scheme as base 1D solver, and the O6-4 time splitting 
   algorithm from Table~\ref{table:TimeSplittingIntegrators}.
   The phase-space mesh has $(\N_x,\N_y) = (256,512)$ cells, and the 
   time-step is $\Delta t = 0.5$.
   The maximum Courant parameters along $x$ and $v$, according 
   to~\eqref{eq:CxCv}, are $(C_x,C_v) \approx (49,9.4)$.
   }
   \label{fig:testVP.bump-on-tail.snapshots}
\end{figure}

In Figure~\ref{fig:testVP.bump-on-tail.snapshots} we show four consecutive 
snapshots of the solution obtained with the F22-CS as base 1D solver, 
and the 4th-order Runge-Kutta-Nyström (RKN) time splitting algorithm O6-4 by 
Blanes and Moan~\cite{BlanesMoan2002} 
(see Table~\ref{table:TimeSplittingIntegrators}).
We employed a phase-space discretization of $(\N_x,\N_v) = (256,512)$ cells, 
and a constant time-step $\Delta t = 0.5$.
By the end of the simulation ($t=22$) the BGK vortex is fully developed, and 
travels at a velocity $v_\phi \approx 3.5$.
Despite the relatively coarse mesh and the large time-step employed (only 44 
steps to the final solution), the fine features of the vortex are accurately 
captured by our numerical scheme.

\begin{figure}[htb!]
   \centering
   \includegraphics[width=\textwidth]{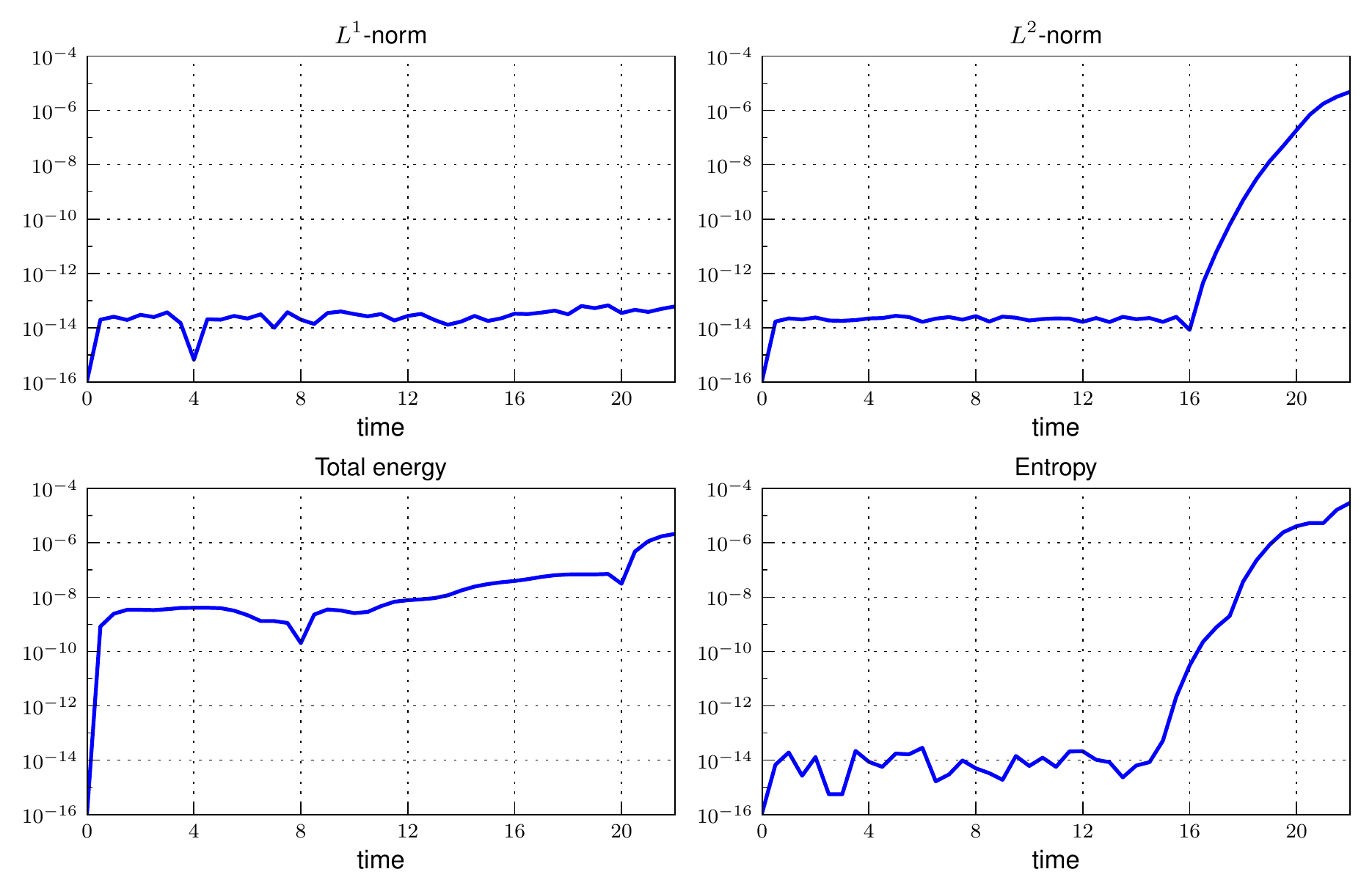}
   \caption{1D-1V Vlasov-Poisson system, `bump-on-tail': relative error in the 
   discrete conserved quantities.
   We solve the initial value problem~\eqref{eq:tests.VlasovPoisson.Eq} with 
   $L=10\pi/3$ and $V=8$, with periodic boundary conditions along $x$ and $v$ 
   and initial condition~\eqref{eq:tests.VlasovPoisson.bump-on-tail}, until 
   the final time $T=22$.
   We employ the F22 scheme as base 1D solver, and the O6-4 time splitting 
   algorithm from Table~\ref{table:TimeSplittingIntegrators}.
   The phase-space mesh has $(\N_x,\N_y) = (256,512)$ cells, and the 
   time-step is $\Delta t = 0.5$.
   The maximum Courant parameters along $x$ and $v$, according 
   to~\eqref{eq:CxCv}, are $(C_x,C_v) \approx (49,9.4)$.
   The conserved quantities are computed according to equations 
   \eqref{eq:tests.LinearVlasov.L1-norm},
   \eqref{eq:tests.LinearVlasov.L2-norm},
   \eqref{eq:tests.VlasovPoisson.TotEn} 
   and~\eqref{eq:tests.LinearVlasov.Entropy}, respectively.
   }
   \label{fig:testVP.bump-on-tail.conservations}
\end{figure}

In Figure~\ref{fig:testVP.bump-on-tail.conservations} we plot the relative 
errors in the conservation of the discrete invariants.
As usual, the $L^1$-norm is only affected by round-off errors, while the 
other invariants are also affected by the truncation error of the numerical 
scheme.
The error in total energy does not show important secular growth, thanks to 
the stability of the symplectic time-splitting integrator.
The $L^2$-norm and the entropy are conserved within machine precision until 
time $t=16$ and $t=15$, respectively; at later times the error appears to grow 
linearly with time, but much faster than in the linear Landau damping case, 
presumably because of the fast rate of filamentation of the nonlinear vortex 
structure.

We now repeat the same simulation with a finer phase-space mesh, which has 
$(\N_x,\N_y) = (1024,1024)$ cells, until the final time $t=16$.
This permits us to fully resolve the phase-space features for the entirety of 
the simulation: we obtain a `reference' solution by employing the 6th-order 
RKN integrator O11-6 (see Table~\ref{table:TimeSplittingIntegrators}) with a 
very small time-step $\Delta t=0.05$, which ensures that all invariants are 
properly conserved (so far as roundoff error allows).
We compute the errors in the other simulations with respect to this reference 
solution.

\begin{figure}[htb!]
   \centering
   \includegraphics[width=0.75\textwidth]{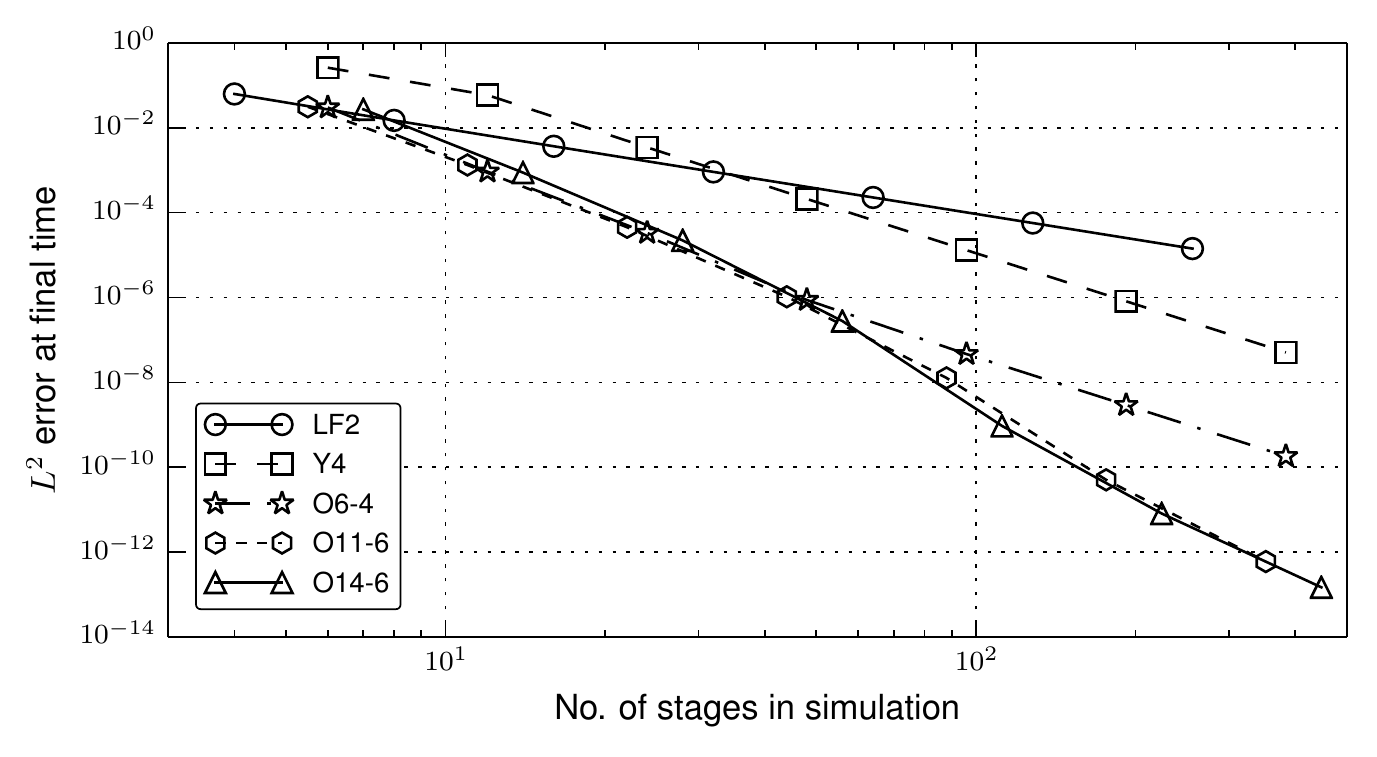}
   \caption{1D-1V Vlasov-Poisson system, `bump-on-tail': comparison of 
   symmetric splitting time integration schemes.
   We solve the initial value problem~\eqref{eq:tests.VlasovPoisson.Eq} with 
   $L=10\pi/3$ and $V=8$, with periodic boundary conditions along $x$ and $v$ 
   and initial condition~\eqref{eq:tests.VlasovPoisson.bump-on-tail}, until 
   the final time $T=16$.
   All simulations employ the F22 scheme as base 1D solver, and a phase-space 
   discretization of $(\N_x,\N_y) = (1024,1024)$ cells.
   The $y$ axis reports the $L^2$-norm of the error at the final time, with 
   respect to a `fully converged' numerical solution;
   the $x$ axis reports the total number of stage evaluations during the 
   simulation, to which the computational cost of the scheme is linearly 
   proportional.
   A brief description of the different integrators, complete with references, 
   is given in Table~\ref{table:TimeSplittingIntegrators}.
   }
   \label{fig:testVP.bump-on-tail.efficiency}
\end{figure}

Figure~\ref{fig:testVP.bump-on-tail.efficiency} reports an `error-work' 
diagram for all the time-splitting integrators from 
Table~\ref{table:TimeSplittingIntegrators}: the relative $L^2$-norm of the 
error at $t=16$ is plotted against the total number of stages in the 
simulation.
For error levels above $10^{-6}$ the most efficient integrators are O6-4 and 
O11-4, with the latter only marginally better; for error levels below 
$10^{-6}$ the most efficient integrator is O14-6, but O11-4 is very close.
Table~\ref{table:testVP.convergence} reports a refinement analysis for three 
of these integrators, complete with the algebraic convergence rates.

\begin{table}[htb!]
   \centering
		{\setlength{\bigstrutjot}{1.5pt}
		\begin{tabular}{|c|cc|cc|cc|}
		\hline
		\bigstrut[t]
		 & \multicolumn{2}{c|}{\bf{LF2}} & 
		   \multicolumn{2}{c|}{\bf{O6-4}} & 
		   \multicolumn{2}{c|}{\bf{O11-6}}\\
		\cline{2-7}
		\bigstrut[t]
		$\bm{\Delta t}$ & $\bm{L^2}$ \bf{error} & \bf{Order} 
		                & $\bm{L^2}$ \bf{error} & \bf{Order}
		                & $\bm{L^2}$ \bf{error} & \bf{Order} \\
		\hline
		\hline
		\bigstrut[t]
			$6.40$ & & & & & 
			$3.12\times 10^{-2\phantom{0}}$ & ---
		\\
			$3.20$ & & & 
			$3.00\times 10^{-2\phantom{0}}$ & --- & 
			$1.33\times 10^{-3\phantom{0}}$ & $4.56$
		\\
			$1.60$ & & & 
			$9.22\times 10^{-4\phantom{0}}$ & $5.02$ & 
			$4.52\times 10^{-5\phantom{0}}$ & $4.87$
		\\
			$0.80$ & 
			$6.25\times 10^{-2}$ & --- & 
			$3.33\times 10^{-5\phantom{0}}$ & $4.79$ & 
			$1.04\times 10^{-6\phantom{0}}$ & $5.44$
		\\
			$0.40$ & 
			$1.49\times 10^{-2}$ & $2.07$ & 
			$8.75\times 10^{-7\phantom{0}}$ & $5.25$ & 
			$1.28\times 10^{-8\phantom{0}}$ & $6.35$
		\\
			$0.20$ & 
			$3.66\times 10^{-3}$ & $2.02$ & 
			$4.65\times 10^{-8\phantom{0}}$ & $4.23$ & 
			$5.06\times 10^{-11}$ & $7.98$
		\\
			$0.10$ & 
			$9.11\times 10^{-4}$ & $2.01$ & 
			$2.92\times 10^{-9\phantom{0}}$ & $3.99$ & 
			$6.00\times 10^{-13}$ & $6.40$
		\\
			$0.05$ & 
			$2.28\times 10^{-4}$ & $2.00$ & 
			$1.83\times 10^{-10}$ & $4.00$ & &
		\\
			$0.025$ & 
			$5.69\times 10^{-5}$ & $2.00$ & & & &
		\\
			$0.0125$ & 
			$1.42\times 10^{-5}$ & $2.00$ & & & & \\
		\hline
		\end{tabular}}
   \caption{1D-1V Vlasov-Poisson system, `bump-on-tail': refinement analysis 
   for three different time-splitting integrators.
   We solve the initial value problem~\eqref{eq:tests.VlasovPoisson.Eq} with 
   $L=10\pi/3$ and $V=8$, with periodic boundary conditions along $x$ and $v$ 
   and initial condition~\eqref{eq:tests.VlasovPoisson.bump-on-tail}, until 
   the final time $T=16$.
   All simulations employ the F22 scheme as base 1D solver, and a phase-space 
   discretization of $(\N_x,\N_y) = (1024,1024)$ cells.
   The table reports the $L^2$-norm of the error at the final time, for 
   progressively smaller time-step size $\dt$.
   The errors are computed with respect to the reference numerical solution, 
   obtained with the O11-6 integrator and $\Delta t=0.05$.
   The `Order' columns refer to the algebraic order of convergence, computed 
   as the base-2 logarithm of the ratio of two successive error norms.
   A brief description of the different integrators, complete with references, 
   is given in Table~\ref{table:TimeSplittingIntegrators}.
   }
   \label{table:testVP.convergence}
\end{table}

\subsubsection{Bump-on-tail instability: long-time integration}
\label{sec:NumericalTests.VlasovPoisson.bump-on-tail.long-time}

We now investigate the ability of our Vlasov-Poisson solver to properly 
dissipate small features in the solution (e.g., filamentation) when their sizes 
fall below the mesh size.
For this purpose we extend the bump-on-tail simulation, previously described 
in Section~\ref{sec:NumericalTests.VlasovPoisson.bump-on-tail}, until the 
final time $T=1000$.
As we had pointed out in our comment on 
Figure~\ref{fig:testVP.bump-on-tail.conservations}, the solution is fully 
resolved on the phase-space mesh only until time $t\approx 16$; therefore, the 
long-time evolution that we present hereafter is corrupted by the spurious 
viscosity introduced by our numerical scheme.
(We recall that the Vlasov-Poisson system is not dissipative.)

Nevertheless, by virtue of its high order numerical diffusion, we wish our 
solver to recover, to some extent at least, some steady-state solution (in a 
moving reference frame) to the Vlasov-Poisson system.
Such a solution should be `compatible' with the initial conditions, meaning 
that it has a certain number of invariants very close to the initial values.
The rationale behind this requirement is the observation that, in the presence 
of a small amount of collisionality (Coulomb collisions between electrons) the 
Boltzmann-Poisson system evolves rapidly toward a stationary solution to the 
Vlasov-Poisson system, and then only very slowly toward an equilibrium 
(Maxwellian) distribution.

In order to ease the comparison of this `instability saturation' with other 
mesh-based methods, we have repeated the same simulation using three different 
versions of the CS as our base 1D solver: P4 (polynomial, 4th order), P6 
(polynomial, 6th order), and F22 (FFT-based).
All three simulations employed the 4th-order Runge-Kutta-Nyström (RKN) time 
splitting algorithm O6-4 by Blanes and Moan~\cite{BlanesMoan2002} 
(see Table~\ref{table:TimeSplittingIntegrators}), and the spectral Poisson 
solver described in \ref{sec:Appendix-D}.
As in the previous simulation, we employed a phase-space discretization of 
$(\N_x,\N_v) = (256,512)$ cells, and a constant time-step $\Delta t = 0.5$.

\begin{figure}[htb!]
   \centering
   \includegraphics[width=\textwidth]{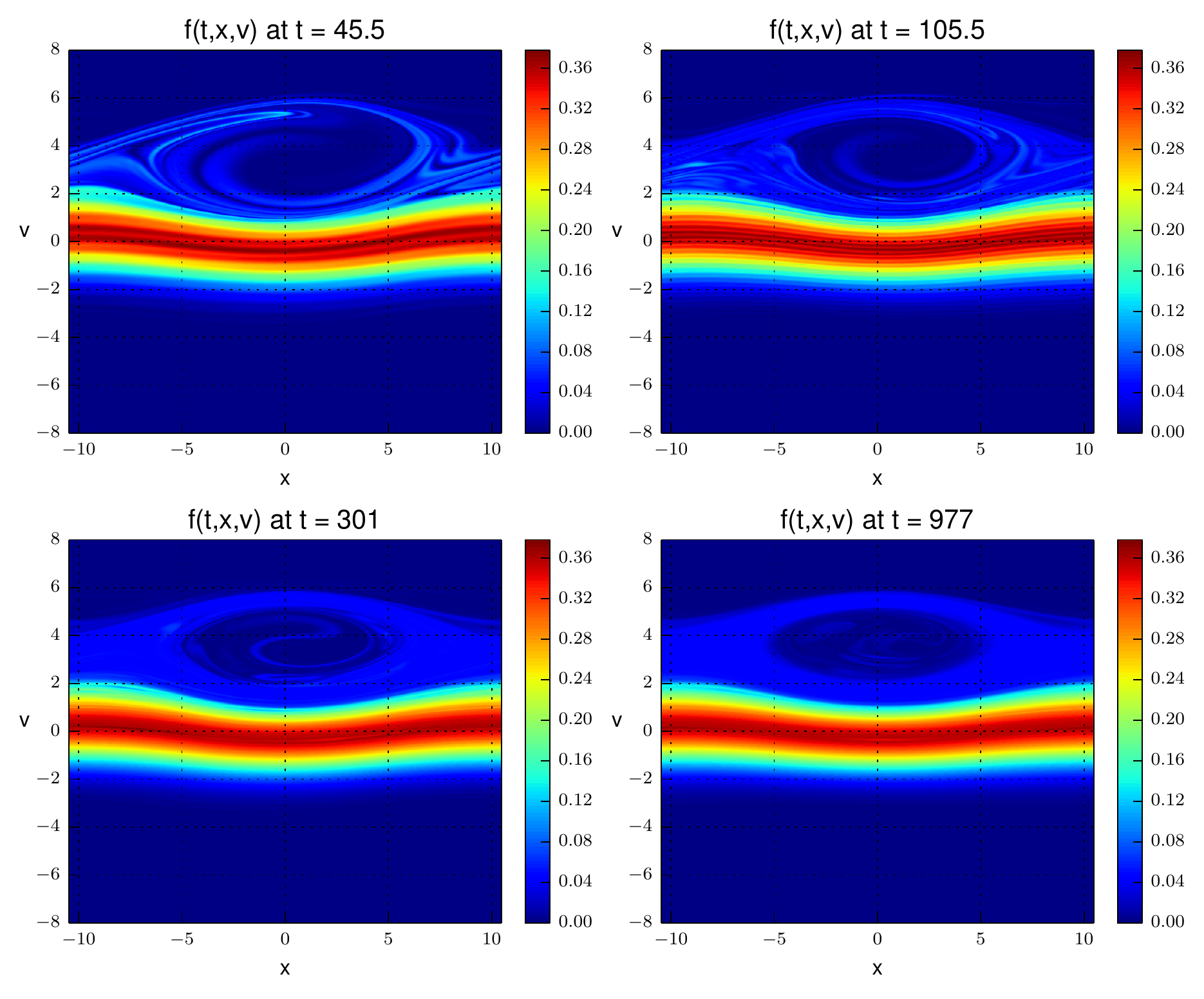}
   \caption{1D-1V Vlasov-Poisson system, `bump-on-tail', long time integration: 
   numerical solution at successive time instants.
   We solve the initial value problem~\eqref{eq:tests.VlasovPoisson.Eq} with 
   $L=10\pi/3$ and $V=8$, with periodic boundary conditions along $x$ and $v$ 
   and initial condition~\eqref{eq:tests.VlasovPoisson.bump-on-tail}, until 
   the final time $T=1000$.
   We employ the F22 scheme as base 1D solver, and the O6-4 time splitting 
   algorithm from Table~\ref{table:TimeSplittingIntegrators}.
   The phase-space mesh has $(\N_x,\N_y) = (256,512)$ cells, and the 
   time-step is $\Delta t = 0.5$.
   The maximum Courant parameters along $x$ and $v$, according 
   to~\eqref{eq:CxCv}, are $(C_x,C_v) \approx (49,9.4)$.
   }
   \label{fig:testVP.bump-on-tail.long-time.snapshots}
\end{figure}

In Figure~\ref{fig:testVP.bump-on-tail.long-time.snapshots} we show four 
consecutive snapshots of the solution obtained with the F22 scheme, at time 
instants $t_k\in\{45.5,105.5,301,977\}$.
(Since the BGK vortex is moving with a positive phase-speed 
$v_\phi\approx 3.5$, we have selected time instants when its center is near 
$x=0$.)
We can observe two different filamentation processes: one, very strong, occurs 
around the BGK vortex and is rapidly dissipated between $t=45.5$ and $t=105.5$; 
the other, much slower, is essentially linear Landau damping of the bulk 
Maxwellian, and it is only dissipated between $t=105.5$ and $t=301$.
Beyond $t=301$ the solution appears to be slowly settling to a `pure' BGK mode, 
although some minor structures are still visible at $t=977$.

\begin{figure}[htb!]
   \centering
   \includegraphics[width=\textwidth]{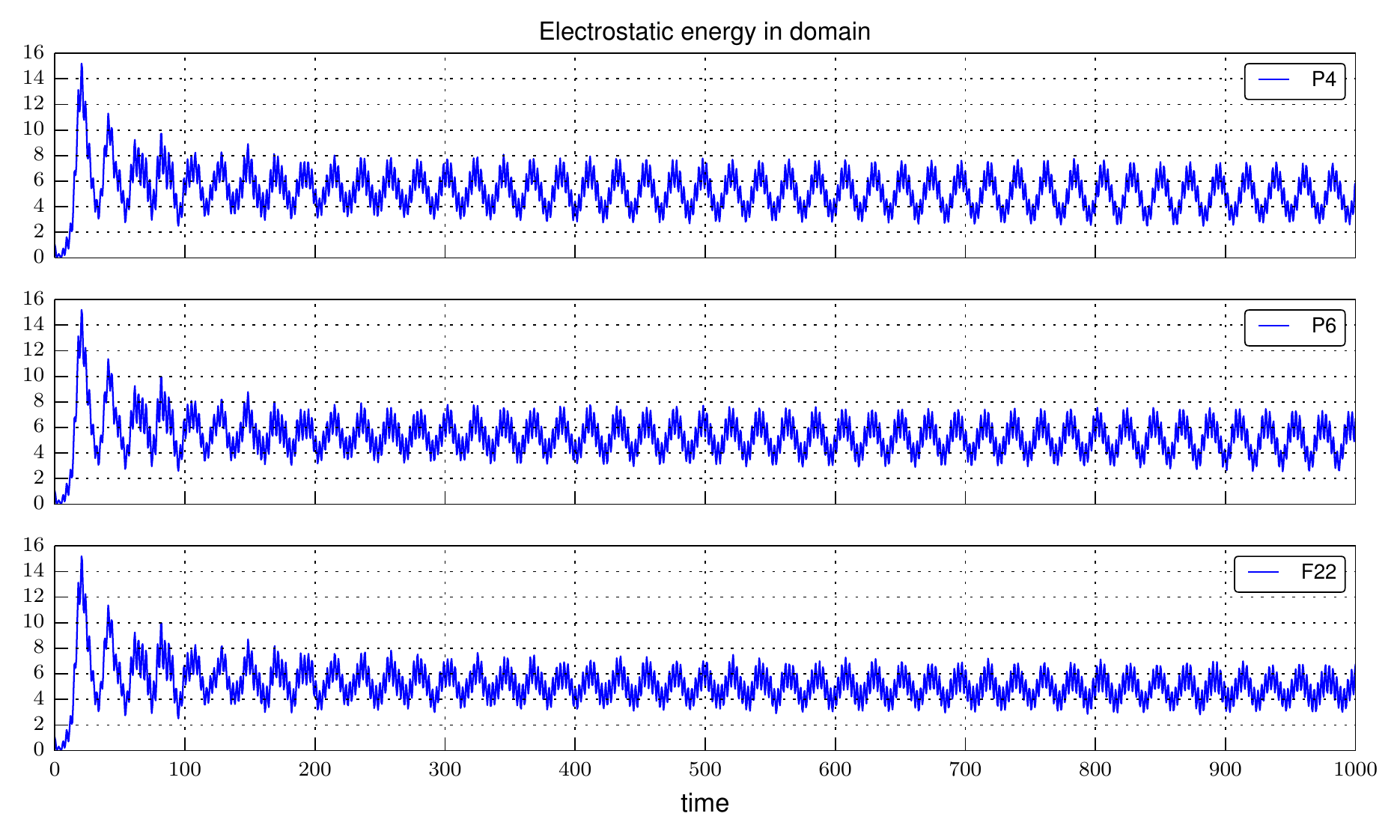}
   \caption{1D-1V Vlasov-Poisson system, `bump-on-tail', long time integration: 
   evolution of the electrostatic energy (normalized with respect to its 
   initial value) for three different versions of the CS.
   We solve the initial value problem~\eqref{eq:tests.VlasovPoisson.Eq} with 
   $L=10\pi/3$ and $V=8$, with periodic boundary conditions along $x$ and $v$ 
   and initial condition~\eqref{eq:tests.VlasovPoisson.bump-on-tail}, until 
   the final time $T=1000$.
   We compare three identical simulations employing 
   different 1D constant advection solvers: P4 and P6 are based on polynomial 
   interpolation (see Section~\ref{sec:HighOrderCS.implementation.polynomial}) 
   of order 4 and 6, respectively, and F22 is based on filtered trigonometric 
   interpolation (see Section~\ref{sec:HighOrderCS.implementation.trigonometric}). 
   All simulations employ the O6-4 time splitting algorithm from 
   Table~\ref{table:TimeSplittingIntegrators}, and the spectral Poisson solver 
   described in \ref{sec:Appendix-D}.
   The phase-space mesh has $(\N_x,\N_y) = (256,512)$ cells, and the 
   time-step is $\Delta t = 0.5$.
   The maximum Courant parameters along $x$ and $v$, according 
   to~\eqref{eq:CxCv}, are $(C_x,C_v) \approx (49,9.4)$.
   }
   \label{fig:testVP.bump-on-tail.long-time.electric-energy}
\end{figure}

In Figure~\ref{fig:testVP.bump-on-tail.long-time.electric-energy} we show the 
time evolution of the electrostatic energy for all three simulations.
For easier comparison with similar test-cases using a different domain size or 
mode number~\cite{Nakamura1999,Arber2002,Crouseilles2010}, we normalize the energy with respect to its initial value.
The results appear qualitatively very similar, but no steady-state is ever 
obtained for any of the schemes: the electrostatic energy shows small-amplitude 
plasma oscillations, superimposed on larger and slower oscillations which we 
identify as a non-linear `breathing' of the BGK mode.
The amplitude of such breathing oscillations is largest for P4 and smallest 
for F22, and for all simulations it is does not vary considerably in time.
We attribute this high degree of stability to the use of a 4th order 
symplectic time integrator.

\begin{figure}[htb!]
   \centering
   \includegraphics[width=\textwidth]{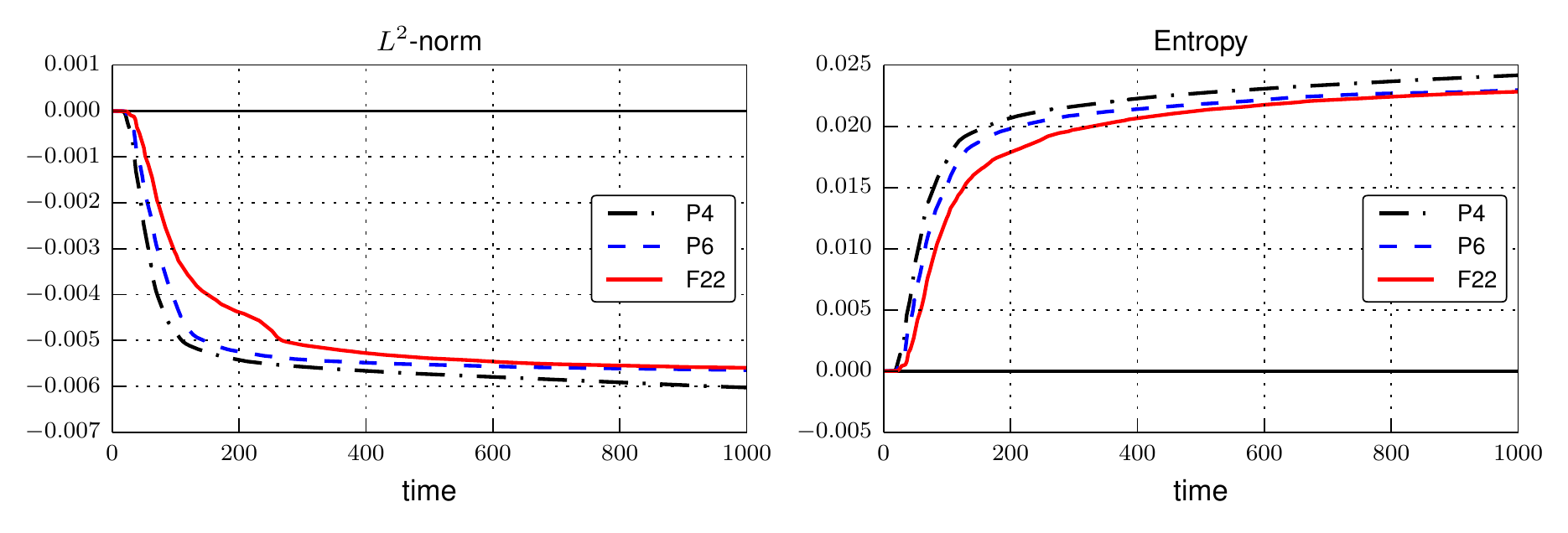}
   \caption{1D-1V Vlasov-Poisson system, `bump-on-tail', long time integration: 
   relative errors in $L^2$-norm (left) and entropy (right) for three 
   different versions of the CS.
   We solve the initial value problem~\eqref{eq:tests.VlasovPoisson.Eq} with 
   $L=10\pi/3$ and $V=8$, with periodic boundary conditions along $x$ and $v$ 
   and initial condition~\eqref{eq:tests.VlasovPoisson.bump-on-tail}, until 
   the final time $T=1000$.
   We compare three identical simulations employing different 1D constant 
   advection solvers: P4 and P6 are based on polynomial interpolation (see 
   Section~\ref{sec:HighOrderCS.implementation.polynomial}) of order 4 and 6, 
   respectively, and F22 is based on filtered trigonometric interpolation (see 
   Section~\ref{sec:HighOrderCS.implementation.trigonometric}). 
   All simulations employ the O6-4 time splitting algorithm from 
   Table~\ref{table:TimeSplittingIntegrators}, and the spectral Poisson solver 
   described in \ref{sec:Appendix-D}.
   The phase-space mesh has $(\N_x,\N_y) = (256,512)$ cells, and the 
   time-step is $\Delta t = 0.5$.
   The maximum Courant parameters along $x$ and $v$, according 
   to~\eqref{eq:CxCv}, are $(C_x,C_v) \approx (49,9.4)$.
   }
   \label{fig:testVP.bump-on-tail.long-time.L2norm&entropy}
\end{figure}

In Figure~\ref{fig:testVP.bump-on-tail.long-time.L2norm&entropy} we show the 
relative errors in the discrete conservation of $L^2$-norm and entropy.
As expected, the P4 scheme has the largest errors, and F22 has the smallest. 
Interestingly, P6 and F22 show the same trend asymptotically; this suggests 
that, if the Convected Scheme has sufficient resolution power, it will recover 
this same final BGK state, with an error that depends more on the cell size 
than on the order of accuracy.

\section{Conclusions}
    \label{sec:Conclusions}

The Convected Scheme (CS), a semi-Lagrangian algorithm for solution of 
transport equations, has been extended to 
arbitrarily high order of accuracy with minor changes in the basic scheme.
The original CS consists of an approach to remapping moving cells containing 
(phase) fluid to a fixed mesh after (one or more) time steps of the scheme.
The remapping involves finding the fractional overlaps of `moving cells' with 
fixed cells.
Since the fractions are by definition positive and sum to one, positivity and 
local conservation of density are achieved automatically.
The extended CS consists of including small changes in the cells' displacement 
and hence small changes in the fractions.
The desirable features of the basic scheme are thus retained in the extended 
scheme.

A semi-analytical theory for the high-order CS was described, based on a 
modified equation analysis, which requires the approximation of ${\N-2}$ 
spatial derivatives, with $\N$ the order of the truncation error.
Two different approaches to computing those derivatives were proposed, one 
based on polynomial interpolation, the other on filtered trigonometric 
interpolation.
For both approaches, details were provided for constructing arbitrarily 
high-order algorithms, which are strictly conservative and 
positivity-preserving.

The present work focused on the Vlasov-Poisson system, which 
describes the evolution of the velocity distribution function of a collection 
of charged particles subject to reciprocal Coulomb interactions.
The Vlasov equation was split into two constant advection equations, one in 
configuration space and one in velocity space, and high order time accuracy 
was achieved by proper composition of the operators.
Our algorithms were thoroughly tested on problems of increasing complexity: 
constant advection to analyze our CS solvers, 2D rotating advection 
to assess the time splitting error, 1D-1V linear Vlasov to study 
filamentation and preservation of steady-state solutions.
Finally, the solution of the 1D-1V Vlasov-Poisson system was illustrated on 
classic problems in kinetic theory (linear Landau damping and `bump-on-tail'
instability).

In future we intend to implement high order treatment of boundary effects and 
collisions.
A large number of important applications which were studied in the 
past~\cite{Hitchon1999} (including sheath and presheath formation, with and 
without collisions, and heating effects in radio-frequency discharges) could 
profitably be revisited with a high order scheme.
Two and three dimensional discharge simulations with high order codes will be 
the next major challenge after that.

\section*{Acknowledgments}
    \label{sec:Acknowledgments}
One of the authors (Y.G) is grateful to Matthew F. Causley and David C. Seal 
for fruitful discussions about the numerical methods and the test cases 
presented in this paper.
Additionally, since all the algorithms were implemented in Python and Cython 
code with the aid of several free open-source libraries (Sympy, Numpy, Scipy 
and Matplotlib), Y.G. would like to acknowledge the precious work of their
developers.

This work has been supported in part by the Michigan State University 
Foundation through grant SPG-RG100059, by the Air Force Office of Scientific 
Research through grants FA9550-11-1-0281, FA9550-12-1-0343 and 
FA9550-12-1-0455, and by the National Science Foundation through grant 
DMS-1115709.

\appendix
\renewcommand{\thesection}{Appendix \Alph{section}}
\section{Construction of a Fourier filter for the Spectral-CS}
\label{sec:Appendix-A}

The pseudo-spectral algorithm in 
Section~\ref{sec:HighOrderCS.implementation.trigonometric} computes high-order 
corrections for the Convected Scheme, which require multiple derivatives of 
the solution, by means of a simple and efficient FFT/IFFT procedure.
In order to obtain a stable numerical scheme for the Vlasov equation, we want 
to introduce an amount of numerical diffusion that is strictly sufficient to 
dissipate under-resolved features in the solution.
For this purpose, we multiply the Fourier spectrum of the high-order 
corrections by the Fourier spectrum of the regularized Shannon kernel (RSK) 
described in~\cite{Sun2006}.
The spatial representation of the RSK filter is
\begin{equation*} 
  K(X) = \frac{\sin \Par{\pi X}}{\pi X}
  \exp\Par{-\frac{X^2}{2\sigma^2}},
  \qquad X\in\mathbb{R},
\end{equation*}
where the normalized spatial variable is $X:=x/\Delta x$, and 
$\sigma\in\mathbb{R}^+$ determines the width of the Gaussian envelope.
The Fourier spectrum of $K(X)$ is fully contained in $[-\pi,\pi]$, it is 
constant and equal to one in a region around $0$, and it smoothly decays to 
zero as the normalized wavenumber goes to $\pm\pi$.
The `filtering' of the high frequency modes is stronger for smaller values of 
$\sigma$.
In the numerical examples presented in Section~\ref{sec:NumericalTests} we 
always use $\sigma=4$.

Since $K(X)$ is bandlimited, it is sufficient to sample it at the Nyquist 
rate, at the locations $X_i = i+\frac{1}{2}$ with $i\in\mathbb{N}$.
In order to avoid finite-window ringing, we should sample $K(X)$ over a 
symmetric interval $[-W,W]$ with $W \gg \sigma$.
In double precision we choose $W \approx 9\sigma$.
After computing the discrete Fourier transform of our $2W$ samples, the 
coefficients so obtained may have non-zero imaginary component because of a 
shift in the spatial coordinate; accordingly, we simply neglect any phase 
information, and we store only the magnitude $\hat{K}_r \in\mathbb{R}$ of our 
$2W$ Fourier modes.
Given the wavenumbers $\xi_r$ and the amplitudes $\hat{K}_r$, we construct a 
cubic spline representation $\hat{K}(\xi\Delta x)$ over the whole interval 
$[-\pi,\pi]$, which we can then sample at will at any frequency value.

\section{Non-dimensionalization}
\label{sec:Appendix-B}

The numerical examples in Section~\ref{sec:NumericalTests} employed a common 
non-dimensional version of the Vlasov-Poisson 
system~\eqref{eq:VlasovPoisson-e} for electrons in a uniform ion background. 
In that setting the $\vect{x}$ coordinates are measured 
in electron Debye lengths, the $\vect{v}$ coordinates are normalized with 
respect to the electron thermal velocity, and time $t$ is given in units 
of (electron oscillation periods)/$2\pi$.

More rigorously, such a non-dimensionalization process starts with defining a 
\emph{reference} number density $\overline{n}$ [m$^{-3}$] and a 
\emph{reference} temperature $\overline{T}$ [K], which are then 
used as normalization constants.
It is common practice to use the average electron values in the domain at 
$t=0$ as a reference.
From the macroscopic quantities $\Par{\overline{n},\overline{T}}$ one 
computes the following characteristic plasma properties:
\begin{itemize}
  \setlength{\itemsep}{0pt}
  \item The plasma oscillation (angular) frequency 
        $\omega_p = \sqrt{\bigl.\overline{n}\,q_e^2 \!
        \bigm/ \!\Par{\eps_0 m_e} \bigr.}$ [s$^{-1}$],
  \item The Debye length 
        $\lambda_D = \sqrt{\left.\eps_0 k_B \overline{T} 
        \middle/ \Par{\overline{n}\,q_e^2} \right.}$ [m],
  \item The electron thermal velocity 
        $v_{th} = \sqrt{\left.k_B \overline{T} \middle/ m_e\right.}$ [m/s],
\end{itemize}
which are then used to normalize independent variables and dependent 
quantities:
\begin{enumerate}
  \item $t$ is normalized by $(\omega_p)^{-1}$,
  \item $\vect{x}$ is normalized by $\lambda_D$,
  \item $\vect{v}$ is normalized by $v_{th} = \lambda_D \omega_p$,
  \item $\vect{E}$ is normalized by 
      $\overline{E} = \left. m_e\lambda_D(\omega_p)^2 \middle/ |q_e| \right.$,
  \item $f$ is normalized by 
        $\overline{f} = \left. \overline{n} \middle/ (v_{th})^3 \right.$,
\end{enumerate}
and of course the ion and electron densities are normalized by $\overline{n}$.
Finally, we obtain the non-dimensional system
\begin{equation*}
  \Bigl( \partial_t + \vect{v}\cdot\gradx - \vect{E}\cdot\gradv \Bigr)
       f(t,\vect{x},\vect{v}) = 0,
  \qquad \gradx\cdot\vect{E}(t,\vect{x})  = 
         n_0 - \int_{\mathbb{R}^3} f(t,\vect{x},\vect{v}) d\vect{v}.
\end{equation*}
Note that if $\overline{n}$ is chosen to equal the background ion density in 
the domain, then $n_0 \equiv 1$.

\section{1D periodic domain}
\label{sec:Appendix-C}

Assume $\vect{x} := [x,y,z]^T$ and $\vect{v} := [v_x,v_y,v_z]^T$.
If the distribution function $f$ depends on the $x$ configuration coordinate 
only, so that $\partial f/\partial y = \partial f/\partial z = 0$, 
the non-dimensional Vlasov-Poisson system reduces to
\begin{equation*}
  \Par{ \frac{\partial}{\partial t} + v_x\frac{\partial}{\partial x} 
        - E_x\frac{\partial}{\partial v_x} } f(t,x,\vect{v}) = 0,
  \qquad \frac{\partial E_x(t,x)}{\partial x} = 
         n_0 - \int_{\mathbb{R}^3} f(t,x,\vect{v}) d\vect{v}.
\end{equation*}
Here, the integral on the right-hand side of Gauss' law can be rewritten as
\begin{equation*}
  \int_{\mathbb{R}^3} f(t,x,\vect{v}) d\vect{v} = 
  \int_{\mathbb{R}} \Par{\int_{\mathbb{R}^2} f(t,x,v_x,v_y,v_z) dv_y dv_z} dv_x 
  = \int_{\mathbb{R}} f_x(t,x,v_x) dv_x,
\end{equation*}
where the \emph{1D distribution function} $f_x$ is obtained from the original 
distribution $f$ by integration over the velocity variables $v_y$ and $v_z$.
Similarly, the Vlasov equation can be integrated over $v_y$ and $v_z$ to get 
the (non-dimensional) 1D Vlasov-Poisson system
\begin{equation*}
  \Par{ \frac{\partial}{\partial t} + v_x\frac{\partial}{\partial x} 
        - E_x\frac{\partial}{\partial v_x} } f_x(t,x,v_x) = 0,
  \qquad \frac{\partial E_x(t,x)}{\partial x} = 
         n_0 - \int_{\mathbb{R}} f_x(t,x,v_x) dv_x.
\end{equation*}
In the body of this work the $x$ subscripts were removed for improved 
readability.

In a 1D periodic domain, periodicity of the electric field requires the system 
to be globally neutral in charge, i.e.\ the average densities of ions and 
electrons in the domain must be identical and constant in time.

\section{Poisson solver}
\label{sec:Appendix-D}

After each advection step in configuration space, we want to obtain the 
instantaneous electric field $E(x)$ at each grid location $x=x_i$, by 
satisfying a discrete form of Gauss' law
\begin{equation*}
  \frac{\partial E(x)}{\partial x} = n_0 - \int_\mathbb{R} f(x,v) dv,
  \qquad x \in[a,b],
\end{equation*}
with spectral accuracy.
For brevity we have dropped the dependence on the time variable $t$, and we 
refer to the right hand side of this equation as the net charge density 
$\rho(x)$.
Since the domain is periodic and the mesh is uniform, it is natural to resort 
to a pseudo-spectral Fourier method~\cite{Boyd2001}.

We note that the solution $E(x)$ is unique up to an additive constant.
In order to maintain the total energy constant in time, we must impose the 
average value of the electric field in the domain be zero; equivalently, we can 
impose periodic boundary conditions on the electrostatic potential $\phi(x)$.
In a pseudo-spectral solver, this is easily achieved by setting to zero the 
constant Fourier mode of the electric field.

First, we need to approximate the integral on the right hand side to obtain 
the electron density $n(x_i)$, and in so doing we have to satisfy a discrete 
form of the charge neutrality condition $\int_a^b \rho(x)dx = 0$.
This is simply achieved by using the midpoint rule 
$n_i = \sum_j f_{i,j} \Delta v$, which is consistent with the discrete mass 
conservation property of the Convected Scheme, because
\begin{equation*}
  (b-a)n_0 = \sum_{i,j} f_{i,j} \Delta x \Delta v =
  \sum_i \Biggl(\sum_j f_{i,j} \Delta v\Biggr) \Delta x = \sum_i n_i \Delta x .
\end{equation*}
When applied to the integration of periodic functions over an entire period on 
a uniform grid, the midpoint rule has `optimal' order of 
convergence~\cite{Kurganov2009}; in particular, it is spectrally accurate for 
smooth functions.

We now assume that $E(x)$ and $\rho(x)$ are approximated by the trigonometric 
polynomials $E_T(x)$ and $\rho_T(x)$, which interpolate the $N$ grid values 
$\{E_i\}$ and $\{\rho_i\}$.
Since the mesh is uniform, the coefficients of such polynomials are obtained 
by taking the discrete Fourier transform (DFT) of the grid values, so that
\begin{equation*}
  \begin{alignedat}{2}
  E_T(x) &= \frac{1}{N}\sum_{r=0}^{N-1} \hat{E}_r\, e^{\,\iu r\xi(x)},
  &\qquad
  \hat{E}_r &= \sum_{i=0}^{N-1} E_i\,\omega^{-ir}, \\
  \rho_T(x) &= \frac{1}{N}\sum_{r=0}^{N-1} \hat{\rho}_r\, e^{\,\iu r\xi(x)},
  &
  \hat{\rho}_r &= \sum_{i=0}^{N-1} \rho_i\,\omega^{-ir}, \\
  \end{alignedat}
\end{equation*}
where $\omega = e^{2\pi\iu/N}$ is the $N^\text{th}$ primitive root of unity, 
and $\iu := \sqrt{-1}$ is the imaginary unit.
$\xi \in [0,2\pi]$ is the canonical spatial coordinate that results from the 
coordinate transformation
\begin{equation*}
  \xi(x) = 2\pi\frac{x-a}{b-a}, \qquad \xi : [a,b] \to [0,2\pi].
\end{equation*}

If we ask that $E_T(x)$ and $\rho_T(x)$ exactly satisfy Gauss' law in the 
domain, we find that the relation $\iu r c\,\hat{E}_r=\hat{\rho}_r$, with 
$c=2\pi/(b-a)$, must hold for each Fourier mode $r$.
Finally, the complete algorithm is simply as follows:
\begin{enumerate}
  \item Compute the charge density in the domain
        \[ \rho_i = n_0 - \sum_j f_{i,j}\, \Delta v
        \]
  \item Compute the discrete Fourier transform of $\{\rho_i\}$ 
        using an FFT algorithm
        \[ \hat{\rho}_r = \sum_{i=0}^{N-1} \rho_i\,\omega^{-ir}
        \]
  \item Compute the $N$ Fourier coefficients of the electric field
        \[ \hat{E}_0 = 0, \qquad 
           \hat{E}_r = \frac{\hat{\rho}_r}{\iu r c} 
           \quad \text{for $r \neq 0$}
        \]
  \item Compute the inverse discrete Fourier transform of $\{\hat{E}_r\}$ 
        using an IFFT algorithm
        \[ E_i = \frac{1}{N} \sum_{r=0}^{N-1} \hat{E}_r\,\omega^{ir}
        \]
\end{enumerate}

\newpage
\section*{}
    \bibliographystyle{elsarticle-num-names-Y}
    \bibliography{GucluChristliebHitchon2013}


\end{document}